\documentclass[]{aa}

\usepackage{graphicx}
\usepackage{multicol}
\usepackage{txfonts}
\usepackage[outdir=./]{epstopdf}
\usepackage{tikz}
\usetikzlibrary{shapes,arrows}
\begin{document} 

\tikzstyle{decision} = [diamond, draw, fill=green!20, 
    text width=7.em, text badly centered, node distance=3.cm, inner sep=0pt]
\tikzstyle{block} = [rectangle, draw, fill=red!20, 
    text width=7em, text centered, rounded corners, minimum height=4em]
\tikzstyle{line} = [draw, -latex']
\tikzstyle{cloud} = [draw, ellipse,fill=red!20, node distance=3cm,
    minimum height=2em]

   \title{The reports of thick discs' deaths are greatly exaggerated}
   \subtitle{Thick discs are NOT artefacts caused by diffuse scattered light}

   \author{S.~Comer\'on
          \inst{1}, H.~Salo\inst{1},
          \and
          J.~H.~Knapen\inst{2,3}
          }

   \institute{University of Oulu, Astronomy Research Unit, P.O.~Box 3000, FI-90014, Finland\\
              \email{seb.comeron@gmail.com}
              \and
              Instituto de Astrof\'isica de Canarias, E-38205 La Laguna, Tenerife, Spain
              \and
              Departamento de Astrof\'isica, Universidad de La Laguna, E-38200 La Laguna, Tenerife, Spain
             }

 
  \abstract{Recent studies have made the community aware of the importance of accounting for scattered light when examining low-surface-brightness galaxy features such as thick discs. In our past studies of the thick discs of edge-on galaxies in the {\it Spitzer} Survey of Stellar Structure in Galaxies -- the S$^4$G -- we modelled the point spread function as a Gaussian. In this paper we re-examine our results using a revised point spread function model that accounts for extended wings out to more than $2\farcm5$. We study the $3.6\mu{\rm m}$ images of 141 edge-on galaxies from the S$^4$G and its early-type galaxy extension. Thus, we more than double the samples examined in our past studies. We decompose the surface-brightness profiles of the galaxies perpendicular to their mid-planes assuming that discs are made of two stellar discs in hydrostatic equilibrium. We decompose the axial surface-brightness profiles of galaxies to model the central mass concentration -- described by a S\'ersic function -- and the disc -- described by a broken exponential disc seen edge-on. Our improved treatment fully confirms the ubiquitous occurrence of thick discs. The main difference between our current fits and those presented in our previous papers is that now the scattered light from the thin disc dominates the surface brightness at levels below $\mu\sim26\,{\rm mag\,arcsec^{-2}}$. We stress that those extended thin disc tails are not physical, but pure scattered light. This change, however, does not drastically affect any of our previously presented results: {\it 1)} Thick discs are nearly ubiquitous. They are not an artefact caused by scattered light as has been suggested elsewhere. {\it 2)} Thick discs have masses comparable to those of thin discs in low-mass galaxies -- with circular velocities $v_{\rm c}<120\,{\rm km\,s^{-1}}$ -- whereas they are typically less massive than the thin discs in high-mass galaxies. {\it 3)} Thick discs and central mass concentrations seem to have formed at the same epoch from a common material reservoir. {\it 4)} Approximately 50\% of the up-bending breaks in face-on galaxies are caused by the superposition of a thin and a thick disc where the scale-length of the latter is the largest.}


   \maketitle
%

\section{Introduction}

Charge-coupled device detectors and space-based telescopes have greatly increased the depth of astronomical imaging. This has opened the door to the detailed study of faint galaxy components such as thick discs and stellar haloes. However, with great observational power come great observational challenges. Indeed, point spread functions (PSFs) often have extended wings that can greatly affect the interpretation of the properties of the above-mentioned galaxy components. In a worst-case scenario, the scattered light from high-surface-brightness features may create artefacts that could be interpreted as faint components that actually do not exist. The uncomfortable reality of diffuse scattered light has been put on the agenda by \citet{JONG08} and \citet{SAN14, SAN15}.

Thick discs are one of the galaxy features whose details require deep imaging to be unveiled, at least in galaxies other than the Milky Way. They were first described by \citet{BURS79} and \citet{TSI79}. Thick discs have a larger scale-height and a lower surface brightness than the thin discs of the galaxies in which they are hosted. The importance of studying them stems from the fact that they are made of old stars, so they are the remains of the early processes that shaped the baryonic component of the universe. However, whether thick discs are a distinct component or whether they are the oldest and most vertically extended tail of the disc star distribution is yet to be known.

Thick discs have been detected in resolved stellar count studies \citep[e.g.,][]{GIL83, MOU05, SETH05, TI05}. Strong evidence for the existence of thick discs also comes from spectroscopic studies that find vertical gradients in disc stellar ages and/or metallicities \citep{YOA08, CO15, CO16, KAS16}. Thus, there is not much support for a position denying the existence of thick discs. However, it is justifiable to question how badly scattered light affects the determination of the properties of thick discs in integrated light studies such as those made by \citet{YOA06} and \citet{CO11B, CO12}. In these papers the surface-brightness profiles of edge-on galaxies are decomposed into two components. They conclude that thick discs, especially those in low-mass hosts -- galaxies with circular velocities $v_{\rm c}\lesssim120\,{\rm km\,s^{-1}}$ -- are more massive in relation to the the total galaxy mass than previously thought. In some galaxies the thick disc could be as massive as the thin disc. Since old stellar populations have larger mass-to-light ratios than their younger counter-parts, this implies that thick discs could contain a fraction of the missing baryons and would somewhat reduce the need for a significant dark matter contribution in the inner kiloparsecs of galaxies. However, the validity of those conclusions is potentially jeopardized by an over-simplistic PSF treatment. Recently, several studies -- such as those by \citet{TRU16} and \citet{PE17} -- have started accounting for the effect of scattered light from high-surface-brightness areas in low-surface-brightness features. The current paper reassesses the effect of extended PSF wings in the results presented in \citet{CO11B, CO12,CO14}.

In this paper we use terms such as the vertical, radial, and axial directions applied to edge-on galaxies. The first two refer to intrinsic coordinates related to the galaxy, while the last one refers to the sky plane. The vertical direction is that perpendicular to the mid-plane of a galaxy. The radial direction is indicated by an outward vector from the centre of the galaxy projected into the galaxy mid-plane. The axial direction is the mid-plane projection of a vector pointing away from the galaxy centre in the sky plane. The vertical coordinates are denoted by $z$, the radial coordinates are denoted by $r$, and the axial coordinates are denoted by $x$. The coordinate axis that is both perpendicular to $x$ and $z$ is denoted as $y$. All the luminosities in this paper are given in the AB magnitude system.

The paper is structured as follows: In Sect.~\ref{psfsection} we describe the properties of the PSF of the instrument that we used (IRAC on-board the {\it Spitzer} Space Telescope). Section~\ref{procedure} is a detailed and technical description of the process used to fit the various vertical and axial surface-brightness profiles, taking into account the gravitational interaction of the two stellar discs, the effect of an extended PSF, and the presence of light from a central mass concentration (CMC). It builds upon the methods developed in our earlier papers \citep{CO11B, CO12, CO14} and combines a detailed account of the overall procedure with a number of new developments. In Sect.~\ref{sample} we present our sample. In Sect.~\ref{results} we revisit some of the old results in \citep{CO11B, CO12, CO14} with our newer fits accounting for an extended PSF. We finally summarise our findings and present our conclusions in Sect.~\ref{summary}.

\section{The S$^4$G point spread function}

\label{psfsection}

\begin{figure}
\begin{center}
  \includegraphics[width=0.48\textwidth]{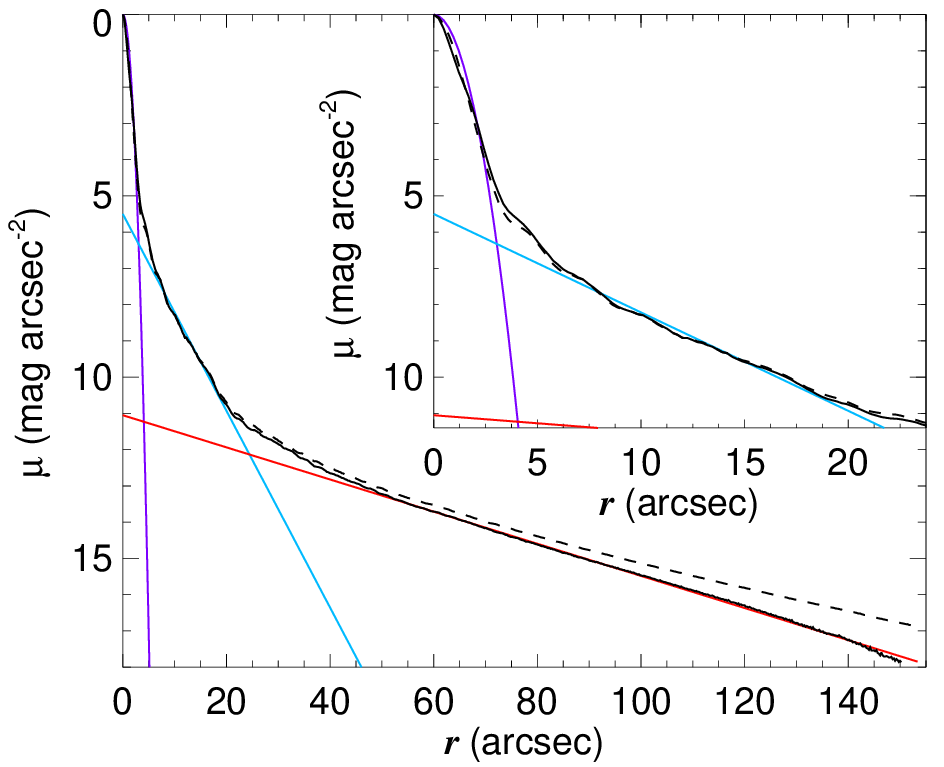}
  \end{center}
  \caption{\label{psf} Surface brightness as a function of the radius of the symmetrised $3.6\mu{\rm m}$ PSFs obtained from the data in \citet{HO12}. The continuous black line corresponds to the warm mission and the dashed line corresponds to the cryogenic mission. The coloured lines represent a Gaussian profile with a $2\farcs1$ FWHM (purple), an exponential function with a $4\farcs0$ scale-length (blue), and an exponential function with a $24\farcs5$ scale-length (red). The {\it inset} displays a radially zoomed version of the central regions of the PSFs. The zero point of the vertical axis is normalised to unity at the centre of the PSF.}
\end{figure}

Our previous work in \citet{CO11A, CO11B, CO12, CO14, CO15} contains decompositions of edge-on galaxies into their thin and thick disc components. Some of these studies also account for a CMC or a third stellar disc component. Our decompositions were made with mid-infrared images from the Spitzer Survey of Stellar Structure in Galaxies\footnote{The S$^4$G data products can be accessed through IRSA at \url{http://irsa.ipac.caltech.edu/data/SPITZER/S4G/}} \citep[S$^{4}$G;][]{SHETH10, MU13, QUE15}. The S$^4$G is a survey of 2352 nearby galaxies in $3.6\mu$m and $4.5\mu$m made with the IRAC camera \citep{FA04} on board the {\it Spitzer} Space Telescope. Most of the S$^4$G images were taken during the warm phase of the {\it Spitzer} mission, after the coolant was spent. However, the images of 597 S$^4$G galaxies are actually cryogenic archival images. The processed S$^4$G images have a pixel size of $0\farcs75$.

\citet{CO11A, CO11B, CO12, CO14} accounted for the PSF by considering a Gaussian function with a $2\farcs2$ full width at half maximum (FWHM). Tom Jarrett provided the S$^4$G collaboration with an extended $3.6\mu$m PSF that has been carefully discussed in \citet{SA15}. The symmetrised version of this PSF model can be approximated as the superposition of two functions; first an almost Gaussian core with a $2\farcs1$ FWHM and second, extended exponential wings that dominate the PSF at radii larger than $5^{\prime\prime}$. Tom Jarrett's PSF model is $15^{\prime\prime}$ in radius and was used -- incorrectly as discussed in Sect.~\ref{mock} -- in \citet{CO15}.

However, a $15^{\prime\prime}$ model PSF is too small for a proper modelling of the faintest galaxy components as demonstrated by \citet{JONG08}, \citet{SAN14, SAN15}, \citet{TRU16}, and \citet{PE17}. Indeed, ideally the PSF model should be at least as large as the observed galaxy. This is why here we use the IRAC PSF described in \citet{HO12}{\footnote{This document can be accessed from \url{http://irsa.ipac.caltech.edu/data/SPITZER/docs/irac/calibrationfiles/psfprf/}}}. The PSF model is slightly over $2\farcm5$ in radius. In this paper we study the galaxies in the $3.6\mu{\rm m}$ band (Sect.~\ref{preliminary}), so we use the $3.6\mu{\rm m}$ PSF model. We use the warm mission model because most of the S$^4$G images have been taken in this mode. Moreover, the differences between the cryogenic and the warm PSFs are small.

The S$^4$G PSF varies from frame to frame. This is because S$^4$G images are made of two sets of exposures with different orientations. Because the IRAC PSF has features that break its axis-symmetry, the superposition of images with varying orientations yields a different PSF for each of the S$^4$G frames. Fortunately, \citet{SA15} have shown that the use of a symmetrised PSF does not significantly affect the decompositions. This roughly holds even with our more extended PSF model, as shown in Sect.~\ref{symmetryzed}. As a consequence, from now on, when talking about the S$^4$G PSF we will refer to its symmetrised version.

The symmetrised versions of the $3.6\mu{\rm m}$ PSFs in \citet{HO12} are shown in Fig.~\ref{psf}. As in Tom Jarrett's PSF, the core is close to Gaussian with a $2\farcs1$ FWHM and the wings are exponential and have a $4\farcs0$ scale-length. Even more extended exponential wings with a $24\farcs5$ scale-length become evident at radii larger than $30^{\prime\prime}$. We note that an exponential PSF decay is not nearly as bad as the power laws described in \citet{SAN14}. Here we note that according to \citet{SAN15} ``no PSF shows a decline that is steeper than a $r^{-2}$ power-law'' which would imply an infinite integrated PSF light at infinity! Following the discussion of the {\it Hubble} Space Telescope ({\it HST}) PSFs in \citet{SAN14} we speculate that the reason for a relatively rapidly decaying IRAC PSF is the lack of atmospheric effects.

\section{Fitting procedure}

\label{procedure}

We have developed a comprehensive fitting procedure which allows us to derive the masses of different components - gaseous, thin and thick disks, and CMC - from deep images of highly inclined galaxies. The procedure allows for the description of two stellar discs in hydrostatic equilibrium, accounting for the light from the CMC, as well as correction for the effects of the PSF. As we developed the overall procedure over several years, the description of elements of it is fragmented in the literature \citep{CO11B, CO12, CO14}. Here, we describe the overall procedure in enough detail to allow reproduction or further development, while adding the new element of proper PSF correction. Readers interested primarily in the scientific findings derived from this can skip Sect.~\ref{procedure}.

\subsection{Preliminary considerations}

\label{preliminary}

The fits were done using the $3.6\mu{\rm m}$ images from the S$^4$G and its early-type galaxy extension \citep{SHETH13} and are presented in Appendices~\ref{twodiscap} and \ref{onediscap}. We used the science-ready images as provided by the S$^4$G Pipeline~1 \citep{MU15}. We also used the masks created in the S$^4$G Pipeline~2 \citep{MU15} as a starting point for an aggressive masking. This masking was made manually and expanded the masked regions around extended objects and saturated stars. We also masked many faint point sources that were ignored by Pipeline~2.

Before starting the actual fitting procedure, images were rotated so the mid-plane of the studied galaxy lied horizontally. This was done in most cases using position angles (PAs) from HyperLeda\footnote{http://leda.univ-lyon1.fr/} \citep{MA14}. In some cases the HyperLeda values were not accurate enough -- the rotation did not result in a close to horizontal orientation for the galaxy mid-plane -- and we used orientations derived from ellipticity profiles in the S$^4$G Pipeline~4 \citep{SA15}. Finally, for a few galaxies in the S$^4$G early-type extension, orientations were obtained from our own ellipse fits.

The steps that we followed to produce structural decompositions of edge-on galaxies are:
\begin{enumerate}
 \item We fitted the vertical surface-brightness profiles of galaxies with the superposition of a thin and a thick disc in hydrostatic equilibrium. We also accounted for the gravitational effect of a gas disc.
 \item We fitted the axial surface-brightness profiles of galaxies with the superposition of a function describing the disc -- an exponential function with possible multiple breaks integrated along the line of sight -- and one describing the CMC -- a S\'ersic function.
 \item We reran the first step accounting for the presence of CMC light and an accurate description of the disc axial surface-brightness distribution.
 \item We produced axial surface-brightness profiles for the heights dominated by the thin and the thick discs. Those profiles were fitted in a similar way as in step~2.
 \item The different components' masses -- gas disc, thin disc, thick disc, and CMC -- were calculated using some assumptions about the mass-to-light ratios.
\end{enumerate}

These five steps are explained in detail below.
  
\subsection{Step 1: Vertical surface-brightness profile fits}

The vertical surface-brightness profile fits are made in a similar way as in \citet{CO11B, CO12}. Here we explain again our method and  point out small differences between what is done in this paper and what was done in our previous work.

\subsubsection{The set of differential equations}

We assume the stellar discs of galaxies to be made of two discs in hydrostatic equilibrium that are allowed to interact gravitationally with each other. At a given distance from the galaxy centre, each of the two discs is assumed to be vertically isothermal, that is, they have a single vertical velocity dispersion at all heights. We also consider the presence of a non-stellar disc, that is, a gas disc with mass but no mid-infrared luminosity. In such a system, the vertical density distribution for a point at a given distance from the galaxy centre is described by
\begin{equation}
\label{narayan}
\frac{d^2\rho_i}{dz^2}=\frac{\rho_i}{\sigma_i^2}\left[-4\pi G\left(\rho_t+\rho_T+\rho_g\right)+\frac{dK_{\rm DM}}{dz}\right]+\frac{1}{\rho_i}\left(\frac{d\rho_i}{dz}\right)^2 
\end{equation}
following the formalism by \citet{NA02}. Here $\rho$ is the volume density and the subindices $t$, $T$, and $g$ denote the thin, the thick, and the gas disc. The subindex $i$ denotes any of the three above-mentioned discs. The velocity dispersion of the discs is denoted by $\sigma$. The $K_{\rm DM}$ term accounts for something behaving like a dark matter halo.

This formalism provides a physically motivated function as opposed to the sometimes ad-hoc analytic functions that have been used elsewhere \citep[e.g.,][]{KRUIT88, YOA06, CO11C}. The downside of our method is that the fitted function is not analytical and needs to be solved by numerical integration.

\subsubsection{Assumptions made during the fit}

\label{assumptions}

The main goal of our structural decompositions is to obtain reasonable estimates of the thin and thick disc masses. Hence, here we discuss the assumptions and approximations based on their effect in the determination of the ratio of disc masses, $\mathcal{M}_{\rm T}/\mathcal{M}_{\rm t}$.

Equation~\ref{narayan} describes volume densities as a function of height, which are not observables. What is observed is the surface brightness as a function of height. Thus, the use of Eq.~\ref{narayan} requires knowing the mass-to-light ratio of the different components -- $\Upsilon_i$ -- and an understanding of the effects of a line-of-sight integration. It also requires assumptions on the dark matter distribution.

In \citet{CO11B} we discussed three different star formation histories (SFHs) reported in the literature for the thin and the thick discs of the Milky Way and obtained mass-to-light ratios using the spectral energy distributions by \citet{BRU03} calculated using the ``Padova 1994'' stellar evolution prescriptions \citep{AL93, BRES93, FAG94A, FAG94B, GI96} and a Salpeter initial mass function \citep{SAL55}. In this paper we use the mass-to-light ratios derived from the Milky Way SFH found in \citet{NY06}. Their SFH model assumes a short initial burst of star formation followed by an exponentially decaying star formation rate. The time-scale of the decay is set to $\tau=8\,{\rm Gyr}$ for the thin disc and $\tau=5\,{\rm Gyr}$ for the thick disc. The resulting ratio of the mass-to-light ratios of the thick and thin disc at $3.6\mu{\rm m}$ --which we need for our fit -- is $\left(\Upsilon_{\rm T}/\Upsilon_{\rm t}\right)_{3.6\mu{\rm m}}=1.2$. SFHs other than that in \citet{NY06} could yield different $\left(\Upsilon_{\rm T}/\Upsilon_{\rm t}\right)_{3.6\mu{\rm m}}$ values. This, however, is not as critical as it might sound because $\mathcal{M}_{\rm T}/\mathcal{M}_{\rm t}$ roughly scales with $\left(\Upsilon_{\rm T}/\Upsilon_{\rm t}\right)_{3.6\mu{\rm m}}$ \citep[see Section.~4.3.1 in][]{CO12}. $\left(\Upsilon_{\rm T}/\Upsilon_{\rm t}\right)_{3.6\mu{\rm m}}=1.2$ is the smallest and most conservative of the mass-to-light ratio values derived from the SFHs explored in \citet{CO11B}, which implies that using this value may even underestimate $\mathcal{M}_{\rm T}/\mathcal{M}_{\rm t}$. We note that although we use a single $\left(\Upsilon_{\rm T}/\Upsilon_{\rm t}\right)_{3.6\mu{\rm m}}$, the mass-to-light ratio certainly varies from galaxy to galaxy depending on, for example, the galaxy mass and environment.

Throughout this paper we ignore the dark matter term in Eq.~\ref{narayan} ($dK_{\rm DM}/dz=0$). Experiments made in \citet{CO11B, CO12} show that omitting this term introduces a small bias that causes $\mathcal{M}_{\rm T}/\mathcal{M}_{\rm t}$ to be slightly overestimated. The more sub-maximal a disc is, the more $\mathcal{M}_{\rm T}/\mathcal{M}_{\rm t}$ would be overestimated (typically by $\sim10\%$). Ignoring the effect of something behaving like a dark matter halo allows us to assume that the galaxy has a cylindrical symmetry, which leads to a much simpler numerical treatment.

Line-of-sight integration is not a problem when applying Eq.~\ref{narayan} as long as the scale-heights of the thin and the thick discs remain roughly constant and the scale-lengths of both discs are similar. If those conditions are fulfilled the vertical density and luminosity profiles of the disc remain the same at all radii once a multiplicative factor is accounted for. The profiles at different radii are in this case geometrically similar (at least they would be so in the absence of a significant dark matter term, which is one of the assumptions that we describe above). Thus, fitting the result of a line-of-sight projection would imply no inaccuracy. If the above-mentioned disc scale-length and scale-height conditions are not fulfilled, disc locations along the line of sight have different ratios of the thick and thin disc mass surface densities that depend on their distance from the galaxy centre. The resulting observed surface-brightness profile is in this case a weighted average of the vertical luminosity density profiles where the regions closest to the centre of the galaxy have the largest weight. In \citet{CO12} we showed that if thin discs had a break and/or a different scale-length than the thick disc, the fitted $\mathcal{M}_{\rm T}/\mathcal{M}_{\rm t}$ ratios would typically be overestimated by $\sim10\%$.

To sum up, our assumptions regarding the ratio of thick to thin disc mass-to-light ratios -- $\left(\Upsilon_{\rm T}/\Upsilon_{\rm t}\right)_{3.6\mu{\rm m}}$ -- probably bias the ratio of the disc masses -- $\mathcal{M}_{\rm T}/\mathcal{M}_{\rm t}$ -- towards low values whereas our assumptions on similar disc scale-lengths and scale-heights, and on the omission of dark matter haloes, bias $\mathcal{M}_{\rm T}/\mathcal{M}_{\rm t}$ towards large values.

\subsubsection{Surface-brightness profiles perpendicular to the galaxy mid-plane}

\label{production}

Four surface-brightness profiles perpendicular to the mid-plane were made for each galaxy in bins with axial ranges $0.2\,r_{25}<\left|x\right|<0.5\,r_{25}$ and $0.5\,r_{25}<\left|x\right|<0.8\,r_{25}$. The radius of the $25\,{\rm mag\,arcsec^{-2}}$ level in the $B$-band, $r_{25}$, was obtained from HyperLeda. We thus ignored the central regions which are those most affected by the CMC. The CMC breaks the cylindrical symmetry and implies a component unaccounted for in the formalism set in Eq.~\ref{narayan} (although it could formally be included as part of the $dK_{\rm DM}/dz$ term).

We produced the vertical surface-brightness profiles for each axial bin by averaging at each height the number of counts in the bin.

We found the mid-plane by folding the profiles and finding the fold position that would minimise the difference between the upper and lower parts of the profile (in magnitudes). This was done with a precision of 0.1\,pixels and using a dynamic range of $\Delta\mu=4.5\,{\rm mag\,arcsec^{-2}}$. The mid-plane found with this method was then used to create a symmetrised vertical surface-brightness profile to be fitted with Eq.~\ref{narayan}.

\subsubsection{Boundary conditions, free parameters, and normalisations of the fitted function}

The vertical surface-brightness profiles are fitted with a function derived from a set of three second-order differential equations (for the thin, thick, and gas discs). Six boundary conditions are required to solve the set of equations. Trivially, three of them require that the maxima of the vertical density profiles of the three discs are to be found at the mid-plane, which results in
\begin{equation}
\left.\frac{d\rho_{i}}{dz}\right|_{z=0}=0.
\end{equation}
The other conditions are the densities of the three coupled discs at the mid-plane, $\left.\rho_{i}\right|_{z=0}$. These mid-plane densities are transformed into mid-plane surface brightnesses thanks to an assumed mass-to-light ratio and due to the integration along the line of sight. The ratio of two of the $\left.\rho_{i}\right|_{z=0}$ values is fitted in our approach whereas the third one is deduced, as shown below.

Naively speaking, when fitting a surface-brightness profile with the products of Eq.~\ref{narayan}, six parameters need to be fitted. Those parameters are the three mid-plane densities, $\left.\rho_{i}\right|_{z=0}$, and the vertical velocity dispersions of the discs, $\sigma_i$.

A way to remove free parameters is to make assumptions about the gas content of galaxies. One could do so by transforming the 21\,cm hyperfine transition line flux into an atomic gas mass. A conversion factor between the atomic gas mass and the total cold gas mass would also be necessary. However, this approach would only provide an integrated gas mass, without giving any information about its distribution in the plane of the observed galaxies. Another approach -- used here -- is to assume a reasonable gas surface density (this time we talk about a surface density as seen from the galaxy pole and not on an edge-on view). Here we assume that the gas surface density is 20\% of that of the thin disc (this does not yet fix the $\left.\rho_{\rm g}\right|_{z=0}$ value). This value is larger than that found in the Milky Way \citep[][]{BAN07}, but has been chosen because our sample is dominated by galaxies with masses smaller than that of the Milky Way which typically have a larger gas fraction than our Galaxy \citep[see for a recent study][]{SAIN16}. We also set the vertical velocity dispersion of the gas disc to be one third of that of the thin disc -- $\sigma_{\rm g}=1/3\sigma_{\rm t}$ -- which is similar to the values measured in the Solar neighbourhood \citep[][and references therein]{BAN07}. Errors on the assumed gas disc properties do not introduce large errors in $\mathcal{M}_{\rm T}/\mathcal{M}_{\rm t}$, as explained in the Sects.~4.1 and 4.3 in \citet{CO12}.

More free parameters can be removed by normalising the surface-brightness profiles. If the observed surface-brightness profile is normalised to its mid-plane value, we are in fact assuming that
\begin{equation}
\left(\Upsilon_{\rm t}\right)_{\rm 3.6\mu{\rm m}}\left.\rho_{\rm t}\right|_{z=0}+\left(\Upsilon_{\rm T}\right)_{\rm 3.6\mu{\rm m}}\left.\rho_{\rm T}\right|_{z=0}=\,{\rm constant}.
\end{equation}
In this case $\left.\rho_{\rm t}\right|_{z=0}$ and $\left.\rho_{\rm T}\right|_{z=0}$ are no longer independent and $\left(\left.\rho_{\rm T}\right|_{z=0}\right)/\left(\left.\rho_{\rm t}\right|_{z=0}\right)$ is fitted.

Another free parameter can be removed by vertically scaling the surface-brightness profiles. We rescale the vertical axis so that
\begin{equation}
\left.I\right|_{z=z_{\rm c}}=f\left.I\right|_{z=0} 
\label{stretch}
\end{equation}
where $I$ is the surface brightness (in linear scale) and $f$ is a fraction that goes between 0.1 and 0.4 as discussed in Sect.~\ref{actual}. The height $z_{\rm c}$ is the height at which the observed profile -- to which the synthetic profile is compared -- has a surface brightness equal to that of the mid-plane multiplied by a factor $f$. The effect of the scaling is to link $\sigma_{\rm t}$ and $\sigma_{\rm T}$ so only $\sigma_{\rm T}/\sigma_{\rm t}$ remains as a free parameter. In practice, this condition causes the observed and the fitted profiles to cross each other at a height $z_{\rm c}$.

After making the assumptions about the gas disc and the normalisations explained above, we are left with two free parameters, $\left(\left.\rho_{\rm T}\right|_{z=0}\right)/\left(\left.\rho_{\rm t}\right|_{z=0}\right)$ and $\sigma_{\rm T}/\sigma_{\rm t}$, and a condition that links the gas disc surface density to the thin disc surface density as seen from the galaxy pole. 

The coupled equations in Eq.~\ref{narayan} were solved using the Newmark-beta method \citep{NEW59} with the parameters $\beta=0.25$ and $\gamma=0.5$. The Newmark-beta method is a second-order numerical integration method. The gas mid-plane density had to be found iteratively by setting a guess value $\left.\rho_{\rm g}\right|_{z=0}$ and rerunning the integration with new $\left.\rho_{\rm g}\right|_{z=0}$ values until the gas surface-density condition was met.

The calculated surface-density profiles were transformed into synthetic surface-brightness profiles using $\left(\Upsilon_{\rm T}/\Upsilon_{\rm t}\right)_{\rm 3.6\mu{\rm m}}$.

\subsubsection{PSF convolution of synthetic surface-brightness profiles}

\begin{figure*}
\begin{center}
  \includegraphics[width=0.98\textwidth]{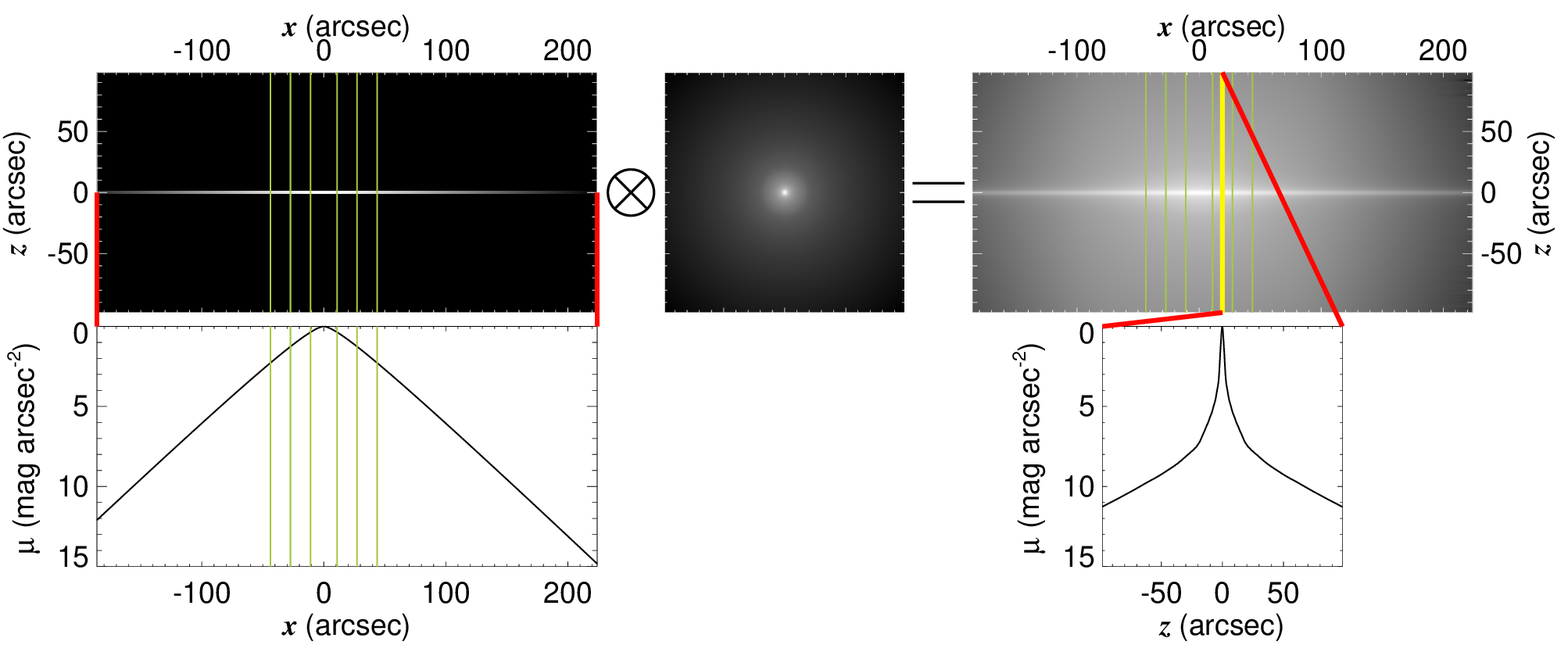}
  \end{center}
  \caption{\label{mock1d} Creation of the 1D mock PSF for the axial bin $0.2\,r_{25}<x<0.5\,r_{25}$ of ESO~533-4. The {\it top-left} panel shows an axial profile that corresponds to the edge-on view of an infinitely thin disc with an exponential scale-length of $14\farcs7$ \citep[obtained from][]{SA15}. The axial surface-brightness profile is shown in the {\it bottom-left} panel. The vertical green lines represent the axial bins to be studied. The image in the {\it top-left} panel is convolved with the 2D PSF ({\it top-middle}), which results in the convolved image ({\it top-right}). A vertical cut (yellow line) of the convolved image at the position of the middle of the axial bin under study results in the 1D mock PSF ({\it bottom-right}).}
\end{figure*}

\begin{figure}
\begin{center}
  \includegraphics[width=0.48\textwidth]{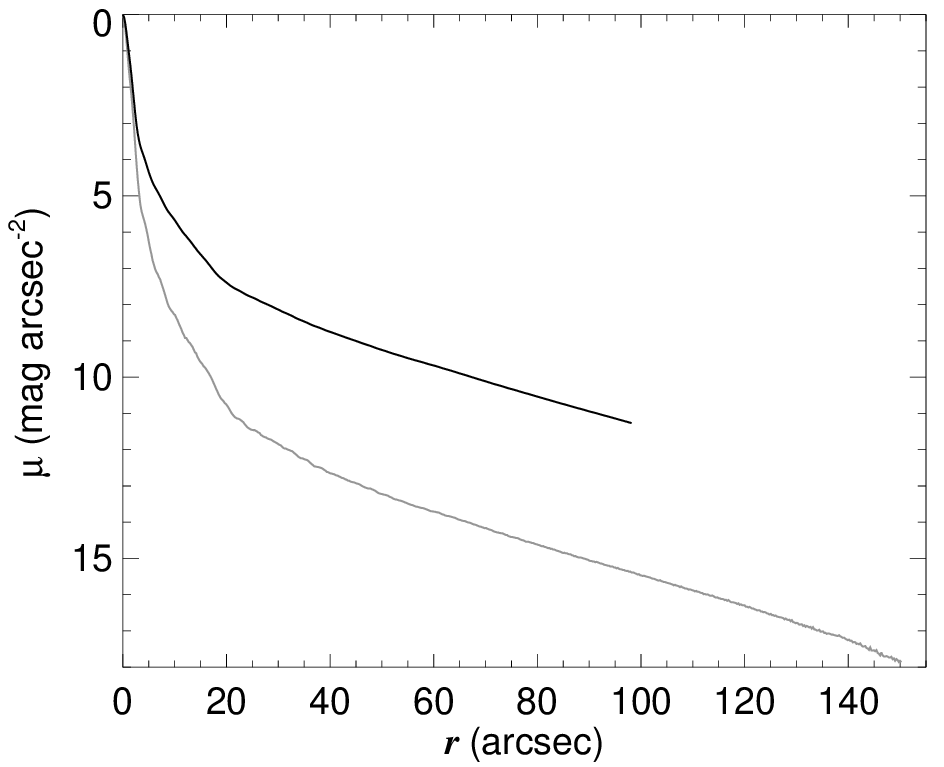}
  \end{center}
  \caption{\label{comparison} Vertical surface brightness of the 1D mock PSF for the $0.2\,r_{25}<\left|x\right|<0.5\,r_{25}$ bins of ESO~533-4 (black) compared to the radial surface-brightness profile of the symmetrised IRAC PSF (grey). The zero point of the vertical axis is arbitrary.}
\end{figure}

\label{mock}

The synthetic surface-brightness profiles have to be convolved with the IRAC PSF so that they can be compared to the observed ones.

A technical difficulty resides in the fact that a one-dimensional (1D) model profile has to be convolved with a two-dimensional (2D) PSF. A way to solve the issue would be to assume a given axial surface-brightness profile, build a 2D model image using the axial and the vertical profiles, and then perform the PSF convolution. However, this would be impractical because 2D convolutions are time-consuming. A mathematically equivalent approach is to convolve the vertical 1D surface-brightness profiles with a 1D mock PSF that would mimic the effect of the 2D convolution. Such a mock PSF can be created by 2D-convolving a 1D axial surface-brightness profile with the 2D PSF as seen in Fig.~\ref{mock1d}. A vertical cut of the resulting image at the position of the middle of the axial bin under study provides a first approximation to the sought 1D mock PSF assuming that the axial profile of the observed galaxy does not change much with height. In our case the axial profile used is that of an edge-on disc with a face-on exponential profile, and with the scale-length taken as that of the thinner of the discs in the \citet{SA15} decompositions. When no decomposition was available in \citet{SA15} or when that was deemed to poorly fit the galaxy, a scale-length of $20^{\prime\prime}$ was used instead. The inaccuracies in the mock PSF introduced by the choice of the scale-length are small. Additionally, the final vertical surface-brightness profile fits -- explained in Sect.~\ref{finalvertical} -- use a more sophisticated approach to model the axial surface-brightness profiles.

For angularly small galaxies, we truncated the 2D PSF so its radius was 1.8 times $r_{25}$. In that way, the recommendation by \citet{SAN14} to work with a PSF that is at least 1.5 times larger than the radius of the object is respected. For larger galaxies we used the whole 2D PSF radius. Galaxies larger than $100^{\prime\prime}$ -- a quarter of the sample -- cannot be studied with a PSF 1.5 larger than their size (the 2D PSF radius is slightly over $2\farcm5$). \citet{SAN14} studied the {\it HST} PSF and recommends using a PSF model that is ``at least 1.5 times as extended as the vertical distance of the edge-on galaxy''. Assuming that all space-based PSFs behave similarly -- and the IRAC PSF seems to do so because it falls faster than $r^{-2}$ -- this prescription allows using our 2D PSF on all but maybe a dozen of our largest galaxies. In all cases, the PSF radius is at least several times larger than the scale-height of the discs. Also, in all cases, the PSF radius is at least as large as the distance between the galaxy mid-plane and the maximum height considered in the vertical surface-brightness profile fits, $z_{\rm l}$ (see Sect.~\ref{actual} for a definition).

Fig.~\ref{comparison} shows the comparison between a radial cut to the 2D PSF and the mock 1D PSF created for a specific galaxy in our sample. A second 1D mock PSF made from a 2D PSF truncated at a radius of $38\farcs4$ was also created. We call this PSF the ``small 1D PSF'' as opposed to the more extended ``large 1D PSF''.

The creation of a mock 1D PSF was not necessary in \citet{CO11A, CO11B, CO12} because we were assuming a Gaussian PSF. Indeed, when applying the procedure explained above to a Gaussian PSF, the resulting mock PSF is the same Gaussian PSF. This is seen also in Fig.~\ref{comparison} where the inner Gaussian parts are essentially identical.

In \citet{CO15} we used a radial cut to the 2D PSF to convolve our 1D vertical surface-brightness profiles, which in light of what is explained above is incorrect. In \citet{CO15} the effect of the PSF extended wings is very underestimated. For example, for ESO~533-4 a height of $z\sim5^{\prime\prime}$ is not the height above which 90\% of the light comes from the thick disc as stated in our previous work. Instead, based on our analysis here we can say that $z\sim5^{\prime\prime}$ is approximately the height above which the thick disc surface brightness is larger than the thin disc surface brightness.

A second technical difficulty due to PSF convolution comes when it has to be combined with the vertical stretching of the surface-brightness profile. The synthetic profiles are first stretched according to the condition Eq.~\ref{stretch}, but once the convolution is applied they no longer cross at $z=z_{\rm c}$ due to the changes in the shape of the profile. Instead, the observed and the synthetic profiles cross at $z=z^{\prime}_{\rm c}$. To counter that, the $z$ axis of the synthetic profile has to be rescaled so Eq.~\ref{stretch} is fulfilled after the convolution. This was done by multiplying the $z$ axis of the synthetic profile by a factor equal to the ratio between $z_{\rm c}$ and the height at which the convolved synthetic and the observed profiles cross $z=z^{\prime}_{\rm c}$. This was done iteratively until the relative difference between $z_{\rm c}$ and $z^{\prime}_{\rm c}$ became smaller than $1/10000$. 

\subsubsection{The actual fit}

\label{actual}

\begin{figure}
\begin{center}
\begin{tikzpicture}[node distance = 2.2cm, auto]
    \node [block] (init) {Vertical surface-brightness profile with $\Delta\mu=4.5\,{\rm mag\,arcsec^{-2}}$};
    \node [block, below of=init, node distance=2.8cm] (fit) {Four fits with $f=0.1$, $f=0.2$, $f=0.3$, and $f=0.4$ using \textsc{idl}'s \textsc{curvefit} and the small 1D PSF};
    \node [block, left of=fit, node distance=3cm] (update) {$\Delta\mu=\Delta\mu+0.5\,{\rm mag\,arcsec^{-2}}$};
    \node [decision, below of=fit, node distance=3.7cm] (decide) {Do any of the fits have $\mu_{\rm rms}<0.1\,{\rm mag\,arcsec^{-2}}$?};
    \node [decision, below of=decide, node distance=4.1cm] (good) {Is $\Delta\mu>4.5\,{\rm mag\,arcsec^{-2}}$?};
    \node [block, below of=good, node distance=3.4cm] (rerun) {Rerun with $\Delta\mu=\Delta\mu-0.5\,{\rm mag\,arcsec^{-2}}$ and the large 1D PSF};
    \node [block, below of=rerun, node distance=2.5cm] (result) {{\bf Fit}};
    \node [block, left of=result, node distance=3.cm] (noresult) {{\bf No good fit}};

    \path [line] (init) -- (fit);
    \path [line] (fit) -- (decide);
    \path [line] (decide) -| node [near start] {yes} (update);
    \path [line] (update) |- (fit);
    \path [line] (decide) -- node [midway]{no}(good);
    \path [line] (good) -- node [midway]{yes}(rerun);
    \path [line] (rerun) -- (result);
    \path [line] (good) -| node [near start] {no} (noresult);

\end{tikzpicture}
\end{center}
\caption{\label{flow} Flow diagram describing the fitting procedure of the vertical surface-brightness profiles as explained in Sect.~\ref{actual}.}
\end{figure}
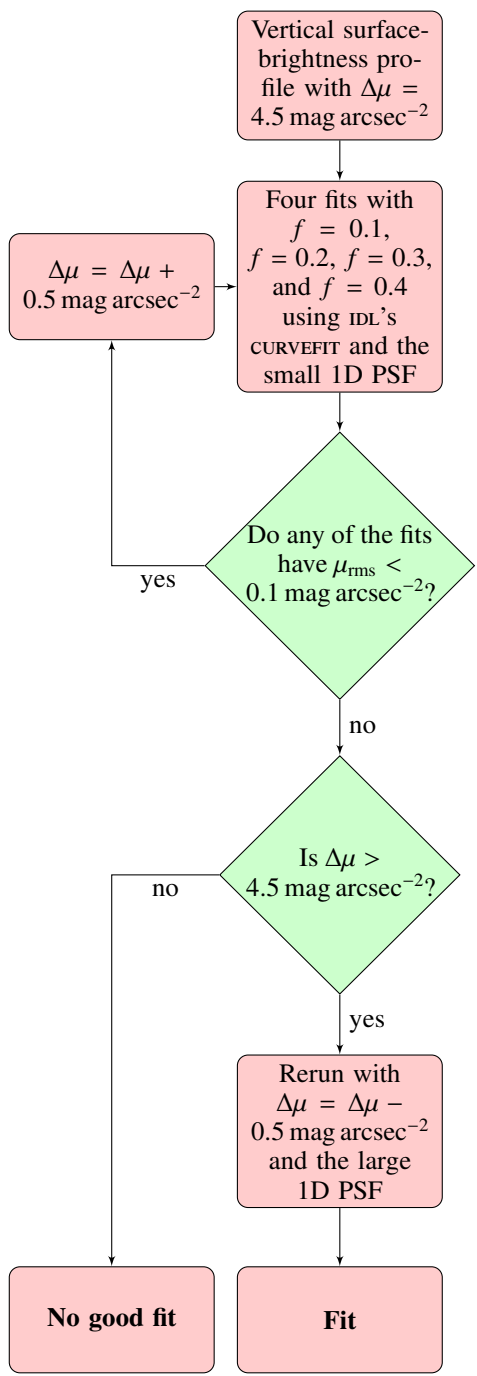

\begin{figure*}
\begin{center}
  \includegraphics[width=0.98\textwidth]{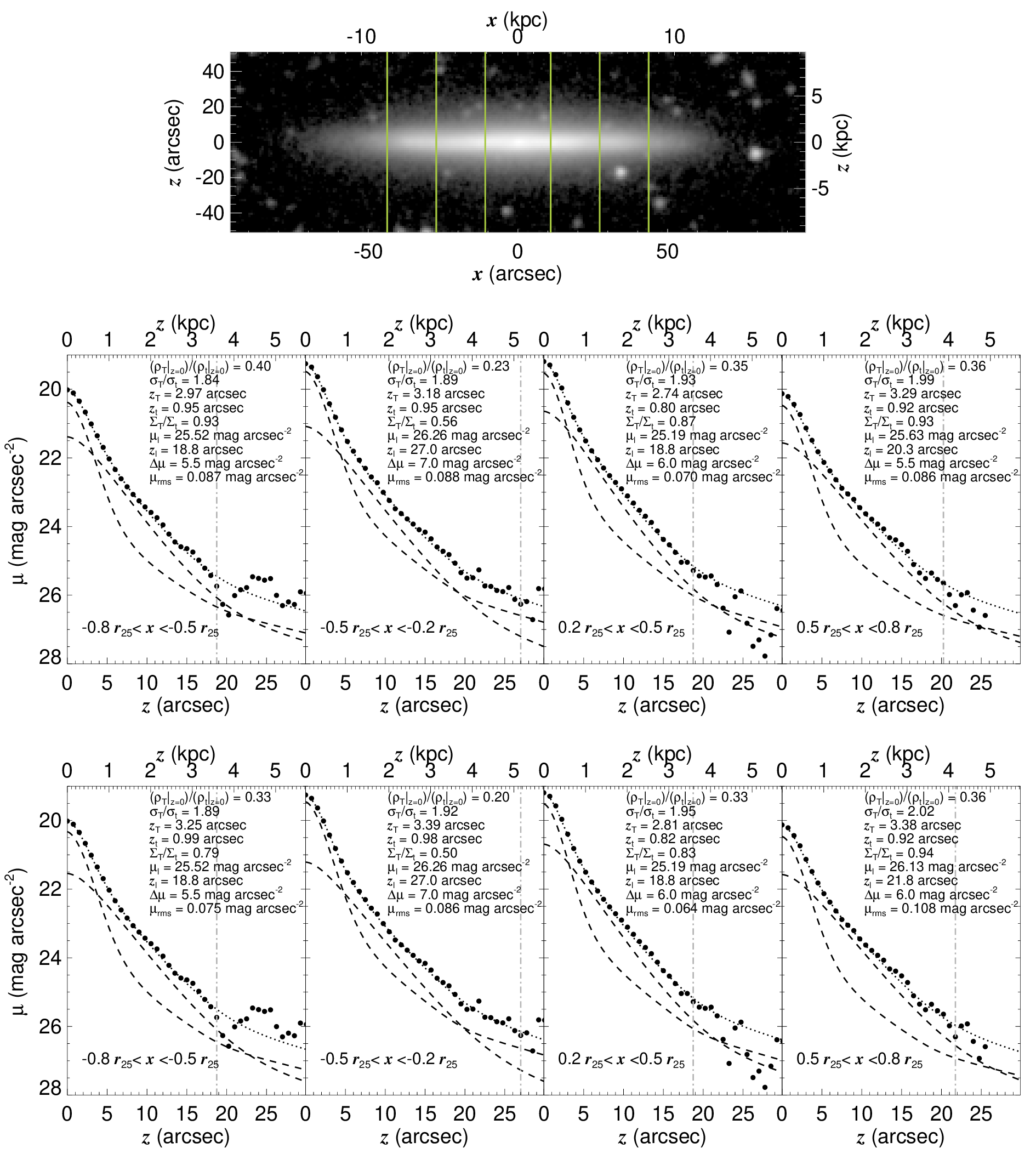}

  \end{center}
  \caption{\label{verfitnobul} {\it Top} panel: $3.6\mu{\rm m}$ image of ESO~533-4. The four axial bins for which surface-brightness profiles were produced are marked with vertical green lines. {\it Middle} panels: surface-brightness profiles of the four bins (large filled circles) and fits produced as explained in Sect.~\ref{actual} incorporating full PSF treatment (dotted lines). The dashed lines indicate the thin and thick disc contributions. The vertical dot-dashed grey lines indicate the range of heights considered for each fit. Each of the {\it middle}  panels contains some of the data obtained from the fits: the fitted parameters ($\left(\left.\rho_{\rm T}\right|_{z=0}\right)/\left(\left.\rho_{\rm t}\right|_{z=0}\right)$ and $\sigma_{\rm T}/\sigma_{\rm t}$), the scale-heights of the thick and the thin discs ($z_{\rm T}$ and $z_{\rm t}$, respectively), the thick to thin disc mass ratio in that bin ($\Sigma_{\rm T}/\Sigma_{\rm t}$), the faintest surface brightness level that was considered for the fit ($\mu_{\rm l}$), the maximum height considered for the fit ($z_{\rm l}$), the dynamic range of the fit ($\Delta\mu$), and the root mean square deviation of the fit ($\mu_{\rm rms}$). {\it Bottom} panels: Same as for the {\it middle} panels but this time with the fitting procedure explained in Sect.~\ref{finalvertical} accounting for the presence of the CMC. For this galaxy the CMC contribution is so small that it falls below the $\mu=28\,{\rm mag\,arcsec^{-2}}$ level.}
\end{figure*}

\begin{figure*}
\begin{center}
  \includegraphics[width=0.98\textwidth]{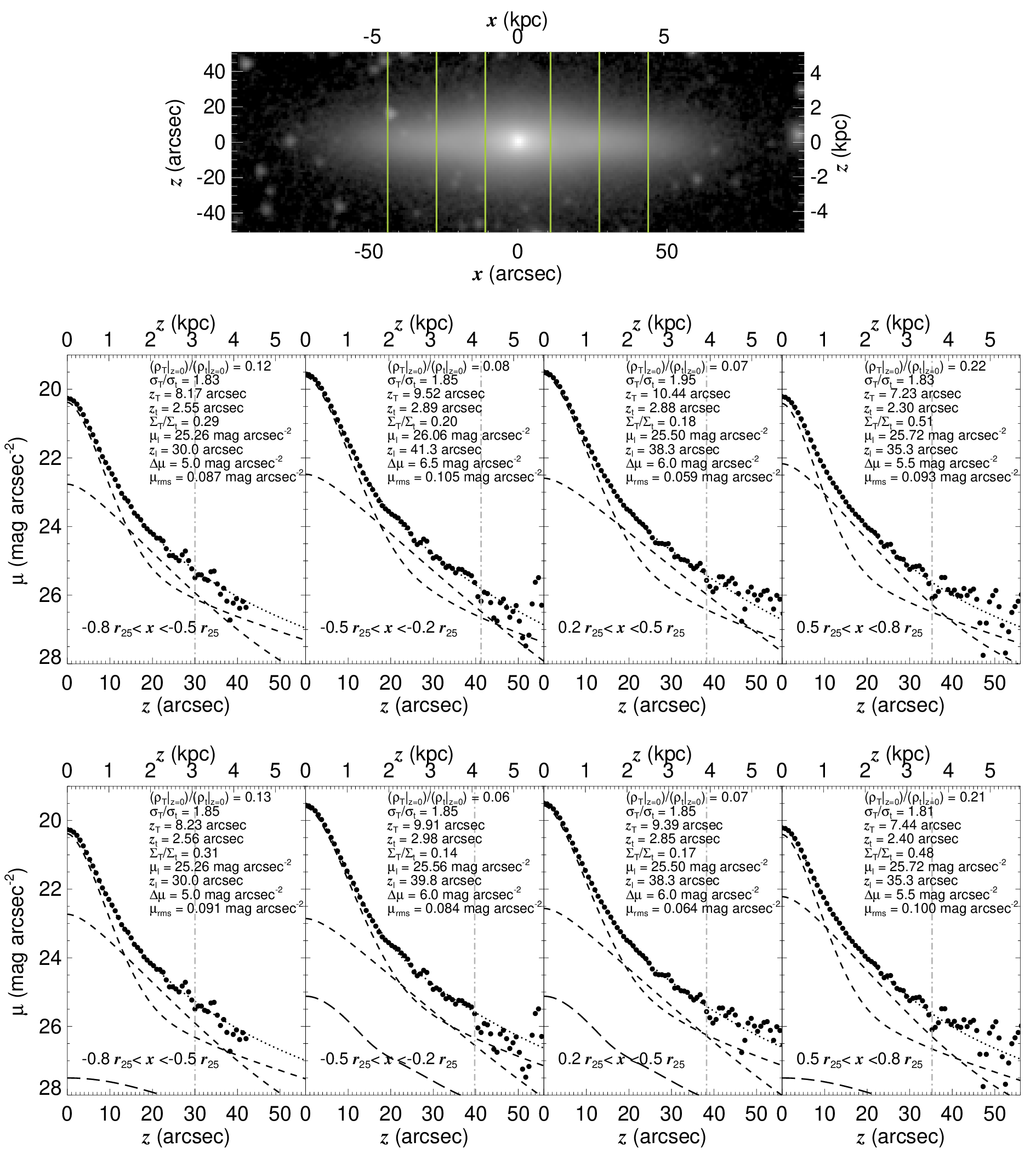}

  \end{center}
  \caption{\label{verfitnobul2} Same as Fig.~\ref{verfitnobul}, but for NGC~6010. In the {\it bottom} panels the long-dashed lines denote the CMC contribution. Even in this case with a relatively large CMC, its influence on the fitted thin and thick disc parameters is fairly small.}
\end{figure*}

Vertical surface-brightness profiles are typically dominated by thin disc light close to the mid-plane. The thick disc appears as an up-bending change of slope at heights where the surface brightness is dimmer than that at the mid-plane by a few magnitudes per arcsecond square. Thus a large dynamic range is required to obtain a meaningful thin and thick disc decomposition. If such a condition is not fulfilled, the fit becomes degenerate because of the lack of information on the thick disc scale-height. In practice, a minimum dynamic range of $\Delta\mu=4.5\,{\rm mag\,arcsec^{-2}}$ is required for a good fit \citep{CO11B}.

The S$^4$G images have a typical nominal depth of $\mu=27\,{\rm mag\,arcsec^{-2}}$. Our experience, however, shows that such a depth can hardly ever be achieved due to the presence of extended wings of the PSF of foreground stars and globular clusters surrounding the galaxy under study that affect the photometry even when aggressive masking is applied. Our experience is that the limiting magnitude is in general around $\mu=26\,{\rm mag\,arcsec^{-2}}$. This, combined with our dynamic range requirement, made us consider, for analysis, only profiles with a mid-plane surface brightness brighter than $\left.\mu\right|_{z=0}=22\,{\rm mag\,arcsec^{-2}}$. Profiles for the galaxies in the sample that do not fulfil this condition are included in the Appendices~\ref{twodiscap} and \ref{onediscap} but are not taken into consideration when measuring galaxy parameters.

The fitting process is summarised in the flow diagram shown in Fig.~\ref{flow}. The observed luminosity profiles had their luminosity normalised to a mid-plane surface brightness of zero magnitudes. They were then fitted with the PSF-convolved profile resulting from Eq.~\ref{narayan} using \textsc{idl}'s \textsc{curvefit}. Here, we used the small 1D PSF to save computing time. The fits were done over the profiles in units of ${\rm magnitude\,arcsec^{-2}}$ using the same weight for each data point. Because fits did usually not converge when done over large dynamical ranges -- unless the initial parameters were manually chosen to be very close the final result -- they were first run over $\Delta\mu=4.5\,{\rm mag\,arcsec^{-2}}$.

The fits were done four times with four vertical stretches of the synthetic profile (Eq.~\ref{stretch}) with $f=0.1$, $f=0.2$, $f=0.3$, and $f=0.4$. The best fit was selected to be that with an $f$ resulting in the smallest root mean square deviation ($\mu_{\rm rms}$). The reason for this approach is that some values of $f$ favoured convergence. Typically different $f$ values give very similar fits. Indeed, the value of $f$ sets a surface brightness level at which the observed and the synthetic profiles must cross. Since Eq.~\ref{narayan} results in profiles that are very close to the observed ones, the observed and synthetic profiles are close to cross each other at all heights. The parameter $f$ could have been made a free parameter but that caused convergence to be difficult.

If the fit resulted in $\mu_{\rm rms}>0.1 \,{\rm mag\,arcsec^{-2}}$ we considered that no good fit could be obtained. This typically occurred for profiles with mid-plane surface-brightness profiles close to the $\left.\mu\right|_{z=0}=22\,{\rm mag\,arcsec^{-2}}$ limit. If $\mu_{\rm rms}<0.1 \,{\rm mag\,arcsec^{-2}}$ the resulting fitted parameters -- $\left(\left.\rho_{\rm T}\right|_{z=0}\right)/\left(\left.\rho_{\rm t}\right|_{z=0}\right)$ and $\sigma_{\rm T}/\sigma_{\rm t}$ -- were used as the initial parameters for a fit with a dynamical range larger by $0.5\,{\rm mag\,arcsec^{-2}}$. This procedure was repeated until the rms of the fit became larger than $\mu_{\rm rms}=0.1\,{\rm mag\,arcsec^{-2}}$. When that happened, the last iteration where $\mu_{\rm rms}<0.1\,{\rm mag\,arcsec^{-2}}$ was rerun using the large 1D PSF. The resulting fit was considered as the best fit for a given vertical surface-brightness profile.

As an example, the fits corresponding to ESO~533-4 and NGC~6010 are shown in the middle rows in Figs.~\ref{verfitnobul} and \ref{verfitnobul2}, respectively. According to the fits, due to the extended PSF wings, the thin disc again dominates the surface-brightness profiles at large heights. For ESO~533-4 this happens at or just below surface brightness levels of $\mu\sim26\,{\rm mag\,arcsec^{-2}}$. For this particular galaxy, the fitted $\left(\left.\rho_{\rm T}\right|_{z=0}\right)/\left(\left.\rho_{\rm t}\right|_{z=0}\right)$ values are larger than in \citet{CO15} -- where the PSF treatment was incorrect and thus the effect of the PSF wings was largely underestimated -- and the fitted $\sigma_{\rm T}/\sigma_{\rm t}$ are smaller. This results in a similar thin to thick surface-density ratio, $\Sigma_{\rm T}/\Sigma_{\rm t}$. The same happens when we compare the ESO~533-4 fits here with those in \citet{CO12}, where the PSF was treated as a Gaussian core. For NGC~6010, the thin disc dominates the surface brightness close to the mid-plane and again at large heights at surface brightness levels slightly below $\mu\sim26\,{\rm mag\,arcsec^{-2}}$.

In general, we find that even though the thin disc is less vertically extended than the thick disc, it dominates the emission at large heights because of the light scattered from the mid-plane to large heights. This effect would remain unnoticed if we had not accounted for the extended wings of the PSF.

\subsubsection{Differences with respect to the fits in our previous papers}

Here we explain the main differences between the fitting procedure in this paper and that in our previous studies of thick discs in large samples, namely \citet{CO11B, CO12}.

\begin{enumerate}
\item In \citet{CO12}, the surface-brightness profiles were produced using an average of the S$^4$G 3.6 and 4.5$\mu{\rm m}$ images. The rationale was to attempt to reach lower surface brightness levels than with a single filter. Our experience, however, is that the limiting factor with deep photometry is the PSF wings of point-sources near the target galaxy, so using the combination of two bands did little to improve the fits. Here we only consider the 3.6$\mu{\rm m}$ data, just as we did in \citet{CO11B}.

\item In \citet{CO12}, the vertical surface-brightness profiles were smoothed to increase the signal-to-noise ratio. This was done adaptively so bright regions had a small smoothing. Such a smoothing is not used here because even if it might marginally improve the photometry, it harms the spatial resolution.

\item When we started working with mid-infrared surface-brightness profiles we were concerned that dust could have a significant impact in the decompositions. That is why, in \citet{CO11B} we added to the fitting scheme an extra loop to account for some mid-plane dust attenuation. However, the dust effect was found to be ``small or negligible'' so dust is not accounted for here.

\item In \citet{CO11B}, we fitted the profiles with both a gas disc and no gas disc and found that the results were similar. In \citet{CO12} we assumed a gas disc with a face-on surface mass density equal to 20\% of that of the thin disc for galaxies with a morphological type $T\geq1$. For lenticular galaxies, $T<1$, we assumed no gas disc. However, experience has shown that it is hard to assign a meaningful $T$-type to edge-on galaxies, especially when they have small angular sizes. For example ESO~533-4 is an S0$^{\rm o}$ according to the analysis of S$^4$G images in \citet{BU15}, but is an Sc according to HyperLeda. When studied with an integral field unit \citep{CO15}, ESO~533-4 shows significant star formation in the mid-plane which argues against an S0 classification. Because of the huge uncertainty in the $T$-type classification we feel uneasy about making strong statements on what is an S0 and what is not. Therefore, in this paper, we apply a flat rate to the face-on gas surface mass density, namely 20\% that of the thin disc.
\end{enumerate}

\subsection{Step 2: Vertically averaged axial surface-brightness profiles}

Here we explain how we produce the axial surface-brightness profiles and how those are decomposed into their disc and CMC components.

\subsubsection{Surface-brightness profiles parallel to the galaxy mid-plane}

\label{axial}

We assume galaxies to be roughly bi-symmetric and produced a single quadrant image by folding the galaxies with respect to the mid-plane and the $x=0$ axis. Axial surface-brightness profiles were produced from these folded images by averaging the light from $z=0$ to $z=z_{\rm u}$ where $z_{\rm u}$ is the average of the heights at which $\mu=26\,{\rm mag\,arcsec^{-2}}$ according to the valid fits of the vertical surface-brightness profiles. The definition of $z_{\rm u}$ is admittedly somewhat arbitrary. However, this height is a reasonable compromise between the need to include as much light as possible in the profile and the need to include little background noise.

The profiles were logarithmically binned in their axial direction. That is, each data point corresponds to an average of the  surface brightness over an axial range 1.03 times wider than that of the previous data point. The first data point of each profile corresponds to the size of an S$^4$G pixel or $0\farcs75$.

\subsubsection{Functions chosen for the axial profile fit}

\label{axialfunctions}

When seen face-on, discs are generally found to have roughly exponential radial surface-brightness profiles \citep{deVA59}. Those profiles can have down-bending \citep{FREE70} and/or up-bending breaks \citep{POH06, ER08}. Breaks result in piece-wise exponential profiles. Unbroken profiles are called Type~I profiles. Type~II and III profiles denote  profiles with down-bending and up-bending breaks, respectively. Profiles can have both down-bending and up-bending breaks, causing some galaxies to have classifications such as, for example, Type~II+III+II. A CMC often sits at the centre of a disc. The CMC can be a classical bulge, a pseudo-bulge \citep{KOR93}, or even a superposition of both called a composite CMC \citep{FAL06, ER15}. In some views, boxy/peanut/X-shape features are also a part of the CMC although -- because they are a part of the bar -- they can also be considered as belonging to the disc \citep{ATH05, LAU14, LAU17, SA17}. It is customary to parametrise the CMC with a S\'ersic function \citep{SER63},
\begin{equation}
\label{sersic}
I_{\rm CMC,0}(r)=I_{\rm CMC,0}\,{\rm e}^{-\left(r/r_{\rm CMC}\right)^{\frac{1}{n}}}.
\end{equation}
In the above expression $I_{\rm CMC,0}$ corresponds to the central surface brightness of the CMC, $r_{\rm CMC}$ denotes its scale radius, and $n$ is the so-called S\'ersic index that describes the shape of the CMC.

Type~I discs have a face-on surface brightness defined by
\begin{equation}
 I_{\rm disc}(r)=I_{\rm disc,0}\,{\rm e}^{-r/h}
\end{equation}
where $I_{\rm disc,0}$ is the central disc surface brightness and $h$ is the disc scale-length.

We assume that broken discs have a face-on surface-brightness profile that can be parametrised by the generalisation of the broken exponential function in \citet{ER08} proposed in \citet{CO12} to describe discs with more than one break. The function is
\begin{equation}
\label{faceon}
 I_{\rm disc}(r)=S\,I_{\rm disc,0}\,{\rm e}^{-r/h_1}\prod_{i=2}^{i=m}\left\{\left[1+{\rm e}^{\alpha_{i-1,i}\left(r-r_{i-1,i}\right)}\right]^{\frac{1}{\alpha_{i-1,i}}\left(\frac{1}{h_{i-1}}-\frac{1}{h_i}\right)}\right\}
\end{equation}
where $S$ is a normalisation factor set so that $I_{\rm disc}(r=0)=I_{\rm disc,0}$,
\begin{equation}
 S^{-1}=\prod_{i=2}^{i=m}\left\{\left[1+{\rm e}^{-\alpha_{i-1,i}r_{i-1,i}}\right]^{\frac{1}{\alpha_{i-1,i}}\left(\frac{1}{h_{i-1}}-\frac{1}{h_i}\right)}\right\}.
\end{equation}
The break radii between the segments $i-1$ and $i$ are denoted by $r_{i-1,i}$, whereas $\alpha_{i-1,i}$ parametrises the sharpness of the transition between the segments $i-1$ and $i$, and $h_i$ denotes the exponential scale-length of the segments. The number of exponential segments is denoted by $m$. Because here we observe edge-on discs, the axial surface-brightness profiles are described by the integration of Eq.~\ref{faceon} along the line of sight. This integral is then averaged over the height of study
\begin{equation}
\label{edgeon}
 J_{\rm disc}(x)=\frac{\int_{-\infty}^{\infty}I_{\rm disc}(\sqrt{x^2+y^2})\,dy}{2z_{\rm u}}.
\end{equation}
The assumption that all the disc light is comprised between heights $z=-z_{\rm u}$ and $z=z_{\rm u}$ is implicit in the above equation.

For galaxies that have a CMC, $J_{\rm disc}$ is insufficient to describe the surface-brightness profile. In those cases we need to account for the CMC contribution:
\begin{equation}
 J(x)=J_{\rm disc}(x)+J_{\rm CMC}(x),
\end{equation}
where $J_{\rm CMC}(x)$ corresponds to the vertical average of the S\'ersic function at a given $x$,
\begin{equation}
\label{sersic2}
J_{\rm CMC}(x)=\frac{\int_{-z_{\rm u}}^{z_{\rm u}}I_{\rm CMC}(\sqrt{x^2+z^2})\,dz}{2z_{\rm u}}.
\end{equation}
Here we have assumed that CMCs are intrinsically spherical because pseudo-bulges and boxy/peanut/X-shaped CMCs are parametrised as part of the discs, as explained below.

\subsubsection{Fitting the axial surface-brightness profiles}

\label{axialactual}

\begin{figure}
\begin{center}
  \includegraphics[width=0.48\textwidth]{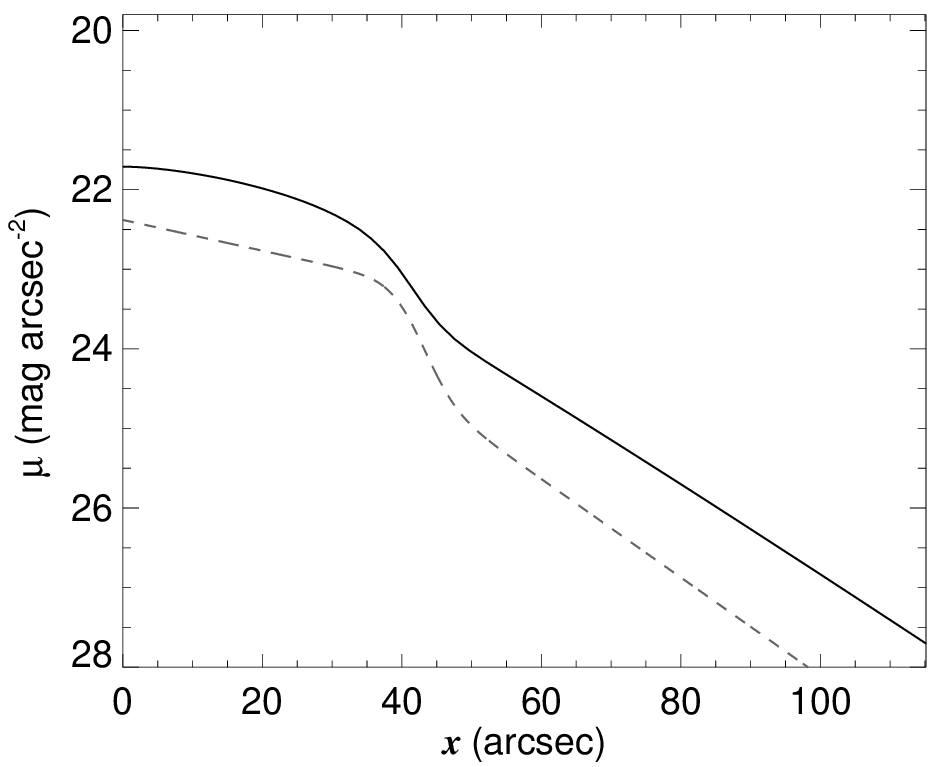}
  \end{center}
  \caption{\label{soft} Comparison of the fitted unconvolved edge-on surface-brightness profile for the NGC~3454 disc (dark line) with the one that would be observed if the galaxy were seen face-on (grey dashed line). The surface brightness of the face-on profile is scaled to appear close to the edge-on profile in the plot.}
\end{figure}

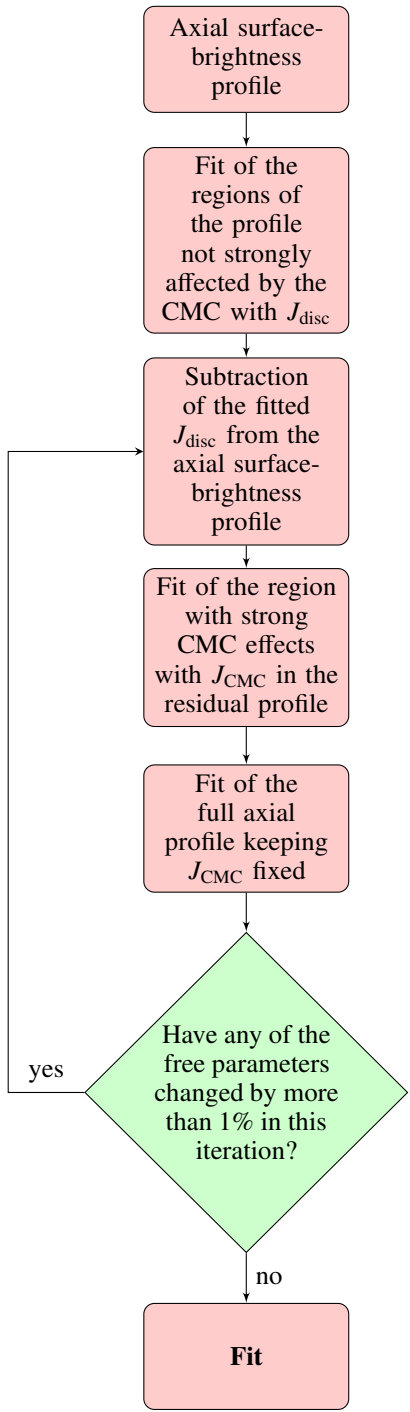
\begin{figure}
\begin{center}
\begin{tikzpicture}[node distance = 2.2cm, auto]
    \node [block] (init) {Axial surface-brightness profile};
    \node [block, below of=init, node distance=2.4cm] (fit) {Fit of the regions of the profile not strongly affected by the CMC with $J_{\rm disc}$};
    \node [block, below of=fit, node distance=2.8cm] (subtr) {Subtraction of the fitted $J_{\rm disc}$ from the axial surface-brightness profile};
    \node [block, below of=subtr, node distance=2.6cm] (bfit) {Fit of the region with strong CMC effects with $J_{\rm CMC}$ in the residual profile};
    \node [block, below of=bfit, node distance=2.4cm] (dfit) {Fit of the full axial profile keeping $J_{\rm CMC}$ fixed};
    \node [decision, below of=dfit, node distance=3.5cm] (good) {Have any of the free parameters changed by more than 1\% in this iteration?};
    \node [block, below of=good, node distance=3.5cm] (end) {{\bf Fit}};

    \path [line] (init) -- (fit);
    \path [line] (fit) -- (subtr);
    \path [line] (subtr) -- (bfit);
    \path [line] (bfit) -- (dfit);
    \path [line] (dfit) -- (good);
    \draw [solid] (good.west) -- +(-1,0) node[midway,above]{yes};
    \draw [-stealth] (good.west) -- +(-1,0) |- (subtr.west);
    \path [line] (good) -- node [midway]{no}(end);

\end{tikzpicture}
\end{center}
\caption{\label{flowaxial} Flow diagram describing the fitting procedure of the axial surface-brightness profiles as explained in Sect.~\ref{axialactual}.}
\end{figure}

\begin{figure}
\begin{center}
  \includegraphics[width=0.48\textwidth]{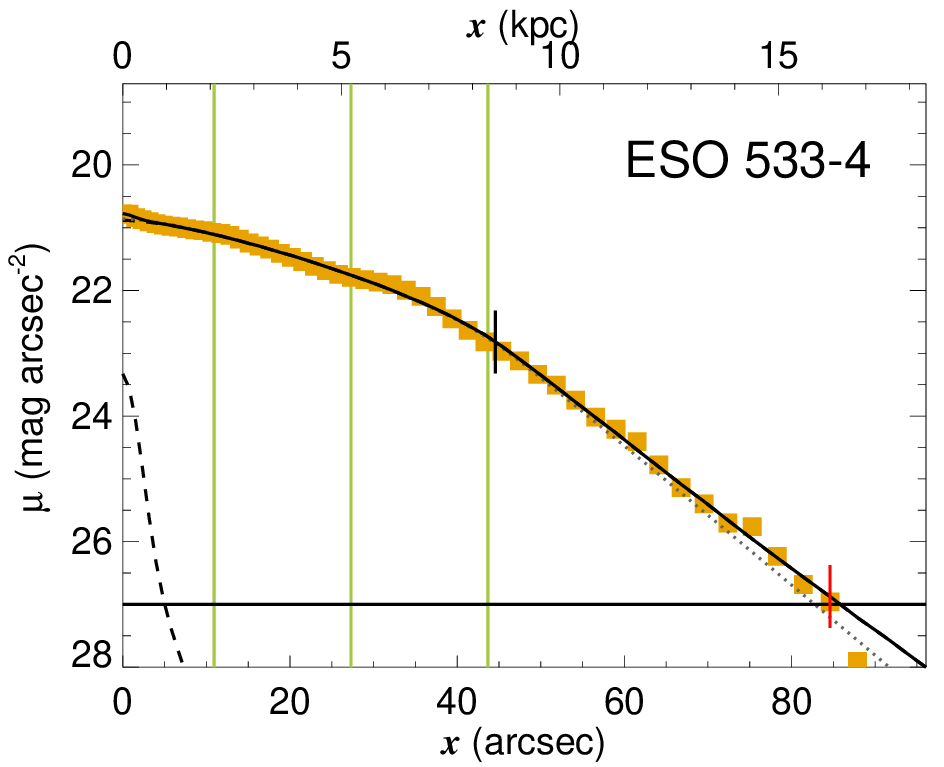}
  \includegraphics[width=0.48\textwidth]{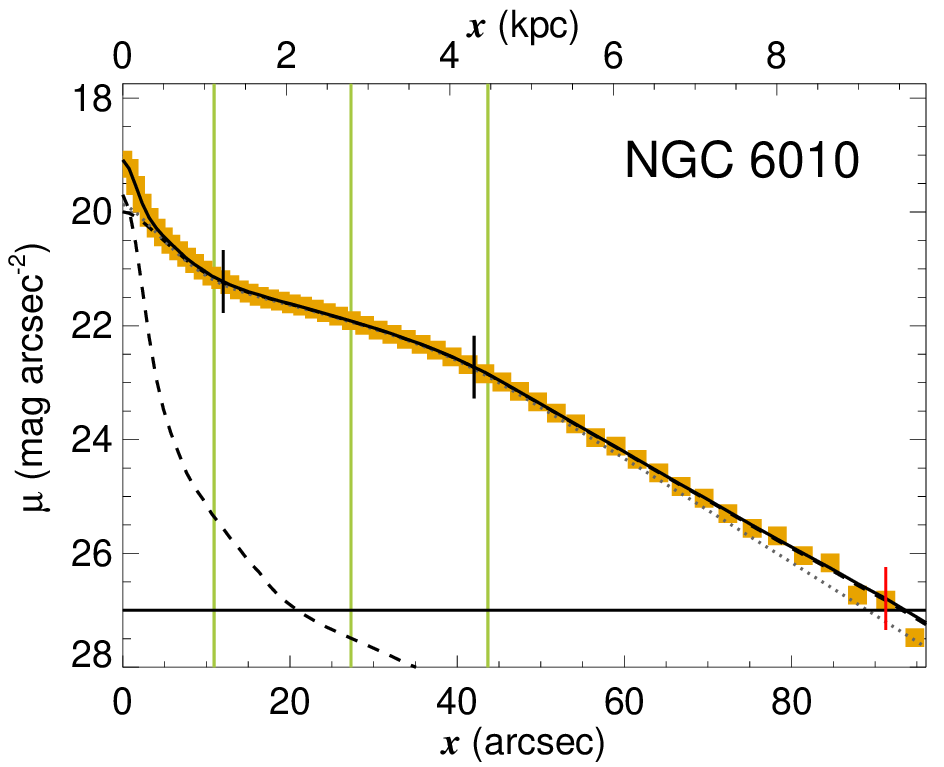}
  \end{center}
  \caption{\label{axialfit} Axial surface-brightness profiles (yellow symbols) and fits (continuous black curve) for ESO~533-4 ({\it top}) and NGC~6010 ({\it bottom}). The fitted disc and CMC contributions are indicated with dashed lines. The small vertical black lines indicate the break radii. The red vertical lines indicate the outermost fitted point. The vertical green lines indicate the axial extent of the $0.2\,r_{25}<\left|x\right|<0.5\,r_{25}$ and $0.5\,r_{25}<\left|x\right|<0.8\,r_{25}$ vertical surface-brightness profiles. The horizontal lines indicate the $\mu=27\,{\rm mag\,arcsec^{-2}}$ level. The grey dotted curves indicate the unconvolved disc fitted model.}
\end{figure}

We plotted the axial surface-brightness profiles and we inspected them to visually choose the number of exponential segments in the disc and discern whether a galaxy has a CMC or not. Identifying the exponential segments is possible because the edge-on line-of-sight projection of the piece-wise exponential profile results in a profile that is very similar to that of a piece-wise exponential (see Fig.~\ref{soft}, where we compare the fitted axial surface-brightness profile of a galaxy with what the radial surface-brightness profiles would look like if the galaxy were seen face-on). The qualitative differences between the face-on and the edge-on profiles are that the central cusp and the breaks soften in projection. The maximum number of segments that we found in a single galaxy is four.

The axial profiles were truncated at a surface brightness of $\mu=27\,{\rm mag\,arcsec^{-2}}$. First guesses of the exponential scale-lengths were obtained by fitting an exponential function to the disc segments identified in the axial profiles. The fits were done over the profiles in units of mag\,arcsec$^{-2}$. We were careful not to include regions strongly affected by a classical bulge in those preliminary fits. Typically, box/peanut/X-shape CMCs appeared as exponential segments with a distinct slope and were fitted as part of the disc. In fact, it is likely that many features that would be identified as a pseudo-bulge in a face-on view are here identified as ``normal'' disc sections. Galaxies with central components not obeying an exponential law were considered to host a bona fide CMC (most likely a classical CMC). Sometimes those central components are unresolved, so some could actually be nuclear star clusters. 

In Sect.~\ref{mock} we discussed how we produced mock 1D PSFs to convolve the synthetic vertical surface-brightness profiles during the fitting process. We also made 1D mock PSFs for the axial profiles. Just as the mock PSF for the vertical profiles was made by PSF-convolving an infinitely thin axial profile, the mock PSF for the axial profiles is created by convolving an infinitely narrow vertical profile. This vertical profile was obtained by averaging the valid vertical profile fits. The number of valid vertical profile fits was between two and four (the reasons for that are explained in Sect.~\ref{sample}).

For galaxies with no CMC, we fitted Eq.~\ref{edgeon} in a straight-forward way using {\sc IDL}'s {\sc curvefit}. The free parameters are $I_{\rm disc,0}$, $h_{\rm i}$, and $r_{\rm i-1,i}$ whereas we fix $\alpha_{i-1,i}=0.5\,{\rm arcsec}^{-1}$, which is representative of what is found in \citet{ER08}. This was done to avoid an excess of free parameters in profiles with two or three breaks and the selected $\alpha$ value was found to work well in \citet{CO12}.

We found that galaxies with a CMC often could not have their CMC and disc component fitted at the same time because of the large number of free parameters. The number of free parameters of the CMC is three ($I_{\rm CMC,0}$, $n$, $r_{\rm CMC}$) and that of the disc varies from two for a Type~I disc to eight for a disc with three breaks. To circumvent the convergence problems we separately fitted the CMC and the disc components as summarised in the flow diagram in Fig.~\ref{flowaxial}. We manually divided the axial range. We first fitted the region little affected by the CMC with Eq.~\ref{edgeon}. The result of this fit was subtracted from the surface-brightness profile. The region with a significant CMC contribution of the residual profile was then fitted using Eq.~\ref{sersic2}. To do so, we produced a 2D image of the CMC using the 2D convolution of Eq.~\ref{sersic} which was then vertically integrated. Then, the full axial profile was fitted keeping the $J_{\rm CMC}$ part fixed. We then produced a new residual profile with the updated $J_{\rm disc}$ and fitted the part of it with a significant CMC contribution with $J_{\rm CMC}$. The iteration process was repeated until all the parameters varied by less than $1\%$.

Fig.~\ref{axialfit} shows the axial surface-brightness profiles and the fits for ESO~533-4 and NGC~6010. The grey dotted curves indicate the disc profiles before the PSF convolution is applied. Naturally, the disc PSF effects in the axial direction are much subtler than in the vertical direction. The PSF effects are so small that it seems unlikely that they could create features mimicking an up-bending break, at least within the surface-brightness limits of the S$^4$G \citep[but see how it could affect deeper surface brightness levels in][]{TRU16, PE17}. The only error caused by not accounting for the PSF would be to slightly overestimate the scale-length of the outermost exponential segment.

\subsection{Step 3: Vertical surface-brightness profile fit accounting for the presence of a CMC}

\label{finalvertical}

There are two reasons for repeating the vertical surface-brightness profile fits from Step~1 (Sect.~\ref{actual}). First, the details of the axial profiles are not well described by a single scale-length. Second, the first round of fits did not account for the presence of a CMC.

Before running the second round of vertical profile fits, we produced a new set of 1D mock PSFs. This time we used $J_{\rm disc}(x)$ to describe the infinitely thin axial profile. Instead of producing the mock PSF from a vertical cut in the 2D convolution of the axial 1D profile, we made it by averaging over the whole axial bin concerned.

The light of the CMC was accounted for by subtracting a 2D model based on the fit in Sect.~\ref{axialactual} from the S$^4$G images before the vertical surface-brightness profiles were obtained.

The improved vertical profile fits were then produced as in Sect.~\ref{actual}. We note however that our approach ignores the gravitational interaction between the CMC and the disc, that is, we do not include it in Eq.~\ref{narayan} when solving the vertical surface-brightness profiles. The effect of the CMC is minimised by deliberately ignoring the axial range $-0.2\,r_{25}<x<0.2\,r_{25}$ where it might even dominate the surface brightness in a few galaxies. At $\left|x\right|>0.2\,r_{25}$, the CMC contribution is often small and/or caused by the PSF convolution of a bright central cusp.

Examples of vertical surface-brightness profile fits accounting for the CMC are shown in the bottom panels in Figs.~\ref{verfitnobul} and \ref{verfitnobul2}. Typically the fits do not change drastically when compared to those made in Sect~\ref{actual}. Furthermore, some of the difference is not due to the inclusion of a CMC component, but is due to the use of a different 1D mock PSF. This is reassuring, since large differences in the fitted models would imply that the CMC has a significant impact and its gravitational effect could not safely be neglected.

The vertical surface-brightness profiles obtained as explained in this section are considered to be the ``final'' fits from which structural galaxy parameters such as disc scale-heights are obtained. We find that, on average, the fits are done over a dynamical range 0.1\,mag larger than those in \citet{CO12} when the overlapping galaxies between the two samples (see Sect.~\ref{sample}) are compared.

\subsubsection{The effect of using a symmetrised PSF on fitting the vertical surface-brightness profiles}

\label{symmetryzed}

\begin{figure*}
\begin{center}
  \includegraphics[width=0.48\textwidth]{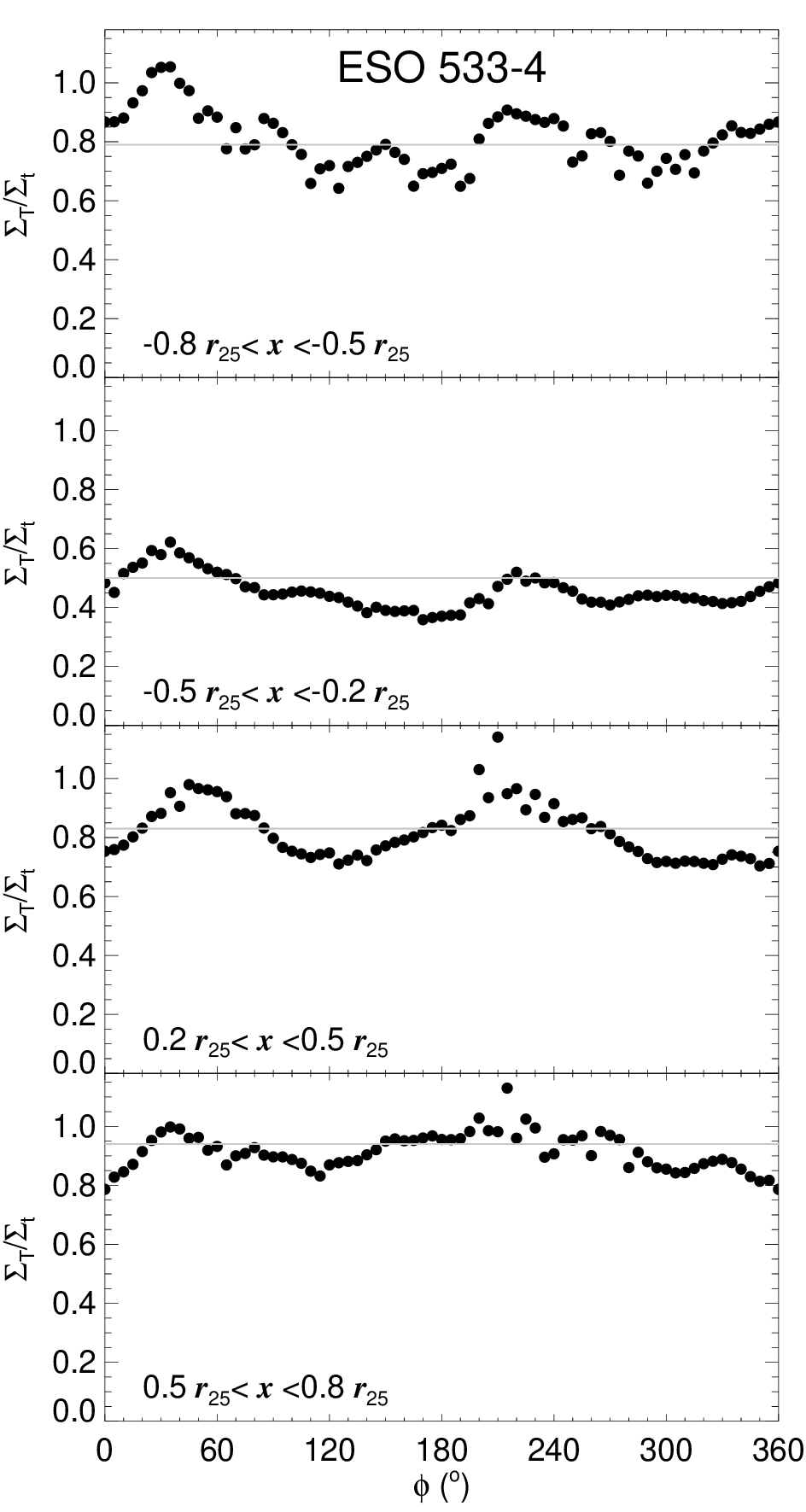}
  \includegraphics[width=0.48\textwidth]{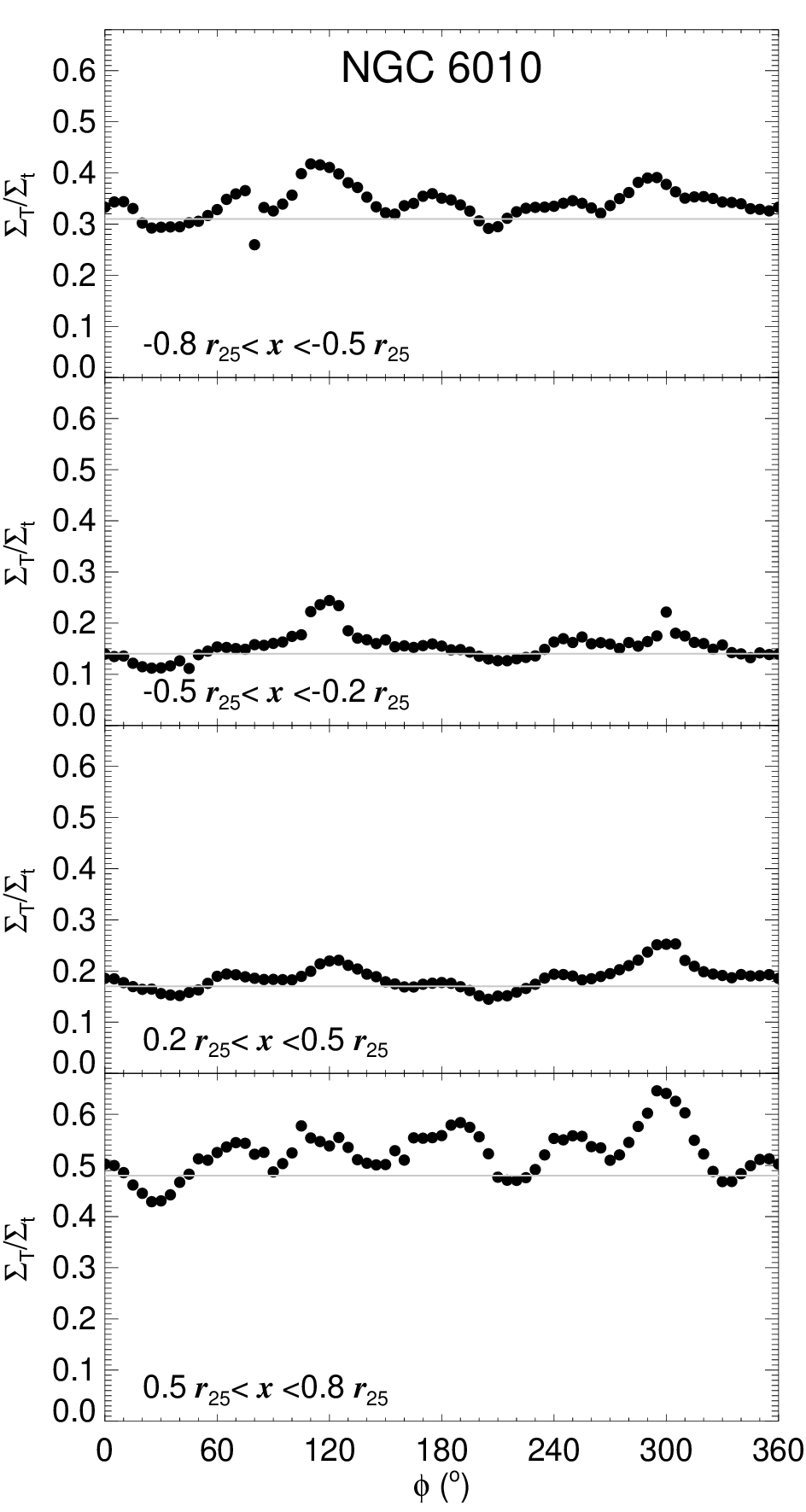}
  \end{center}
  \caption{\label{rotation} {\it Left} column: the ratio of the thick to thin disc face-on surface density -- $\Sigma_{\rm T}/\Sigma_{\rm t}$ -- for the four axial bins of ESO~533-4. Each data point corresponds to a fitted value obtained using a non-symmetrised PSF rotated $\phi$ degrees clockwise with respect to the direction indicated by the mid-plane of ESO~533-4. The grey horizontal lines indicate the $\Sigma_{\rm T}/\Sigma_{\rm t}$ values obtained in the fits made using the symmetrised PSF (same as in the bottom panels in Fig.~\ref{verfitnobul}). {\it Right} column: Same as the {\it left} column but for NGC~6010 and with $\Sigma_{\rm T}/\Sigma_{\rm t}$ values indicated by the horizontal grey line corresponding to Fig.~\ref{verfitnobul2}.}
\end{figure*}

Our fits use a symmetrised version of the IRAC PSF. However, the IRAC PSF has significant departures from rotational symmetry. First, the PSF has diffraction spikes with a six-fold symmetry. Second, the PSF has a bright annulus, likely to be a reflection, $\sim4\arcsec$ in radius centred at $\sim14\arcsec$ of the PSF maximum. This annulus is placed between two of the spikes.

To test the reliability of the fits made with the symmetrised PSF, we fitted the vertical surface-brightness profiles of two galaxies with a non-symmetrised PSF using 72 equispaced orientations (Fig.~\ref{rotation}). The test shows that the obtained thick to thin disc surface density ratios -- $\Sigma_{\rm T}/\Sigma_{\rm t}$ -- have typical deviations of $10-20\%$ with respect to the values computed using a symmetrised PSF. Some of the fit series (like the second and third axial bins for ESO~533-4) show a two-fold symmetry, probably caused by the reflection (the vertical surface-brightness profiles are symmetrised with respect to the mid-plane). Other series of fits, like that of the third axial bin of NGC~6010, show a clear $60\degr$ modulation, likely to be caused by the diffraction spikes.

The uncertainties in $\Sigma_{\rm T}/\Sigma_{\rm t}$ add up to those discussed in Sect.~\ref{assumptions}. It should however be noted that the $10-20\%$ uncertainties estimated from Fig.~\ref{rotation} are a worst-case scenario. Indeed, the S$^4$G science-ready images are made of two sets of exposures with two different orientations. The resulting PSF is necessarily somewhere between the symmetrised PSF and the single-orientation IRAC PSF. Thus the uncertainties estimated from Fig.~\ref{rotation} correspond to a case with similar orientations for the two exposures or to integer multiples of $60\degr$ in the difference between the orientations. In the latter case, the diffraction spikes of the two sets of exposures would overlap, so there would be little smoothing due to the sum of the two PSFs.

\subsection{Step 4: Axial surface-brightness profile fit for the heights dominated by thin and thick disc light}

In Sect.~\ref{axial} we decompose the vertically integrated axial profile of galaxies for the whole height of the disc. Here we decompose the axial profiles vertically integrated over the heights dominated by the thin and the thick disc light.

\label{axial2}

\subsubsection{Production of the axial profiles for thin and thick disc-dominated heights}

\label{zcs}

We define $z_{\rm c1}$ as the height below which the thin disc surface brightness exceeds that of the thick disc according to the fits in Sect.~\ref{finalvertical}. This height is an average over all the axial bins with valid vertical surface-brightness fits. The thin disc typically dominates the vertical profiles at large heights due to the PSF wings, as discussed in Sect.~\ref{actual}. We define $z_{\rm c2}$ as the height above which the thin disc is again brighter than the thick disc. This height is an average over all the valid vertical fits. If $z_{\rm c2}>z_{\rm u}$ we redefine $z_{\rm c2}=z_{\rm u}$ where $z_{\rm u}$ is the average height at which $\mu=26\,{\rm mag\,arcsec^{-2}}$ according to the vertical profile fits in Step~1 (Sect.~\ref{axial}). An axial surface-brightness profile dominated by thin disc was produced by averaging the light between $z=0$ and $z=z_{\rm c1}$. A thick disc-dominated profile was made by averaging the light between $z_{\rm c1}$ and $z_{\rm c2}$. The profiles were logarithmically binned, as done in Sect.~\ref{axial} for the axial profiles for the whole disc. The reason to limit $z_{\rm c2}$ to be smaller than or equal to $z_{\rm u}$ was to reduce the background noise in the axial profile dominated by thick disc light.

Not all galaxies have well defined $z_{\rm c1}$ and $z_{\rm c2}$ values. Some galaxies are thick or thin disc-dominated at all heights. Others have valid $z_{\rm c1}$ and $z_{\rm c2}$ values for a given axial bin, but not for others. No thin- and thick-disc-dominated profiles were produced in these cases.

The height $z_{\rm u}$ was not updated based on the fits in Sect.~\ref{finalvertical}. Indeed, its definition is somewhat arbitrary, so small $z_{\rm u}$ refinements are superfluous.

\subsubsection{Fitting of the axial profiles of the thin and thick discs}

\label{finalaxial}

Thin- and thick-disc axial profiles had their CMC contribution removed. That contribution was estimated from the CMC fits in Sect.~\ref{axialactual}. The axial profiles were then fitted using the disc functions in Sect.~\ref{axialfunctions} following the same procedure as for the galaxies without a CMC in Sect.~\ref{axialactual}. The only difference here is that we did not make any attempt to account for the PSF effects. The reason is that the 1D mock PSFs created for the fits in Sect.~\ref{axialactual} only work under the assumption that the axial profile is made by integrating light between a height and the same height at opposite sides of the mid-plane, for example, $-z_{\rm k}$ and $z_{\rm k}$. Thus, a fit of the thick disc axial profile accounting for the PSF would require a more complicated modelling that is beyond the scope of this paper. This might cause us to slightly overestimate the actual scale-length of the disc segments, but this effect is not likely to be large as discussed in Sect.~\ref{axialactual} (see also Fig.~\ref{axialfit}).

When the thin or the thick disc dominated at all heights we reused the fit made in Sect.~\ref{axialactual}.

\subsection{Step 5: Calculating the mass of the galaxy components}

\label{components}

Mass-to-light ratios are necessary to estimate the mass of the galaxy components. As in \citet{CO14}, we assume the thin disc mass-to-light ratio to be $\left(\Upsilon_{\rm t}\right)_{\rm 3.6\mu{\rm m}}=0.55\,\left(\mathcal{M}_{\bigodot}/L_{\bigodot}\right)_{3.6\mu{\rm m}}$ so the global mass-to-light ratio of a galaxy is $\mu_{3.6\mu{\rm m}}\sim0.6\,\left(\mathcal{M}_{\bigodot}/L_{\bigodot}\right)_{3.6\mu{\rm m}}$ in accordance with \citet{MEIDT14}, once we account for $\left(\Upsilon_{\rm T}/\Upsilon_{\rm t}\right)_{3.6\mu{\rm m}}=1.2$. We assume that $\left(\Upsilon_{\rm CMC}\right)_{3.6\mu{\rm m}}=\left(\Upsilon_{\rm T}\right)_{3.6\mu{\rm m}}$. This is based on the finding that, at least for the Milky Way, the CMC and the thick disc share a similar star formation history \citep{MEL08}.

The intrinsic luminosity of a galaxy was obtained from the $3.6\mu{\rm m}$ apparent magnitudes in the S$^4$G Pipeline~3. For galaxies in the S$^4$G extension, no Pipeline~3 products are available yet (as of June 2017), so the photometry was produced by ourselves. The apparent magnitudes were transformed into an absolute luminosity, $L_{\rm total}$, using the galaxy distances as described in Sect.~\ref{sample}. The intrinsic luminosities were transformed into units of solar luminosities \citep[$\left(L_{\bigodot}\right)_{3.6\mu{\rm m}}=6.06\,{\rm mag}$;][]{OH08}.

An estimate of the light fraction emitted by the CMC and the disc -- $q_{\rm CMC}$ and $q_{\rm disc}$, respectively -- is obtained from the axial fits in Sect.~\ref{axialactual}. This can be transformed into a CMC mass
\begin{equation}
\mathcal{M}_{\rm CMC}=q_{\rm CMC}L_{\rm total}\left(\Upsilon_{\rm CMC}\right)_{3.6\mu{\rm m}}.
\end{equation}

The vertical surface-brightness profile fits in Sect.~\ref{finalvertical} yield thin and thick disc vertical surface-brightness profiles. We use those profiles to compute the thin and thick disc emission in a given axial bin, $L_{{\rm t},i}$ and $L_{{\rm T},i}$. We then obtain thick to thin disc mass ratios
\begin{equation}
\frac{\mathcal{M}_{\rm T}}{\mathcal{M}_{\rm t}}=\frac{\sum_i\,L_{{\rm T},i}}{\sum_i\,L_{{\rm t},i}}\left(\frac{\Upsilon_{\rm T}}{\Upsilon_{\rm t}}\right)_{3.6\mu{\rm m}}
\end{equation}
where the sums are done over all the bins with valid fits. We note that this $\mathcal{M}_{\rm T}/\mathcal{M}_{\rm t}$ definition is not exactly the same as that used in \citet{CO11B, CO12}. In our previous studies, $\mathcal{M}_{\rm T}/\mathcal{M}_{\rm t}$ was calculated as a weighted average of $\Sigma_{\rm T}/\Sigma_{\rm t}$ over the valid axial bin fits.

The total mass of the disc is found as
\begin{equation}
 \mathcal{M}_{\rm disc}=q_{\rm disc}L_{\rm total}\frac{1+\frac{\mathcal{M}_{\rm T}}{\mathcal{M}_{\rm t}}}{1+\frac{\mathcal{M}_{\rm T}}{\mathcal{M}_{\rm t}}\left(\frac{\Upsilon_{\rm t}}{\Upsilon_{\rm T}}\right)_{3.6\mu{\rm m}}}\left(\Upsilon_{\rm t}\right)_{3.6\mu{\rm m}}.
\end{equation}

To obtain the gas disc masses, $\mathcal{M}_{\rm g}$, we fetched the atomic gas (H{\sc i}) fluxes from HyperLeda, $f_{\rm H{\textsc i}}$. We then converted them into an atomic gas mass using the formalism in \citet{ZWAAN97}
\begin{equation}
 \frac{\mathcal{M}_{\rm H{\textsc i}}}{\mathcal{M}_{\bigodot}}=236\,d^2f_{\rm H{\textsc i}}
\end{equation}
where $d$ is the galaxy distance in Mpc, and $f_{\rm H{\textsc i}}$ is the flux of the 21\,cm line in ${\rm mJy\,km\,s^{-1}}$. To account for helium and metals we considered
\begin{equation}
 \mathcal{M}_{\rm g}=1.4\mathcal{M}_{\rm H{\textsc i}}
\end{equation}
which corresponds to a hydrogen abundance $X\approx0.7$. This approach might underestimate $\mathcal{M}_{\rm g}$ in galaxies containing large amounts of molecular gas. Eighteen galaxies have no available gas measurement, so in what follows we set their gas masses $\mathcal{M}_{\rm g}=0$.

We compute the baryonic mass of galaxies as
\begin{equation}
 \mathcal{M}_{\rm baryon}=\mathcal{M}_{\rm g}+\mathcal{M}_{\rm t}+\mathcal{M}_{\rm T}+\mathcal{M}_{\rm CMC}.
\end{equation}

\section{The sample}

\label{sample}

\begin{figure}
\begin{center}
  \includegraphics[width=0.48\textwidth]{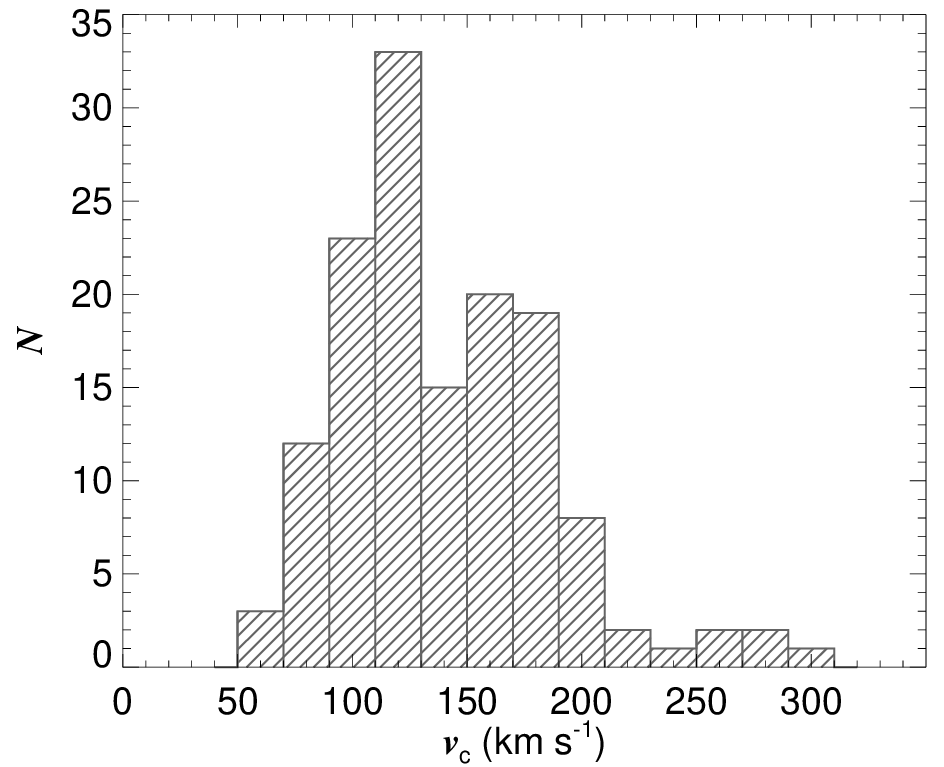}
  \includegraphics[width=0.48\textwidth]{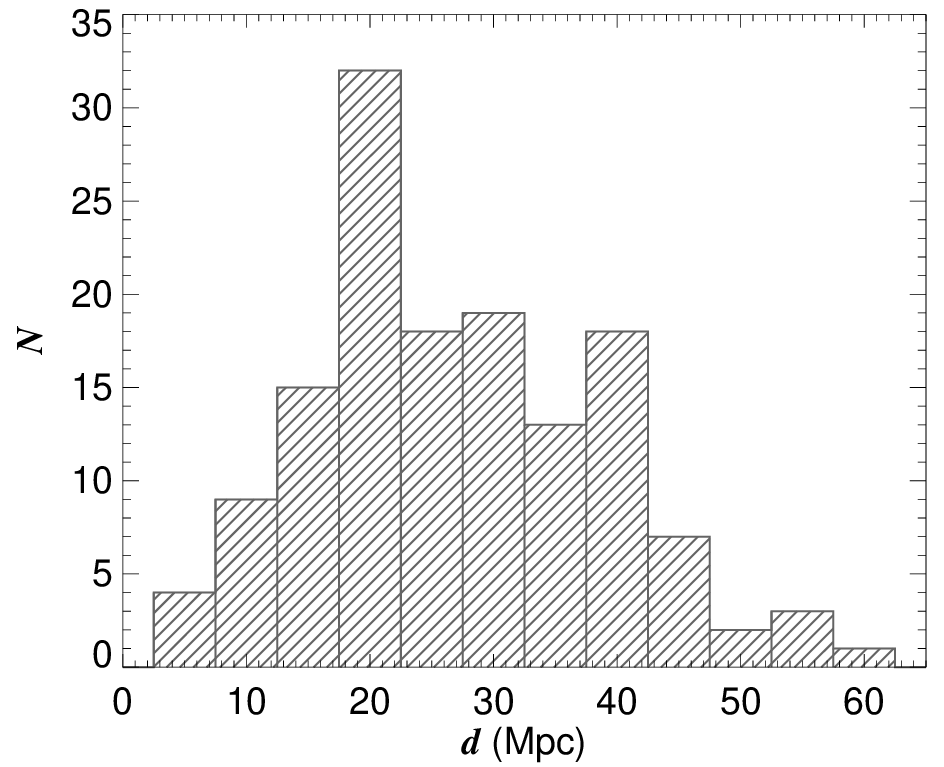}
  \includegraphics[width=0.48\textwidth]{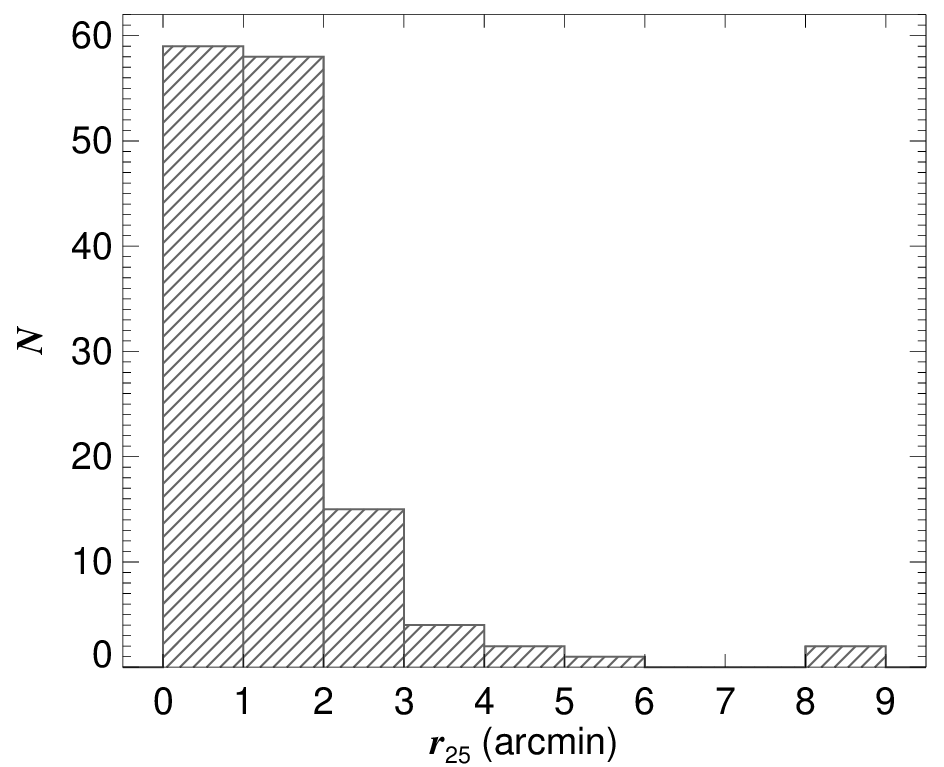}
  \end{center}
  \caption{\label{samplef} {\it Top} panel: distribution of circular velocities, $v_{\rm c}$, of the galaxies in our final sample. The bin width is $20\,{\rm km\,s^{-1}}$. {\it Middle} panel: distribution of distances to the galaxies. The bin width is $5\,{\rm Mpc}$. {\it Bottom} panel: distribution of the isophotal $25\,{\rm mag\,arcsec^{-2}}$ radii in the $B$-band, $r_{\rm 25}$. The bin width is $1\arcmin$.}
\end{figure}

Our final sample consists of 141 edge-on galaxies that were selected as specified below.

We visually pre-selected all the edge-on galaxies in the S$^4$G. We also selected the edge-on galaxies in the early-type galaxy extension of the S$^4$G that had gone through Pipeline~1 -- the production of science-ready images -- by June 2017 (461 out of 465 galaxies). We attempted to avoid galaxies with conspicuous signs of disc structure such as spiral arms and rings, since those indicate a far from edge-on orientation. This is admittedly a subjective selection criterion. Indeed, early-type and distant galaxies have a smoother appearance, which might ease their access to the sample. On the other hand, a few of the galaxies in our final sample -- such as ESO~240-11, NGC~4244, NGC~4437, NGC~4565 -- show signs of what might be disc structure. The visual selection of edge-on galaxies led to a tentative sample of about 260 galaxies out of the 2813 galaxies in the S$^4$G and the processed fraction of its extension.

We run the 260 pre-selected galaxies through the fitting procedure (Sect.~\ref{procedure}). To ensure the quality of the fits, a large dynamical range is desirable. Hence, we discarded galaxies for which at least three of the vertical surface-brightness profiles had a mid-plane surface brightness fainter than $\mu(z=0)=22\,{\rm mag\,arcsec^{-2}}$. We also discarded those galaxies for which we obtained a valid fit for less than two axial bins in the fitting procedures described in Sect.~\ref{actual} and/or Sect.~\ref{finalvertical}. Poor fits were caused by several factors; most corresponded to axial bins that barely made it through the surface-brightness threshold so that they had noisy profiles at large heights. The remaining poor fits typically corresponded to angularly small galaxies, which resulted in the PSF FWHM to be comparable to the width of the axial bins. In this case, point sources superimposed on the disc may significantly affect the photometry, which results in non-fitting profiles. In a few cases the fits were not good due to the presence of neighbouring galaxies or saturated stars. Finally, three galaxies had no valid Sect.~\ref{finalvertical} fits because of the presence of dominant and extended CMC components within which the disc is embedded (NGC~1596, NGC~5084, and NGC~7814).

Throughout our edge-on galaxies study we have used the maximum circular velocity of galaxies, $v_{\rm c}$, as a proxy for their mass. We have thus excluded from the sample those galaxies for which such velocities were not available in the literature. That was the case for five galaxies in the S$^4$G early-type galaxy extension that had gone through our previous selection criteria.

Our 141 galaxy sample is twice as large as that in \citet{CO12}. There are several reasons for this. First, when we compiled our previous sample we only considered a subset of 2132 S$^4$G galaxies which had been processed by the pipelines at that point. Second, we here include the edge-on galaxies from the early-type galaxy S$^4$G extension. Third, we reran our fitting code over edge-on galaxies that did not make it to our previous samples. Some of these galaxies are now included because the new PSF has yielded good fits. Also, the fit quality criteria required to make it into the final sample have been slightly relaxed. For example, we do not require the thick disc to start to dominate the surface brightness at a roughly constant height, we do not exclude galaxies that seem to require a third disc to be fitted all the way down the typical S$^4$G sensitivity ($\mu\sim26\,{\rm mag\,arcsec^{-2}}$), and we do not remove galaxies whose light distributions are compatible with those of a single-disc galaxy. Conversely, six galaxies that were in the \citet{CO12} sample are not studied here. In two cases (NGC~1596 and NGC~5084) this is due to the presence of a very extended CMC. In the other four cases the minimum two good vertical surface-brightness profile fits were not obtained.

As mentioned above, maximum circular velocities of galaxies are needed. For the sake of data uniformity we preferably used data from the comprehensive Extragalactic Distance Database \citep[EDD;][]{TU09}. The EDD contains H{\sc i} line width information from several sources. The $v_{\rm c}$ data for our galaxies come from \citet{SPRIN05}, \citet{THEU06}, and \citet{COUR09} as well as from pre-digital sources compiled by the EDD authors. For galaxies not in the EDD we looked for alternative sources for $v_{\rm c}$, including those with velocities obtained from stellar absorption lines, which are a good alternative for H{\sc i} determination in galaxies with little gas. Those additional $v_{\rm c}$ sources are \citet{KRUMM76}, \citet{BAL81}, \citet{RICH87}, \citet{HAY90}, \citet{DON95}, \citet{MATH96}, \citet{SI97, SI98, SI00}, \citet{RU99}, \citet{KAR04}, \citet{CHUNG04}, \citet{MEY04}, \citet{BED06}, and ATLAS3D \citep{CA11, KRA11}.

Galaxy distances are required to know the physical scales of discs. Again for the sake of data homogeneity we have used distance determinations made using the Tully-Fisher relation because that is the redshift-independent distance determination that was available for the largest number of galaxies. The sources are Cosmicflows 1 and 3 \citep{TU08, TU16}, the SFI++ \citep{SPRIN07}, and distance determinations from \citet{THEU07} averaged over as many bands as possible. Finally, if no Tully-Fisher distance determination was available in the literature, we used the NED Hubble-Lema\^itre flow distance with respect to the cosmic microwave background assuming a Hubble-Lema\^itre constant $H_0=75\,{\rm km\,s^{-1}\,Mpc^{-1}}$.

Some of the properties of the sample are summarised in Fig.~\ref{samplef}. The circular velocities range from $v_{\rm c}=50\,{\rm km\,s^{-1}}$ to $v_{\rm c}=300\,{\rm km\,s^{-1}}$ with most of the galaxies in the $v_{\rm c}=100-200\,{\rm km\,s^{-1}}$ interval. Hence, Milky Way-sized galaxies ($v_{\rm c}>200\,{\rm km\,s^{-1}}$) are relatively rare in our sample (9 of 141 galaxies). The sample contains galaxies out to $d=60\,{\rm Mpc}$ and samples well the $v_{\rm c}=10-40\,{\rm Mpc}$ distance range. The reason why we have galaxies with $d>40\,{\rm Mpc}$ -- the S$^4$G limiting distance -- is the use of different distance indicators in the S$^4$G and here (Hubble-Lema\^itre distances versus Tully-Fisher distances).

The isophotal $25\,{\rm mag\,arcsec^{-2}}$ radii in the $B$-band for the galaxies in our sample ranges between $r_{25}\gtrsim0\farcm5$ and slightly over $r_{25}=8\arcmin$ for NGC~4244 and NGC~4565 (bottom panel in Fig.~\ref{samplef}). The $r_{25}$ values were obtained from HyperLeda. The lower limit of the $r_{25}$ distribution is set by the minimum size for a galaxy to be included in the S$^4$G. Most of the galaxies in the sample -- 132 out of 141 -- have $r_{25}<3\arcmin$.

Our sample is slightly biased against early-type galaxies because we exclude a few galaxies from the S$^4$G early-type galaxy extension for which we were not able to find a $v_{\rm c}$ value. Our sample is also biased against galaxies with a low $v_{\rm c}$ because those typically have a lower peak surface brightness than more massive galaxies \citep[Fig.~6 in][]{YOA06} so they are less likely to go through our mid-plane surface-brightness threshold. Because of these biases our sample is not complete but is nevertheless representative of galaxies in the local universe.

\section{Results and discussion}

\label{results}

In this section we revisit some of the results found in our previous papers \citep{CO11B, CO12, CO14} to check how a proper PSF treatment affects them.

\subsection{Galaxies with one, two, and three disc components}

\begin{figure*}
\begin{center}
  \includegraphics[width=0.98\textwidth]{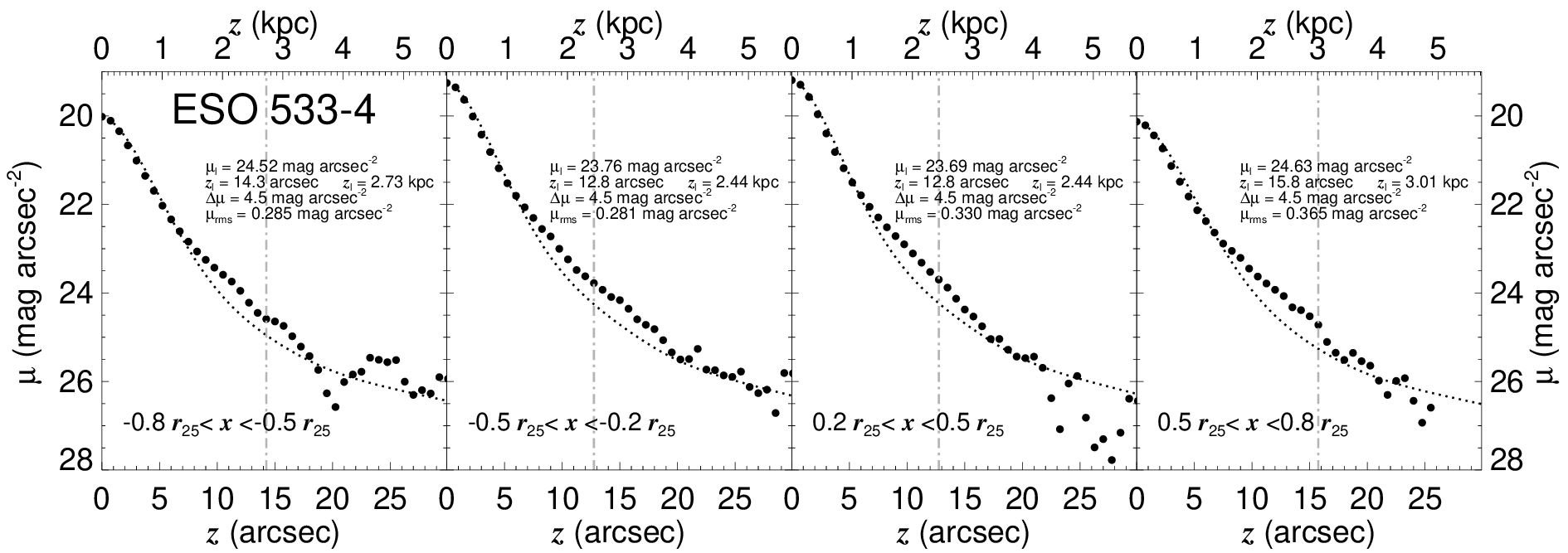}
  \includegraphics[width=0.98\textwidth]{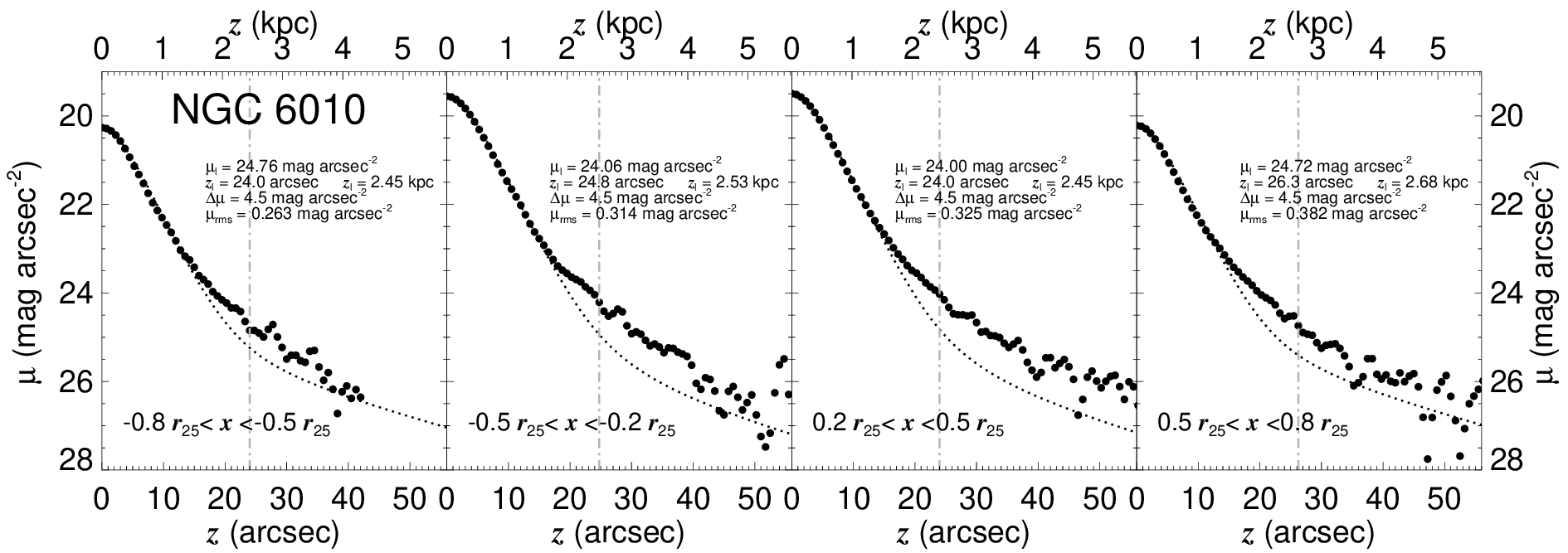}

  \end{center}
  \caption{\label{onedisc} {\it Top}: Surface-brightness profiles of the four axial bins of ESO~533-4 (large filled circles) and fits produced with a single disc component (dotted lines). The vertical dot-dashed grey lines indicate the range of heights considered for each fit. Each of the panels contains some information about the fits: the faintest surface-brightness level that was considered for the fit ($\mu_{\rm l}$), the maximum height considered for the fit ($z_{\rm l}$), the dynamic range of the fit ($\Delta\mu$), and the root mean square deviation of the fit ($\mu_{\rm rms}$). {\it Bottom}: Same as the {\it top} but for NGC~6010.}
\end{figure*}

\begin{table}
 \caption{Number of stellar discs in the sample galaxies.}
 \label{numberstat}
 \centering
 \begin{tabular}{l c c c c}
 \hline\hline
  Number& Number  & Galaxy &$\overline{v}_{\rm c}$&Median($v_{\rm c}$)\\
  of discs&of galaxies &fraction &(${\rm km\,s^{-1}}$)&(${\rm km\,s^{-1}}$)\\
  \hline
  One & 17 & $12\%\pm5\%$&113&102\\
  Two & 116 & $82\%\pm6\%$&141&132\\
  Three & 8 & $6\%\pm4\%$&179&194\\
  \hline
 \end{tabular}

\end{table}

Here we discuss the number of stellar disc components that are identified in the galaxies in our sample according to our fits.

Can the surface-brightness profiles of edge-on galaxies be fitted with a single disc component? Generally speaking the answer is an emphatic no. Indeed, fitting the vertical surface-brightness profiles with a PSF-convolved single disc in hydrostatic equilibrium \citep[a ${\rm sech}^2(z/z_0)$ function where $z_0$ is a scale-height;][]{SPI42} yields poor fits when done over a large dynamical range. An example of a run of our fitting code in Sect.~\ref{actual} over ESO~533-4 and NGC~6010 while fixing the thick disc luminosity to zero -- this means that the observed and the synthetic profile are forced to intersect at a surface brightness somewhere between 0.1 and 0.4 times the mid-plane surface brightness -- is shown in Fig.~\ref{onedisc}. Whereas a two-disc fit of ESO~533-4 is able to fit a dynamical range between $\Delta\mu=5.5\,{\rm mag\,arcsec^{-2}}$ and $\Delta\mu=7.0\,{\rm mag\,arcsec^{-2}}$ depending on the axial bin, a single-disc fit becomes stuck at $\Delta\mu=4.5\,{\rm mag\,arcsec^{-2}}$ -- the minimum dynamical range allowed by our code -- because of the overly large $\mu_{\rm rms}$. The one-disc fit grossly underestimates the high-$z$ luminosity, where the thick disc dominates in a two-disc fit. The fit is even worse for NGC~6010.

The behaviour shown here for ESO~533-4 and NGC~6010 is a good example of what can be seen in many of the remaining galaxies in the sample (124 galaxies in Appendix~\ref{twodiscap}). However, for a significant minority of the galaxies the case for a two-disc structure is not very strong (17 galaxies in Appendix~\ref{onediscap}). Whereas most of the galaxies have upward bends in their vertical profiles at typical surface-brightness levels of $\mu=22-24\,{\rm mag\,arcsec^{-2}}$, these galaxies show no bends (ESO~287-9, ESO~505-3, ESO~548-63, IC~1913, IC~3247, IC~3311, IC~5052, NGC~100, NGC~3501, NGC~4244, NGC~5023, NGC~5073, and UGC~10297) or bends compatible with those caused by the wings of the PSF convolution of a bright thin disc (ESO~443-21, NGC~489, PGC~30591, and UGC~5347).

Does the lack of an obviously detected thick disc component in several galaxies imply that those galaxies have no thick disc? This is not necessarily so. Thick discs become increasingly hard to detect as the host galaxy inclination deviates from an edge-on orientation. In \citet{CO11B} we discussed how decompositions become unreliable for inclinations smaller than $i\approx86\degr$. Figure~9 in the same paper shows how the upward bend in vertical surface-brightness profiles caused by the presence of a thick disc almost disappears at that inclination threshold. NGC~4244 -- a contentious case which might have a subtle thick disc \citep{CO11C} or not \citep{STREICH16} -- has an estimated inclination of $i=84\fdg5$ \citep{OLL96}, which might make this galaxy inadequate for a vertical-structure study. This might be a frequent problem among low-mass galaxies, since those often have ill-defined mid-plane dust lanes \citep{DAL04} and in-plane structures which make an exact determination of their inclination difficult. This could cause a selection bias that explains why galaxies with profiles compatible with a single disc have smaller masses ($\overline{v}_{\rm c}=113\,{\rm km\,s^{-1}}$ and ${\rm median}(v_{\rm c})=102\,{\rm km\,s^{-1}}$) than those that show at least two obvious discs ($\overline{v}_{\rm c}=144\,{\rm km\,s^{-1}}$ and ${\rm median}(v_{\rm c})=134\,{\rm km\,s^{-1}}$). Alternatively, low-mass galaxies might be genuinely less prone to forming a thick disc (or a thin disc within a pre-existing thick disc). This might, for example, be the case for NGC~5023 a galaxy also studied by \citet{STREICH16}, with reported inclinations ranging from $i=87\degr$ \citep{BOT86} to $i=88\fdg5\pm1\fdg3$ \citep{KHAM13}.

Eight galaxies (ESO~79-3, NGC~693, NGC~1381, NGC~4013, NGC~4217, NGC~4452, NGC~5010, and NGC~5403) have high-$z$ light excesses even when fitted with two discs. We interpret those galaxies as having two thick discs or a high-surface-brightness halo component (although in the case of ESO~79-3, this could also be the consequence of the presence of a nearby saturated star). NGC~4013 was already identified in \citet{CO11A} as a three-disc galaxy, but in that study we considered the PSF to be Gaussian. Here we prove that the third disc remains, even when properly accounting for the effect of the thin disc scattered light. Galaxies with three discs are significantly more massive than the rest ($\overline{v}_{\rm c}=179\,{\rm km\,s^{-1}}$ and ${\rm median}(v_{\rm c})=194\,{\rm km\,s^{-1}}$ against $\overline{v}_{\rm c}=138\,{\rm km\,s^{-1}}$ and ${\rm median}(v_{\rm c})=127\,{\rm km\,s^{-1}}$).

Table~\ref{numberstat} summarises the information on the number of discs in the galaxies in our sample. The statistical uncertainties account for 95\% confidence intervals and are calculated using the Wald test criterion \citep{WALD39}. This is the criterion that has been used elsewhere in the literature with 68\% confidence intervals \citep[e.g.,][]{EL90, CO08}.

{\it We find that most of the galaxies in the sample can be fitted with the sum of a thin and a thick disc ($82\%\pm6\%$). However, several galaxies have vertical surface-brightness profiles compatible with a single stellar disc ($12\%\pm5\%$) and a few other galaxies have profiles compatible with three discs ($6\%\pm4\%$).}

\subsection{Galaxy component masses}

\label{masses}

\begin{figure}
\begin{center}
  \includegraphics[width=0.48\textwidth]{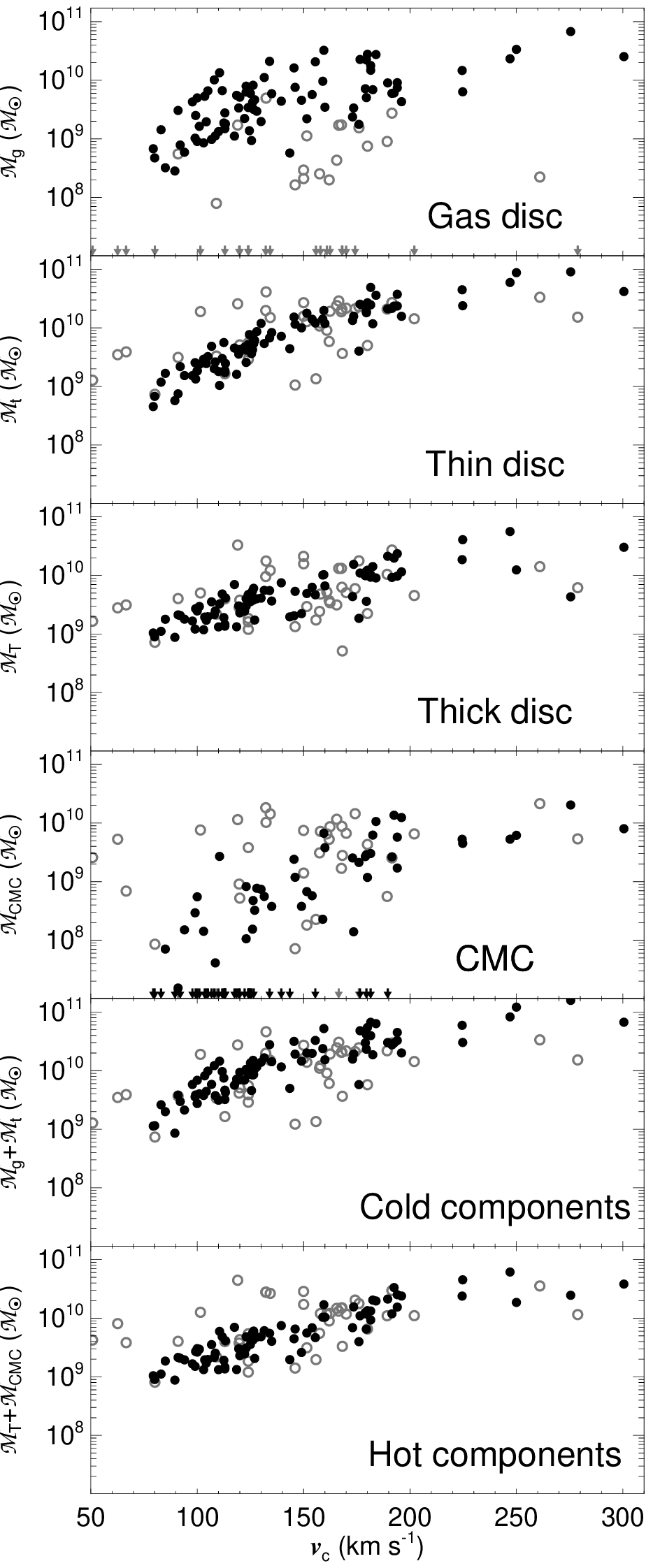}
  
  \end{center}
  \caption{\label{absolute} Masses of the galaxy components as a function of the circular velocity $v_{\rm c}$. From {\it top} to {\it bottom} the components are the gas disc, the thin disc, the thick disc, the CMC, the sum of the cold components (gas and thin disc) and the sum of the hot components (thick disc and CMC). Arrows indicate the velocity of galaxies with no H{\sc i} data in HyperLeda ({\it first} panel) and galaxies with no detected CMC ({\it fourth} panel) with the same colour-coding as the other symbols. Filled black dots correspond to gas-rich galaxies, and  grey circles correspond to gas-poor galaxies.}
\end{figure}

\begin{table}
 \caption{Spearman's rank correlation coefficients and significance in the correlations for the data points in Fig.~\ref{absolute}.}
 \label{spearman}
 \centering
 \begin{tabular}{l c c c c}
 \hline\hline
  Component & \multicolumn{2}{c}{Gas-poor}& \multicolumn{2}{c}{Gas-rich}\\
  &$\rho$ & $p$ & $\rho$ & $p$\\
  \hline
  Gas disc & 0.24 & 0.3 & 0.69 & $9\times10^{-13}$\\
  Thin disc & 0.56 & $2\times10^{-4}$&0.94&$4\times10^{-39}$\\
  Thick disc &0.38 & 0.01&0.83& $5\times10^{-22}$\\
  CMC & 0.20 &0.2& 0.82&$6\times10^{-12}$\\
  Cold components & 0.57 &$10^{-4}$& 0.90 &$3\times10^{-31}$\\
  Hot components & 0.41 &0.007& 0.89&$10^{-28}$\\
  \hline
  \end{tabular}
 \tablefoot{Component mass upper limits (downward pointing arrows in Fig.~\ref{absolute}) not included. Both the Spearman's correlation coefficient $\rho$ and its significance $p$ have been calculated with {\sc idl}'s {\sc r\_correlate}. The significance $p$ is bound between zero and one and low $p$-values indicate a significant correlation.}
\end{table}

\begin{table}
 \caption{Spearman's rank correlation coefficients and significance in the correlations for the data points in the common between \citet{CO12} and this paper (57 galaxies).}
 \label{spearman2}
 \centering
 \begin{tabular}{l c c c c}
 \hline\hline
  Component  &  \multicolumn{2}{c}{Comer\'on et al.} &\multicolumn{2}{c}{This paper}\\
    &  \multicolumn{2}{c}{(2012)} &\multicolumn{2}{c}{}\\
   & $\rho$ & $p$ & $\rho$ & $p$\\
  \hline
  Gas disc	& 0.42 & 0.001			& 0.44 & $5\times10^{-4}$\\
  Thin disc	& 0.93 & $8\times10^{-25}$	& 0.92 & $2\times10^{-24}$\\
  Thick disc	& 0.86 & $2\times10^{-17}$	& 0.78 & $1\times10^{-12}$\\
  CMC 		& 0.79 & $2\times10^{-13}$	& 0.79 & $4\times10^{-13}$\\
  Cold components& 0.83& $3\times10^{-15}$	& 0.86 & $5\times10^{-18}$\\
  Hot components & 0.92& $2\times10^{-24}$	& 0.89 & $2\times10^{-20}$\\
  \hline
  \end{tabular}
 \tablefoot{$\rho$ and $p$ calculated as in Table~\ref{spearman}. The H{\sc i} fluxes in this paper have a different source than in \citet{CO12} for 19 galaxies, hence the slight differences in the gas disc statistical distributions.}
\end{table}

\begin{figure}
\begin{center}
  \includegraphics[width=0.48\textwidth]{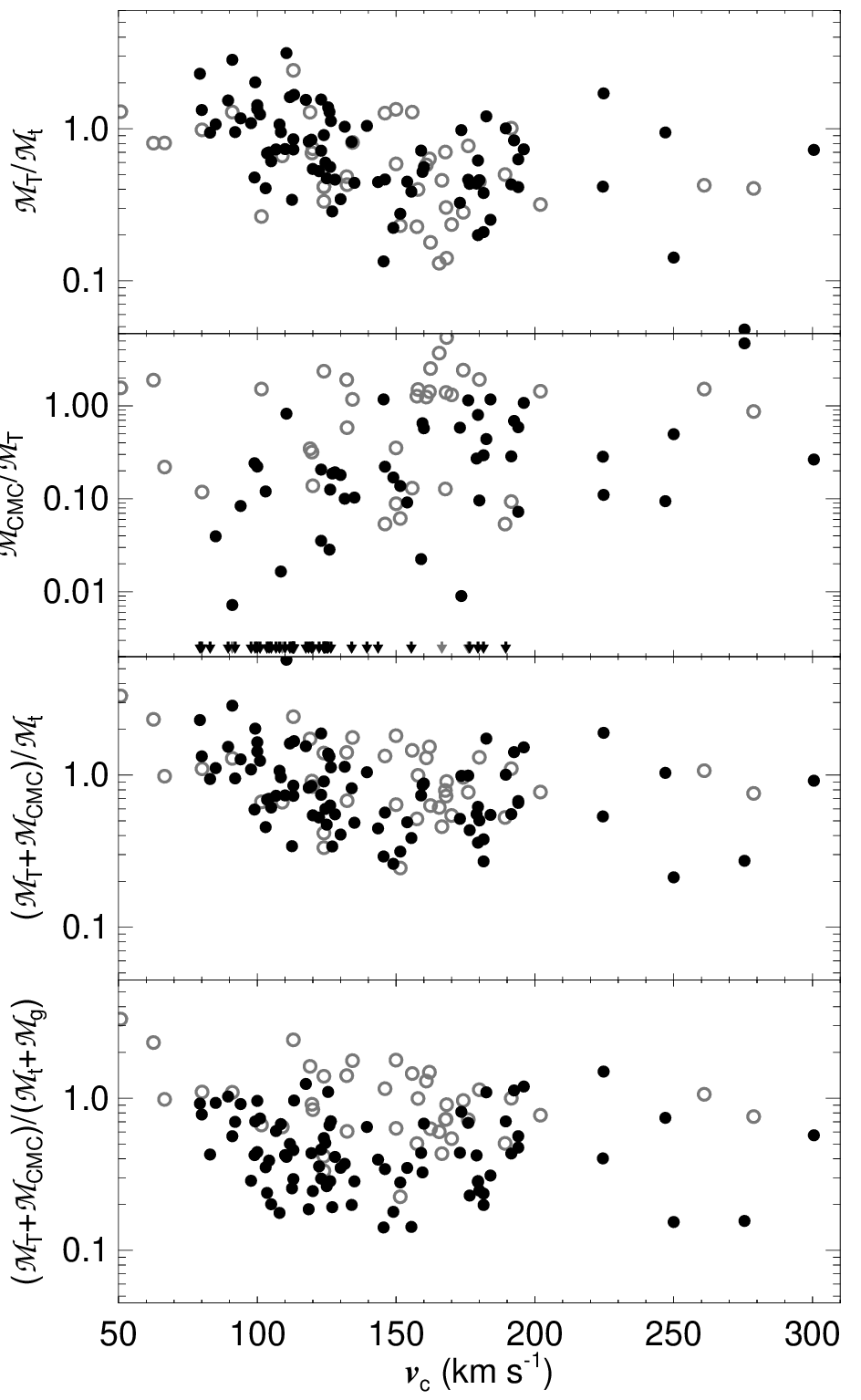}
  
  \end{center}
  \caption{\label{ratios} {\it Top} panel: ratio of the thick and the thin disc masses. {\it Second} panel: ratio of the CMC and the thick disc masses. Arrows indicate the velocity of galaxies with no CMC with the same colour-coding as the other symbols {\it Third} panel: ratio of the hot component and the thin disc masses. {\it Bottom} panel: ratio of the hot component and the cold component masses.  Filled black dots correspond to gas-rich galaxies, and  grey circles correspond to gas-poor galaxies.}
\end{figure}

\begin{figure}
\begin{center}
  \includegraphics[width=0.48\textwidth]{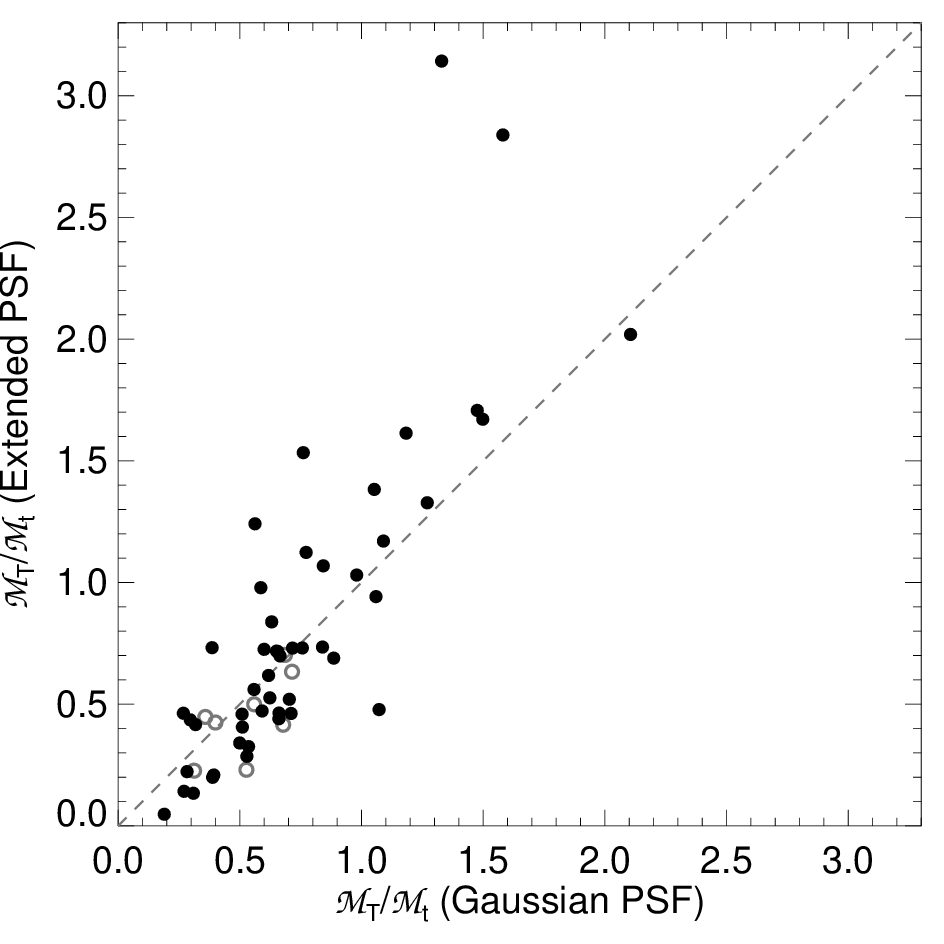}
  
  \end{center}
  \caption{\label{compare} Comparison of the thick to thin disc mass ratios -- $\mathcal{M}_{\rm T}/\mathcal{M}_{\rm t}$ -- calculated with a Gaussian PSF \citep{CO12, CO14} and those calculated (this paper) with the symmetrised version of the PSF provided by \citet{HO12}. The grey line indicates a one-to-one correlation. Filled black dots correspond to gas-rich galaxies, and grey circles correspond to gas-poor galaxies.}
\end{figure}

\begin{figure*}
\begin{center}
  \includegraphics[width=0.98\textwidth]{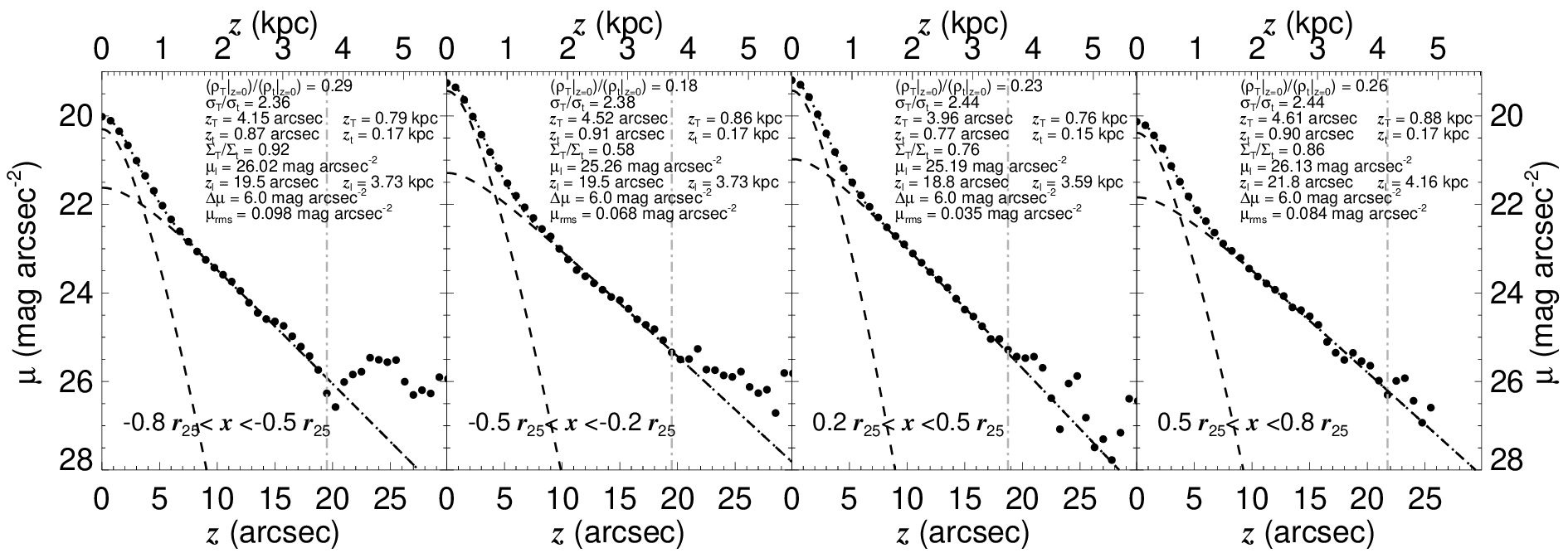}
  
  \end{center}
  \caption{\label{gauss} As the {\it bottom} plots in Fig.~\ref{verfitnobul}, but this time fitting the vertical surface-brightness profiles with a PSF model truncated at a radius of $5\arcsec$.}
\end{figure*}

Here we examine the masses of the galaxy components as obtained from our structural decompositions.

We distinguish two types of galaxy components depending on their vertical thickness which is an indicator of their dynamical coldness/hotness \citep{CO14}. Thin and gas discs have vertical extensions of a few hundred parsecs at most and we consider them as cold components. Thick discs and CMCs have scale-heights of about 1 kpc and are thus dynamically hot compared to the above-mentioned cold components.

The masses of the components obtained in Sect.~\ref{components} for the galaxies with at least two stellar discs (Appendix~\ref{twodiscap}) are plotted in Fig.~\ref{absolute}. All the component masses have a positive correlation with $v_{\rm c}$. The correlations with $v_{\rm c}$ are tighter for the thin and the thick disc masses than for the other components. Forty-five out of 124 galaxies have no fitted CMC. A CMC is found in all galaxies with $v_{\rm c}>200\,{\rm km\,s^{-1}}$. Three quarters of the intermediate-mass galaxies -- 54 out of 72 galaxies in the velocity range $120\,{\rm km\,s^{-1}}<v_{\rm c}<200\,{\rm km\,s^{-1}}$ -- host a CMC. The fraction of low-mass galaxies ($v_{\rm c}\leq120\,{\rm km\,s^{-1}}$) with a fitted CMC is significantly smaller than for larger-mass galaxies (16 out of 43). The correlation between the CMC prevalence and the galaxy mass is well known and has been studied in the S$^4$G sample through the concentration index $C_{82}$ \citep{MU15}.

If we make a distinction between gas-rich and gas-poor galaxies -- we set the limit at $\mathcal{M}_{\rm g}/\mathcal{M}_{\rm baryon}=0.075$ -- we find that two distinct behaviours arise. The thin disc, thick disc, and CMC masses have tight correlations with $v_{\rm c}$ for gas-rich galaxies. The scatter is much larger for gas-poor galaxies. The visual impression is confirmed by the Spearman's rank correlation coefficients for the data points in Fig.~\ref{absolute} that are presented in Table~\ref{spearman}. A reason for this is likely to be the uncertainties in the $v_{\rm c}$ determination. For those gas-poor galaxies that have a $v_{\rm c}$ determination based on rotation curves, the H{\sc i} might be truncated before the rotation curve becomes flat \citep[see][for examples]{CO14}. Other gas-poor galaxies have their $v_{\rm c}$ measured from stellar absorption lines. Those measurements are not as precise as those obtained using H{\sc i}.

There are other reasons to expect a large scatter in gas-poor galaxies. Gas-poor galaxies have little or no star formation so they contain no zero-age stellar populations with almost circular orbits. Thus, the thin disc of gas-poor galaxies is likely to be thicker and less well-defined than that in gas-rich galaxies. It is to be expected that the smaller the difference in scale-height of the two discs, the harder it will be to distinguish them, causing additional scatter. Also, the lack of a star-forming layer and the consequent dust lane causes the orientation of gas-poor galaxies to be hard to assess, which might in turn result in several galaxies in our sample being farther from edge-on than what would be ideal for the fits \citep[see Sect.~3.6.3 and Fig.~9 in][]{CO11B}. Finally, part of the scatter might be related to the morphology-density relation \citep{OEM74}. Early-type galaxies are generally poorer in gas \citep{BOT82} and are statistically found in richer environments than their late-type counterparts. Galaxies in dense groups and in galaxy clusters are prone to suffer ram pressure gas-stripping processes \citep{GUNN72}. Those same galaxies suffer harassment, that is, repeated interactions with other galaxies in the group/cluster \citep{MO96} which, if frequent enough, may result in gas-poor galaxies, not in hydrostatic equilibrium at the time of the observation. Then, the assumptions behind Eq.~\ref{narayan} would not hold. The possible link between scatter in Fig.~\ref{absolute} and environment is supported by the fact that, regarding galaxies with at least two stellar discs, 18 out of 41 ($44\pm15\%$) gas-poor galaxies belong to either the Fornax Cluster Catalogue \citep{FER89} or the Extended Virgo Cluster Catalog \citep{KIMS14}, but only 13 out of 83  ($16\pm8\%$) gas-rich galaxies do.

Our new fits recover the well-known trend that shows that thick discs are more dominant in low-mass galaxies than they are in massive ones \citep[Fig.~\ref{ratios};][]{YOA06, CO11B, CO12}. The ratio between the thick and the thin disc masses is typically in the range $0.2<\mathcal{M}_{\rm T}/\mathcal{M}_{\rm t}<1$ for galaxies with $v_{\rm c}>120\,{\rm km\,s^{-1}}$ but can have much larger values for the low-mass galaxies, up to $\mathcal{M}_{\rm T}/\mathcal{M}_{\rm t}\sim3$. The view on galaxy discs has generally been that thin discs are embedded in a diffuse extended thick component. The low-mass galaxy picture puts the spotlight on the thick disc and suggests them as being a dominant component with a thinner component inside. In a model where thick discs form first from a turbulent gas disc with intense star formation \citep{EL06, BOUR09, CO14}, this is a natural expectation in a universe with down-sizing, that is, a universe where low-mass galaxies take longer to evolve \citep{CO96}. In this picture thin discs in low-mass galaxies are still at their infancy state, whereas those in high-mass galaxies are already mature.

In Fig.~\ref{compare} we compare the $\mathcal{M}_{\rm T}/\mathcal{M}_{\rm t}$ values calculated with a Gaussian PSF in \citet{CO12, CO14} to those calculated using the more sophisticated approach here (57 galaxy overlap, for galaxies with two stellar discs between the two samples). The ratios are comparable, albeit with a considerable scatter. The agreement seems strange since here the thin disc light usually dominates at large heights due to the extended PSF wings whereas in previous fits the thick disc dominated all the way to infinity. However, because of the significant luminosity assigned to the thin disc at large heights, the thick disc scale-heights are smaller than those in our previous work. As a consequence, the mid-plane surface brightnesses of the thick discs are increased. The net effect is that the change of the PSF in the fits has little impact on the $\mathcal{M}_{\rm T}/\mathcal{M}_{\rm t}$ determination. An example of this is seen in Fig.~\ref{gauss} where we show the vertical surface brightness fits for ESO~533-4 made with a model PSF truncated at a radius of $5\arcsec$. This small PSF model only covers the Gaussian core. If we compare the fitted thick disc scale-height values in Fig.~\ref{gauss} with those in the lower row of plots in Fig.~\ref{verfitnobul} we find that for this specific galaxy they are reduced by $\sim20\%$ when the extended PSF wings are taken into account. At the same time, the mass ratio $\mathcal{M}_{\rm T}/\mathcal{M}_{\rm t}$ is affected by less than 10\%.

Because we find that the ratio between the thick and thin disc masses remains roughly unchanged when we use our new model PSF our results confirm that thick discs contain a substantial fraction of the baryonic mass in local galaxies.

In Table~\ref{spearman2} we study the mass of the components in the 57-galaxy overlap between this paper and \cite{CO12} among galaxies with a distinct thin and thick disc. We find that the Spearman correlation coefficients are similar for the data points in the two papers. This suggests that, even though in our previous papers we used an inaccurate PSF model, we could still adequately recover the properties of edge-on discs within the studied surface brightness limit of $\mu\approx26\,{\rm mag\,arcsec^{-2}}$.

We note two additional trends in Fig.~\ref{ratios}. First, the ratio $\mathcal{M}_{\rm CMC}/\mathcal{M}_{\rm T}$ grows with galaxy mass, at least for gas-rich galaxies, with a Spearman correlation coefficient of $\rho=0.44$ and a significance $p=0.003$. Second, the ratio of the hot and cold component masses is roughly constant with $v_{\rm c}$. The Spearman correlation coefficient for the points in this panel is a mere $\rho=-0.067$ with $p=0.5$, suggestive of no or at most a very mild correlation with $v_{\rm c}$. Although both above-mentioned trends have a considerable scatter, in \citet{CO14} this was interpreted as evidence that the thick disc and the CMC stars come from a single material reservoir. In our view, the primordial discs that evolved into present-day thick discs were clumpy for about a Gyr \citep{BOUR07} and in massive galaxies some of the clumps migrated inwards and merged to form a CMC.

{\it We confirm that the ratio of the thick to thin disc mass is the largest in the lowest-mass disc galaxies. In those galaxies the thick disc can host about half of the baryons in a galaxy. The scaling relations between the mass of the galaxy components and the galaxy circular velocity -- a proxy for the total galaxy mass -- are tighter for gas-rich galaxies.}

\subsection{The thin and thick disc scale-heights}

\label{scales}

\begin{table}
 \caption{Parameters of the robust linear regressions of the scale-heights of discs as a function of the circular velocity in Eq.~\ref{linear}.}
 \label{correlations}
 \centering
 \begin{tabular}{l c c c c}
 \hline\hline
  & $N$ & $A$ & $B$ & $\rho$\\
  & & (pc)& $\left[{\rm pc}/\left({\rm km\,s^{-1}}\right)\right]$ &\\
  \hline
  \multicolumn{5}{c}{Thin disc ($\left<z_{\rm t}\right>$)}\\
  \hline
  All galaxies &124& 26&1.23&0.55\\
  Gas-rich &83& -12 & 1.42&0.77\\
  \hline
  \multicolumn{5}{c}{Thick disc ($\left<z_{\rm T}\right>$)}\\
  \hline
  All galaxies & 124 & -262 & 8.64 &0.56\\
  Gas-rich & 83 &  -217 & 7.42 & 0.71\\
  \hline
  \end{tabular}
 \tablefoot{The number of data points used in each fit is denoted by $N$. The Pearson correlation coefficient is denoted by $\rho$.}
\end{table}

\begin{figure}
\begin{center}
  \includegraphics[width=0.48\textwidth]{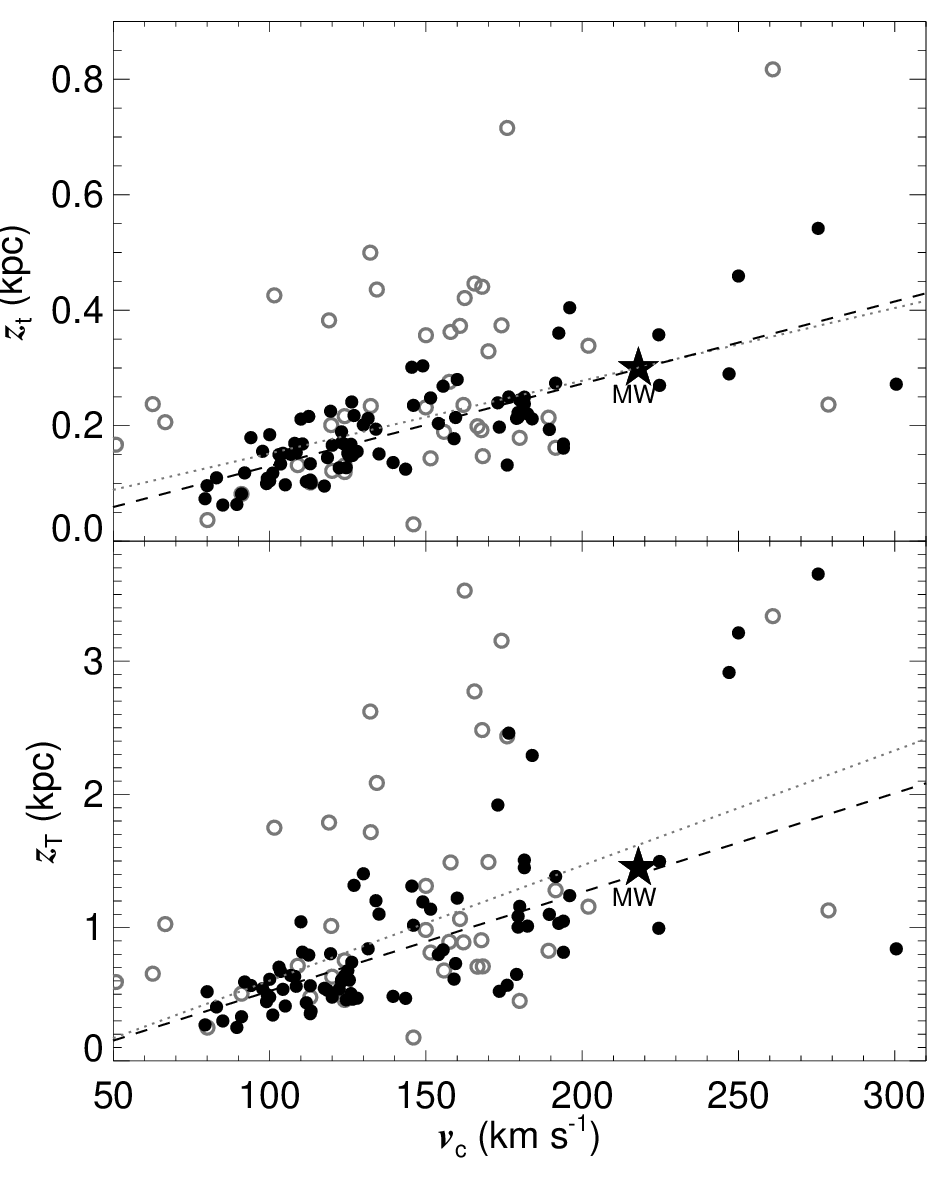}
  \end{center}
  \caption{\label{shs} The thin and thick disc scale-heights ({\it top} and {\it bottom} panels, respectively) for gas-poor and gas-rich galaxies (grey circles and black dots, respectively). The black dashed line corresponds to a robust linear regression made with gas-rich galaxies and the grey dotted line corresponds to a regression done with all galaxies. Parameters of the regression fits are given in Table~\ref{correlations}. The stars indicate the properties of the Milky Way as obtained from \citet{GIL83} for the scale-heights and \citet{BO12} for the circular velocity.}
\end{figure}

\begin{figure}
\begin{center}
  \includegraphics[width=0.48\textwidth]{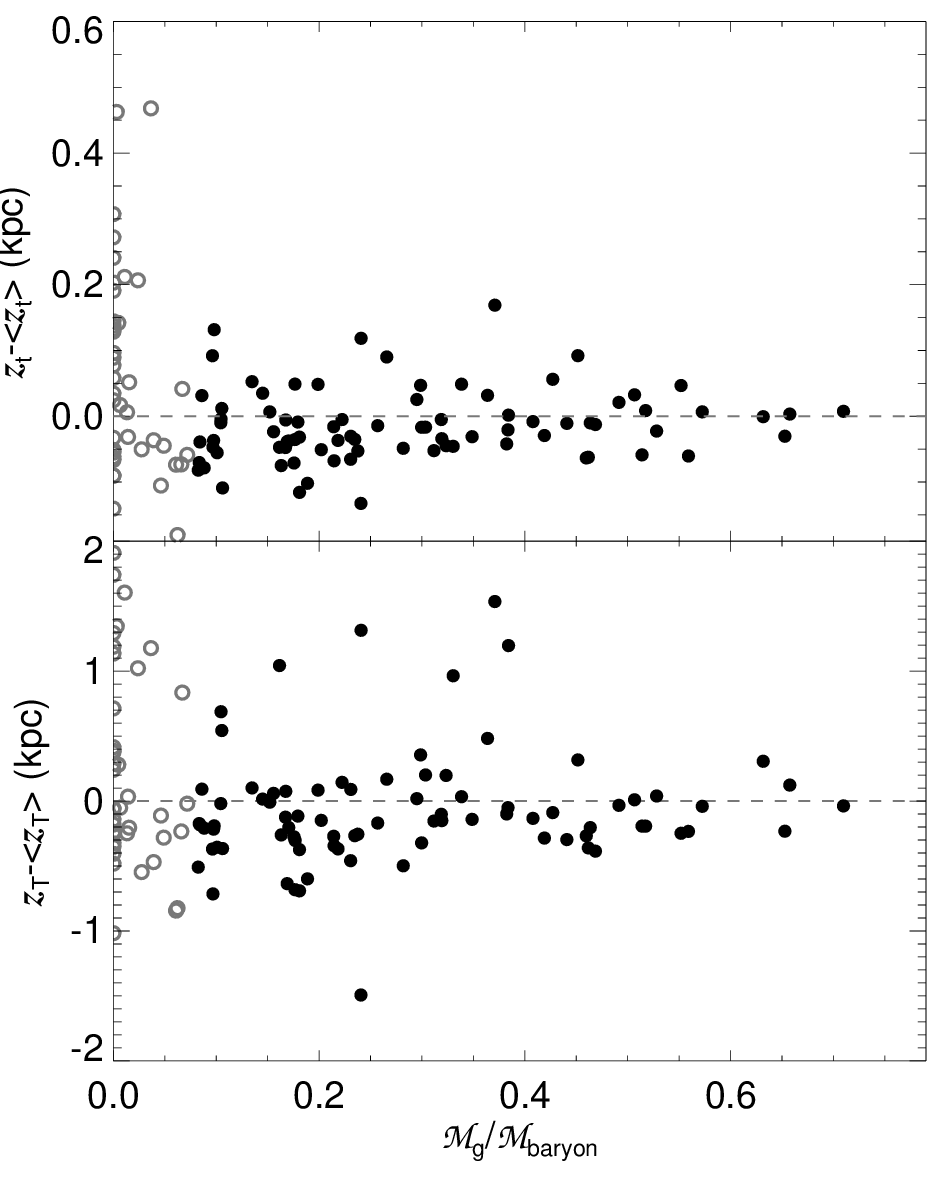}
  \end{center}
  \caption{\label{residual} Residuals for the thin and the thick discs scale-heights with respect to the linear regression ({\it top} and {\it bottom} panels, respectively) as a function of the fraction of baryons in the form of gas. The residuals are calculated with respect to the linear regression made for all galaxies, that is without excluding gas-poor galaxies (grey dotted line in Fig.~\ref{shs}). Filled black dots correspond to gas-rich galaxies, and grey circles correspond to gas-poor galaxies.}
\end{figure}

\begin{figure}
\begin{center}
  \includegraphics[width=0.48\textwidth]{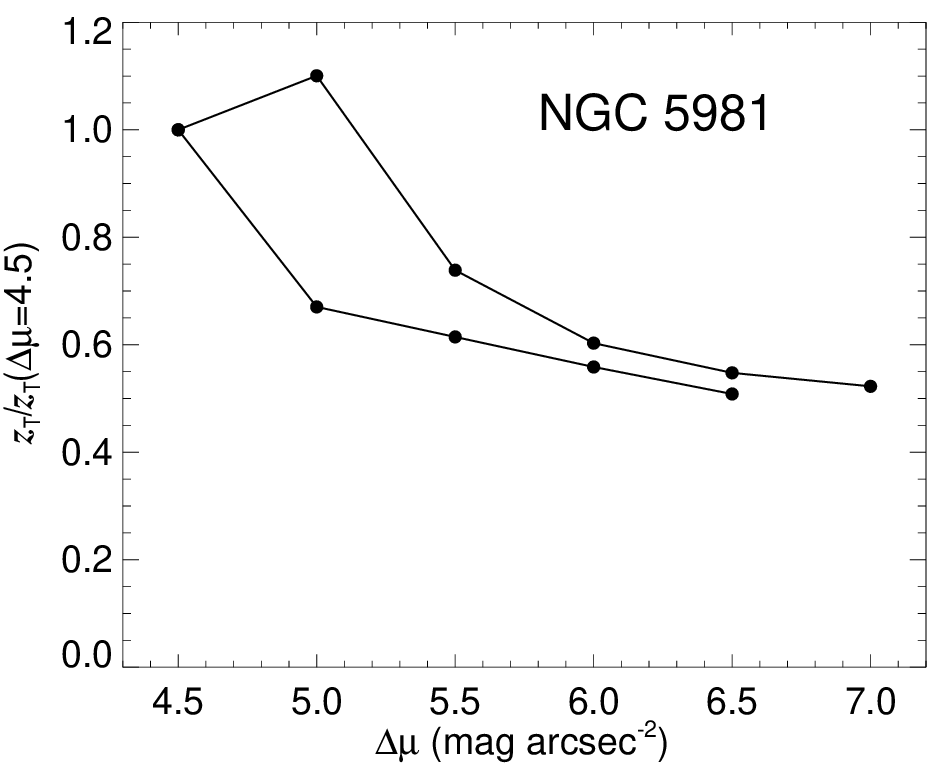}
  \end{center}
  \caption{\label{test} Fitted scale-height of the thick disc of NGC~5981 for the two central axial bins -- those with the largest dynamical range -- compared to the fitted scale-height while using a dynamical range of only $\Delta\mu=4.5\,{\rm mag\,arcsec^{-2}}$. The fits are done for dynamical ranges $\Delta\mu=4.5-7.0\,{\rm mag\,arcsec^{-2}}$ in steps of $0.5\,{\rm mag\,arcsec^{-2}}$.}
\end{figure}

\begin{figure}
\begin{center}
  \includegraphics[width=0.48\textwidth]{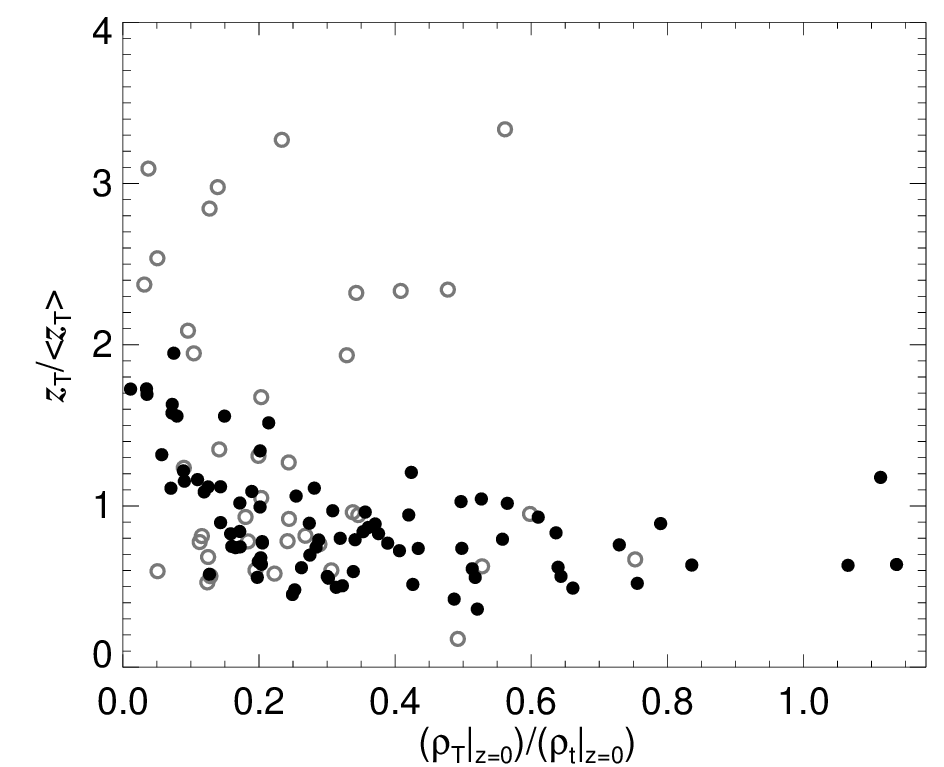}
  \end{center}
  \caption{\label{test2} Thick disc scale-height relative to the fit to Eq.~\ref{linear} as a function of the thick to thin disc mid-plane density ratio $\left(\left.\rho_{\rm T}\right|_{z=0}\right)/\left(\left.\rho_{\rm t}\right|_{z=0}\right)$ averaged over all the axial bins with a valid fit. The considered fit is that for all galaxies (that is, not excluding gas-poor galaxies). Filled black dots correspond to gas-rich galaxies, and grey circles correspond to gas-poor galaxies.}
\end{figure}

Here we study the properties of the thin and thick disc scale-heights as a function of the galaxy circular velocities.

Thin and thick disc scale-heights were obtained from the vertical surface-brightness profile fits in Sect.~\ref{finalvertical}. For each axial bin, a scale-height was calculated by finding the distance between the position where the flux is a factor 100 smaller than at the mid-plane and the position where the flux is $100\times{\rm e}$ times smaller that at the mid-plane in the fitted surface-brightness profiles for each of the two stellar discs. A global scale-height was found by averaging the scale-heights over the axial bins with a valid fit. We denote $z_{\rm t}$ and $z_{\rm T}$ as the thin and thick disc scale-heights, respectively.

We described how the disc scale-heights of the 124 galaxies with at least two well-defined discs relate to the galaxy mass by doing a linear regression
\begin{equation}
\label{linear}
 \left<z_i\right>=A+Bv_{\rm c}
\end{equation}
where $i$ stands for either the thin or the thick disc, $A$ for the scale-height at $v_{\rm c}=0$, and $B$ for the slope. The fits were obtained with {\sc idl}'s {\sc robust\_linefit} routine from the {\sc idl} Astronomy Library\footnote{\url{https://idlastro.gsfc.nasa.gov/}} \citep{LAND93}, an outlier-resistant linear regression algorithm. The results of the regression are shown in Table~\ref{correlations}.

We checked how the fitted relations stand against the Milky Way properties, that is $z_{\rm t}({\rm MW})\sim300\,{\rm pc}$ and $z_{\rm T}({\rm MW})\sim1450\,{\rm pc}$ \citep{GIL83} and $v_{\rm c}\left({\rm MW}\right)\approx218\,{\rm km\,s^{-1}}$ \citep{BO12}. The predicted scale-heights from our regression to the sample of 124 galaxies match the Milky Way scale-heights within 15\%. The agreement is better than 5\% if the fit to gas-rich galaxies is considered.

Fig.~\ref{shs} shows that the scatter in scale-heights is large in gas-poor galaxies compared to that in gas-rich ones. The very tight scaling relation for the thin disc scale-height in gas-rich galaxies is remarkable when one takes into account the amount of assumptions in our rather complicated fits and that the scale-heights are influenced by uncertainties in the galaxy distance determination.

In Fig.~\ref{residual} we show the residuals between the scale-heights and the linear regression in 124 galaxies as a function of the mass fraction of baryons in gas. The typical amplitude of the $z_{\rm t}$ residuals is about 100\,pc and does not vary much with $\mathcal{M}_{\rm g}/\mathcal{M}_{\rm baryon}$ as long as galaxies are gas-rich. The eight galaxies that deviate more than 200\,pc from the linear fit are gas-poor and on the positive side, which is unlikely to be a coincidence. The two largest scale-heights for a thin disc correspond to gas-poor galaxies, namely NGC~1032 and NGC~4370. The scale-heights of these two galaxies are far too large to be moved close to the linear fit by accounting for $v_{\rm c}$ uncertainties. This suggests that part of the large scatter seen for gas-poor galaxies in the plots in Sect.~\ref{masses} and in this subsection might have a physical origin. Visually, the thin disc of NGC~1032 is very diffuse, whereas that of NGC~4370 is barely distinguishable from the thick disc (although the distinction is much clearer in the surface-brightness profile). This and the fact that the seven most extreme outliers are on the positive side in Fig.~\ref{residual} suggests that some discs of gas-poor galaxies may have been dynamically heated, a process which is more efficient in gas-poor galaxies \citep{MOS10}. NGC~4370 is in the Virgo cluster, so harassment is a viable heating mechanism there. NGC~1032 is isolated \citep{TU15} so heating would have been caused by the merger with a companion.

The typical thick disc scale-height departures from the linear regression are of a few hundred parsecs. Contrary to what happens with the thin disc, some of the outliers -- we could define as outliers those galaxies with scale-heights that deviate by more than a kiloparsec from the linear regression -- are gas-rich galaxies. Six out of sixteen outliers are very massive galaxies with circular velocities $v_{\rm c}>200\,{\rm km\,s^{-1}}$ (the total number of galaxies with $v_{\rm c}>200\,{\rm km\,s^{-1}}$ in our sample is nine). In two other outlier galaxies -- NGC~3115 and NGC~4370 -- the thick disc component fits an extended oblate envelope whose axis ratio might be too large to be considered as that of a proper disc. Whereas dynamical heating in a dense environment might suffice to cause the outliers in gas-poor galaxies, other reasons might explain the outlying thick disc scale-height values for gas-rich galaxies. Outlying galaxies are typically those with large masses, that is galaxies with low $\left(\left.\rho_{\rm T}\right|_{z=0}\right)/\left(\left.\rho_{\rm t}\right|_{z=0}\right)$ ratios, so a large dynamical range $\Delta\mu$ is needed to well constrain the thick disc scale-height. For example, the S$^4$G might not be deep enough to sufficiently characterise the thick disc for the two gas-rich galaxies with the tallest thick discs (ESO~240-11 and NGC~4565) for which the mid-plane thick disc density is only $1-5\%$ of that of the thin disc.

To test the plausibility of this speculation we studied the case of NGC~5981, a galaxy fitted over a dynamical range $\Delta\mu=6.5-7.0\,{\rm mag\,arcsec^{-2}}$ in its two central bins. This galaxy has $\left(\left.\rho_{\rm T}\right|_{z=0}\right)/\left(\left.\rho_{\rm t}\right|_{z=0}\right)$ between 0.1 and 0.2. That is, smaller than average in our sample, but still large enough so that regions where the thick disc is relevant can be traced over a couple of magnitudes. We checked how the fitted thick disc scale-height varies as we restrict the fitting dynamical range from $\Delta\mu=4.5,{\rm mag\,arcsec^{-2}}$ down to $\Delta\mu=7.0\,{\rm mag\,arcsec^{-2}}$ in bins of $\Delta\mu=0.5\,{\rm mag\,arcsec^{-2}}$ (Fig.\ref{test}). The plot shows that for this individual galaxy, the scale-height of the thick disc is over-estimated by a factor of almost two if the selected dynamical range is not large enough to properly sample regions with significant thick disc contribution. Furthermore, the galaxies for which the thick disc scale-height deviates the most from the linear regression to Eq.~\ref{linear} are among those that have the smallest relative thick disc mid-plane density, especially for gas-rich galaxies. Thus, we speculate that biases associated with poorly constrained thick disc scale-heights might be, at least in part, explaining the thick disc scale-height outliers. Fitting deeper images of high-mass galaxies is needed to confirm whether these points are indeed statistical outliers, whether they are due to a bias, or whether thick disks are in fact particularly thick in the highest-mass galaxies.

{\it The thin and thick disc scale-heights have tight positive correlations with the galaxy circular velocities. The correlations are even tighter for gas-rich galaxies. Regarding the disc scale-heights, the Milky Way is a typical galaxy.}

\subsection{Disc down-bending and up-bending breaks}

\label{truncations}

\begin{table*}
 \caption{Axial surface-brightness profile classifications.}
 \label{axialclass}
 \centering
 \begin{tabular}{l | c c | c c | c c}
 \hline\hline
  &\multicolumn{2}{c|}{Whole disc}&\multicolumn{2}{c|}{Thin disc dominated heights}&\multicolumn{2}{c}{Thick disc dominated heights} \\
  &\multicolumn{2}{c|}{($N=141$)}& \multicolumn{2}{c|}{($N=105$)} & \multicolumn{2}{c}{($N=104$)}\\
  \hline
  Type~I&  11  &($8\%\pm4\%$)&  7  &($7\%\pm5\%$)&  56  & ($53\%\pm10\%$)\\
  Type~II&  30  &($21\%\pm7\%$)& 21  &($20\%\pm8\%$)& 35 & ($33\%\pm9\%$)\\
  Type~III&  26  &($18\%\pm6\%$)& 11 & ($10\%\pm6\%$)& 10 & ($10\%\pm6\%$) \\
  Type~II+II&  0 & (--) & 1 & ($1\%\pm2\%$) & 0 & (--)\\
  Type~II+III&  18 &($13\%\pm6\%$)& 21 & ($20\%\pm8\%$)&  0 & (--) \\
  Type~III+II&   35 &($25\%\pm7\%$)& 28 & ($27\%\pm8\%$)& 3 & ($3\%\pm3\%$)\\
  Type~III+III&   4  &($3\%\pm3\%$)& 2 & ($2\%\pm3\%$)&  0 & (--)\\
  Type~II+III+II&  4 &($3\%\pm3\%$)&  4 & ($4\%\pm4\%$)& 0 & (--)\\
  Type~II+III+III&  1 &($1\%\pm1\%$)& 0  & (--)& 0 & (--)\\
  Type~III+II+III&  12  &($9\%\pm5\%$)& 10 & ($10\%\pm6\%$) &  0 & (--)\\
  \hline
  \end{tabular}
 \tablefoot{The number of galaxies for which the axial profiles are available is denoted by $N$.}
\end{table*}

\begin{figure}
\begin{center}
  \includegraphics[width=0.48\textwidth]{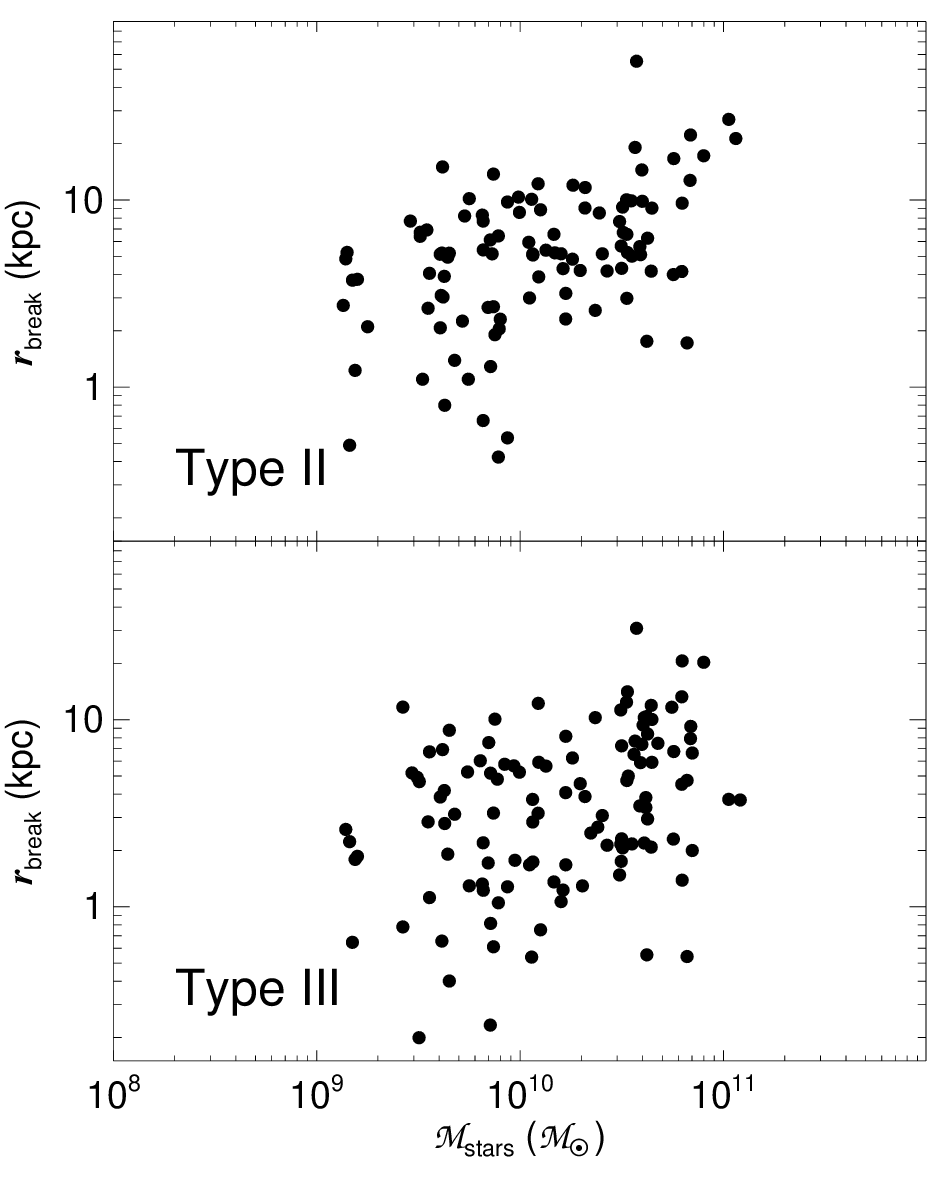}
  \end{center}
  \caption{\label{truncradius} Galactocentric fitted radii of the Type~II and Type~III breaks ({\it top} and {\it bottom} panels, respectively) as a function of the galaxy stellar mass.}
\end{figure}

\begin{figure}
\begin{center}
  \includegraphics[width=0.48\textwidth]{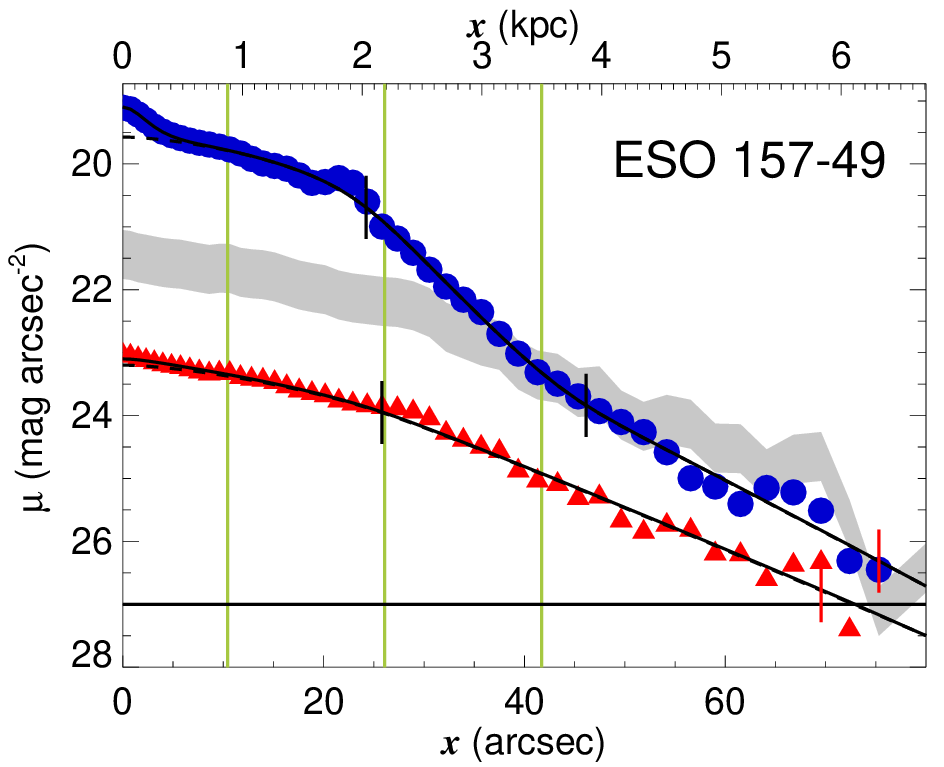}
  \end{center}
  \caption{\label{thinthickaxial} ESO~157-49 axial surface-brightness profiles for the thin-disc-dominated heights (blue circles) and the thick-disc-dominated heights (red triangles). The black continuous lines correspond to the fits to the profiles -- as explained in Sect.~\ref{axial2} -- and the dashed lines indicate the disc contribution. Short vertical lines indicate the break radii. Red vertical lines indicate the outermost fitted point. Green vertical lines indicate the limits of the axial bins used to create the vertical surface-brightness profiles. The horizontal line indicates the $\mu=27\,{\rm mag\,arcsec^{-2}}$ level. The grey region indicates the range of estimates of thick disc contributions at thin disc dominated heights based on the methods described in Sect.~\ref{truncations}.}
\end{figure}

In this subsection we pay attention to the disc breaks as seen in the axial surface-brightness profiles. We do so for the disc of galaxies integrated over all heights and also for thin- and the thick-disc-dominated ranges of heights.

Axial profiles produced by integrating the whole height of the disc are available for the 141 galaxies in our sample (Sect.~\ref{axial}). Among the 124 galaxies with clearly defined thin and thick disc components (see Sect.~\ref{axial2}) thin disc axial profiles are available for 105 galaxies and thick disc axial profiles are available for 104 galaxies (ESO~240-11 is thin disc-dominated at all heights). A summary of the profile classification is shown in Table~\ref{axialclass}.

Roughly speaking our profile classifications are compatible with those in our previous work \citep{CO12}, although we note that the fraction of pure Type~II galaxies is significantly lower here ($46\%\pm12\%$ and $21\%\pm7\%$, respectively; see Sect.~\ref{axialfunctions} for definitions of the profile types).

When compared to face-on galaxy profile classification studies \citep[e.g.,][]{POH06, ER08, CO08, GUT11, LAI14} our sample shows a large abundance of composite profiles -- that is profiles with more than one break. There are several possible reasons for this. First, profiles in edge-on galaxies extend further out than those in face-on galaxies because of the line-of-sight integration, so more breaks can be detected. Second, we chose to describe pseudo-bulges or disky CMCs that do not protrude the disc with a disc section. Third, the classification is slightly subjective and subtle slope changes might be considered as breaks only by a subset of authors (however, breaks in edge-on galaxies are subtler than in face-on ones as shown in Fig.~\ref{soft}). Fourth, flared discs that would not cause any break in a face-on view can be seen as causing a Type~II break in an edge-on view \citep{BOR16}. Finally, in face-on galaxies bumps and ``shoulders'' caused by the bar and prominent star-forming rings are ignored. Excluding breaks caused by those features in an edge-on galaxy is a dangerous exercise. Bars could in principle be taken into account because they are often associated with features such as barlenses that are seen in edge-on galaxies as boxy-peanut CMCs \citep{LAU14, ATH15, LAU17, SA17} -- at least those in centrally concentrated galaxies. We checked whether composite profiles are more prevalent in massive galaxies, that is those with $v_{\rm c}>120\,{\rm km\,s^{-1}}$, which corresponds to a baryonic mass of $\sim10^{10}\,\mathcal{M}_{\bigodot}$ (Fig.~\ref{absolute}). This is about the mass above which barlenses are found \citep{HE15}. We find that composite profiles are found in $57\%\pm10\%$ of the galaxies with $v_{\rm c}>120\,{\rm km\,s^{-1}}$ and in $44\%\pm13\%$ of lower-mass galaxies (two-sigma uncertainties). This mild difference in the two fractions indicates that well-formed bars are not the only reason for composite profiles. This is not completely unsurprising, since the effect of bars on the profiles is deemed to be smaller than in face-on galaxies. For example, in close to end-on galaxies, the bar would appear on top of the CMC and would be included as part of it if the CMC were disc-like or would be simply ignored if the CMC were classical.

We note that Type~III breaks are genuine because they cannot be explained by the extended PSF wings. Indeed, as shown in Fig.~\ref{axialfit}, the extended PSF does not cause up-bending breaks within the range of depths that can be studied with S$^4$G images.

In Fig.~\ref{truncradius} we show the galactocentric radii of the breaks for the vertically integrated disc as a function of the stellar mass of the galaxy. The stellar mass is calculated using our estimated thin disc, thick disc, and CMC masses even for those 17 galaxies where the thin and the thick discs are not well defined. Because there is a factor of 1.2 difference in the mass-to-light ratios of the thin and thick discs, this does not cause huge uncertainties. The plots show a mild correlation between the break radii and the stellar mass of galaxies. The main point of the plot is to check whether our breaks appear at the same radii as those in face-on galaxies. To do so, we can compare it with Fig.~31 in \citet{SA15} where the break radii in \citet{MU13}, \citet{KIM14}, and \citet{LAI14} are compiled in a single plot. Our stellar masses do not correspond exactly to those in \citet{SA15} because they obtain them from \citet{MU15} who in turn used the recipes in \citet{ES12}. The differences between our stellar masses and those in \citet{MU15} are typically of 20\% once the different distance determinations are factored out. The data point cloud in Fig.~\ref{truncradius} roughly spans the same space as that in \citet{SA15}, which indicates that the breaks that we find in edge-on galaxies are the counterparts of those seen in face-on galaxies. The only difference is that we find seventeen breaks at a radius smaller than 1\,kpc. Those breaks would probably be disregarded in a face-on view as being caused by a pseudo-bulge or a nuclear ring.

The profiles of the thick-disc-dominated regions are much simpler than the profiles for the whole disc; in fact, half of them are Type~I profiles, and only three thick-disc-dominated profiles are composite. The simplicity of thick disc profiles can be explained by two factors. First, some of the features that might cause breaks -- bars, lenses, rings -- mostly live in the thin disc. Second, according to some authors, phenomena that redistribute disc material and might create breaks (e.g. bars) are more efficient working on dynamically cold components \citep{VE14, VE16}. Thick discs hardly ever show Type~III breaks. The simplicity of the thick disc axial profiles is consistent with the findings in \citet{LAI16} who find that Type~I profiles are prevalent among low-mass face-on galaxies. Indeed, in these galaxies, thick disc light is likely to dominate the face-on surface-brightness profile because of the large $\mathcal{M}_{\rm T}/\mathcal{M}_{\rm t}$ ratio found in low-mass galaxies (Sect.~\ref{masses}).

Axial profiles for thin-disc-dominated regions are as complex as those for the whole disc and more than half of them are composite. However, thick discs can contribute substantially to the surface brightness, even at the mid-plane. In the case that a thin disc has a shorter scale-length than the thick disc, an up-bending break can result from the thick disc dominating the surface-brightness profile at large radii. This hypothesis was found to be compatible with more than half of the thin disc up-bending breaks in \citet{CO12}. We check this again here. To do so, we extrapolate the light from the thick-disc-dominated regions towards the mid-plane to figure out what the expected thick disc surface brightness is at the thin disc-dominated heights. This is done in two ways. One of them is by calculating the average fraction of thick disc light at the thin-disc-dominated heights on the valid fits to vertical profiles. Another estimate is obtained by calculating the thick disc contribution at the thin-disc-dominated heights if the thin disc were removed from the fitted vertical profiles. The net effect of removing the thin disc is to increase the thick disc scale-height, so its contribution close to the mid-plane is lowered. If the thin disc scale-length is shorter than that of the thick disc, the importance of the thin disc will be lowered in the outskirts of the galaxy, so the thick disc surface brightness contribution should fall somewhere between the two estimates. In Fig.~\ref{thinthickaxial} we show the ESO~157-49 axial profiles for the thin- and thick-disc-dominated heights. The grey area denotes the  estimated range for the thick disc contribution to the thin-disc-dominated regions. We see that the thin disc up-bending break falls within that region, which indicates a likely connection to the thick disc.

We examined each of the galaxies in which the outermost thin disc break is of Type~III to check whether or not those breaks can be linked to the thick disc, just as in the case ESO~157-49. We found that this is so for 20 galaxies, namely: ESO~79-3, ESO~157-49, ESO~443-42, ESO~469-15, IC~335, IC~1553, NGC~1163, NGC~3115, NGC~3454, NGC~4215, NGC~4217, NGC~4251, NGC~4302, NGC~4359, NGC~4544, NGC~4710, NGC~4762, NGC~7241, UGC~7086, and UGC~12692. This is $46\%\pm15\%$ of the 44 galaxies whose thin disc's outermost break is up-bending. The match between the expected and observed surface-brightness levels close to the mid-plane in a large fraction of the galaxies confirms the above-mentioned link between a large fraction of up-bending breaks and thick discs. Since the up-bending breaks seen in edge-on galaxies seem to have statistically similar properties to those in face-on galaxies, the same origin is expected for many of the Type~III breaks in face-on galaxies. Thus, genuine thin disc up-bending breaks are relatively rare.

{\it We find that the axial surface-brightness profiles of thin discs are often composite, whereas those of thick discs are usually simpler. We confirm that in about a half of cases the outermost up-bending break of a thin disc is in fact caused by thick disc light in a situation where the thin disc scale-length is shorter than that of the thick disc.}

\section{Summary and conclusions}

\label{summary}

Studies of the scattered light from high-surface-brightness galaxy features \citep[e.g.,][]{JONG08, SAN14, SAN15} have raised concerns on its effects on the study of low-surface-brightness features such as thick discs and stellar haloes. Because the effects of the extended PSF wings have not been properly assessed in integrated light studies of thick discs, their existence itself could be, in principle, put into question.

In our previous studies we have decomposed the edge-on galaxies in the S$^4$G into their thin disc, thick disc, and CMC components \citep{CO11B, CO12, CO14}. The S$^4$G is a $3.6\mu{\rm m}$ and $4.5\mu{\rm m}$ survey of nearby galaxies made with IRAC on-board the {\it Spitzer} Space Telescope. We used an over-simplistic PSF model that only reproduced the core of the actual IRAC PSF. Given the controversy triggered by the unsettling findings on scattered light it seemed necessary to redo our experiments to check whether the decompositions were drastically affected by the use of an ``incorrect'' PSF model. We used this opportunity to include further galaxies into our sample and to improve our methods. Thus, we now present our study of a sample of 141 galaxies drawn from the S$^4$G and its early-type galaxy extension.

Our profile decompositions assume that the thin and thick discs are two stellar fluids in hydrostatic equilibrium. We also account for the gravitational interaction of a gas disc. By making various normalisations and assumptions we are able to fit surface-brightness profiles perpendicular to the galaxy mid-plane using two free parameters, that is the thick-to-thin disc mass density ratio at the mid-plane, and the ratio of vertical velocity dispersions. Those fits yield as a result the ratio between the thick and thin disc masses. The ratio between the disc and the CMC masses is obtained by fitting axial surface-brightness profiles with the superposition of a broken exponential disc function, integrated along the line of sight, and a S\'ersic function.

The vertical surface-brightness profile fits are noticeably affected by our newly adopted PSF model. In our new decompositions the thin disc light close to the galaxy mid-plane is scattered in such a way that thin disc light is the dominant light source at typical surface-brightness levels below $\mu=26\,{\rm mag\,arcsec^{-2}}$. This typically corresponds to heights of $2-8\,{\rm kpc}$ above the mid-plane, but it can be more in some galaxies. Thick disc luminosity still dominates the regions between $\mu\sim26\,{\rm mag\,arcsec^{-2}}$ and $\mu\sim24\,{\rm mag\,arcsec^{-2}}$ or $\mu\sim23\,{\rm mag\,arcsec^{-2}}$. The fact that the thin discs dominate the surface brightness at large heights by no means indicates that they have any significant mass contribution at large heights. Instead it is an artefact caused by light originating from close to the mid-plane and then re-distributed by the extended wings of the PSF.

The relative weights of the thick and thin disc are not substantially changed by the change in the PSF model. The reason is that in our current approach the low surface-brightness tails of the profiles are fitted by thin disc light, which causes the thick disc scale-heights to be smaller than the in our previous papers. This leads in turn to an increased thick disc mid-plane surface brightness. The combination of an increased mid-plane surface brightness and a reduced scale-height results in a total luminosity that is very similar to what was found when using an ``incorrect'' PSF model. The results from our previous papers are not qualitatively affected by the adopted PSF model and remain valid. We however caution against giving detailed descriptions of galaxies based upon over-simplistic PSF models.

We redo some of the experiments that were presented in our previous work. In no cases do our results differ significantly from what was found previously. We thus confirm that:

\begin{itemize}
 \item Thick discs are nearly ubiquitous. Most of the galaxies in our sample show distinct thin and thick discs. Only seventeen galaxies out of 141 show no clear traces of a thick disc. Those seventeen galaxies are on average less massive than those that show at least two discs.
 \item The ratio between the thick and thin disc masses -- $\mathcal{M}_{\rm T}/\mathcal{M}_{\rm t}$ -- depends on the galaxy circular velocity and thus mass. Low-mass galaxies, with $v_{\rm c}<120\,{\rm km\,s^ {-1}}$, have thick discs that are comparable in mass to their thin counterparts. Thick discs in high-mass galaxies are typically less massive than thin discs.
 \item The ratio between the mass of dynamically cold components -- gas and thin discs -- and that of dynamically hot components -- thick discs and CMCs -- is no dependent on $v_{\rm c}$. This was interpreted in \citet{CO14} as evidence for a common origin for the hot component material.
 \item Eight galaxies in our sample require a third stellar disc component -- or maybe a squashed halo -- to fit their vertical surface-brightness profiles down to the S$^4$G sensitivity level.
 \item Many of the up-bending breaks in face-on galaxies -- around $50\%$ -- are caused by the superposition of a thin and a thick disc where the scale-length of the thick disc is longer than that of the thin disc.
\end{itemize}

Our study indicates that the effects of an extended PSF should be taken into account when studying thick discs because it affects the values of fitted parameters in structural decompositions. However, not considering them is unlikely to cause any dramatic qualitative change in the results, at least for instruments that have an exponentially decaying PSF, and at the surface-brightness levels considered here. Since we find that the scattered light from the thin disc becomes dominant below a surface-brightness level of $\mu\sim26\,{\rm mag\,arcsec^{-2}}$ it definitely seems worth revising works studying the integrated light of stellar haloes. Studies at such depths with careful PSF treatment are now being performed \citep[e.g.,][]{TRU16, PE17}.

We can summarise the findings in this paper by misquoting Mark Twain: \begin{quotation}The reports of thick discs' deaths are greatly exaggerated.\end{quotation} 

\begin{acknowledgements}

We acknowledge Bruce Elmegreen for useful discussions. We thank the referee for helping us to polish the results section.

SC and HS acknowledge support from the Academy of Finland.

JHK acknowledges financial support from the European Union’s Horizon 2020 research and innovation programme under Marie Skłodowska-Curie grant agreement No 721463 to the SUNDIAL ITN network, and from the Spanish Ministry of Economy and Competitiveness (MINECO) under grant number AYA2016-76219-P.

We acknowledge the usage of the HyperLeda database (http://leda.univ-lyon1.fr).

This research has made use of the NASA/IPAC Extragalactic Database (NED) which is operated by the Jet Propulsion Laboratory, California Institute of Technology, under contract with the National Aeronautics and Space Administration.

\end{acknowledgements}

\bibliographystyle{aa}
\bibliography{psf}
\clearpage
\onecolumn
\appendix

\section{Description of the information in the Appendices}

In the Appendix we present information on the fits for the 124 galaxies that we considered to have a distinct thin and thick disc (Appendix~\ref{twodiscap}) and the 17 that we considered not to have two distinct discs (Appendix~\ref{onediscap}). The information is structured as follows.

\subsection*{Top-left: galaxy image}

The {\it top-left} panel presents an image of the galaxy under study. The image is oriented so that the mid-plane of the galaxy is horizontal (Sect.~\ref{preliminary}).

The green vertical lines denote the limits of the four axial bins that were used to prepare the surface-brightness profiles perpendicular to the galaxy mid-plane. Those axial bins are found at $0.2\,r_{25}<\left|x\right|<0.5\,r_{25}$ and $0.5\,r_{25}<\left|x\right|<0.8\,r_{25}$  (Sect.~\ref{production}).

The orange horizontal dashed lines indicate the height of the mean $z_{\rm u}$, where the surface brightness is $\mu\sim26\,{\rm mag\,arcsec^{-2}}$ (Sect.~\ref{axial}).

The red horizontal continuous lines denote the limits of the region where most of the light is emitted by the thick disc, between the mean $z_{\rm c1}$ and the mean $z_{\rm c2}$ (Sect.~\ref{zcs}). The upper limit, $z_{\rm c2}$ often overlaps with $z_{\rm u}$.

\subsection*{Top-right: basic galaxy information}

\begin{itemize}

\hrule
\vspace{0.03cm}
\hrule
\vspace{0.1cm}
  \item RA (J2000.0): Right ascension with a J2000.0 epoch (NED).
  \item Dec (J2000.0): Declination with a J2000.0 epoch (NED).
  \item $r_{25}$: Isophotal $25\,{\rm mag\,arcsec^{-2}}$ radius in the $B$-band (HyperLeda).
  \item PA: Position angle of the galaxy major axis (often from HyperLeda, but see Sect.~\ref{preliminary})
  \item $M_{3.6\mu{\rm m}}{\rm (AB)}$: Absolute $3.6\mu{\rm m}$ magnitude of the galaxy in the AB system, obtained using the galaxy distance $d$ and the apparent luminosities in \citet{MU15}, except for the galaxies in the S$^4$G early-type extension where the apparent luminosities were calculated by ourselves.
  \item $d$: Galaxy distance. The sources are {\it 1)} CF1 \citep{TU08}, {\it 2)} CF3 \citet{TU16}, {\it 3)} SFI++ \citet{SPRIN07}, {\it 4)} \citet{THEU07}, and {\it 5)} Hubble-Lema\^itre flow distances with respect to the cosmic microwave background assuming a Hubble-Lema\^itre constant of $H_0=75\,{\rm km\,s^{-1}}$. The recession velocities are taken from the NED.
  \item $v_{\rm c}$: Circular velocity. The sources are {\it 1)} \citet{SPRIN05}, {\it 2)} \citet{THEU06}, {\it 3)} \citet{COUR09}, {\it 4)} pre-digital measurements compiled in the EDD, {\it 5)} \citet{KRUMM76}, {\it 6)} \citet{BAL81}, {\it 7)} \citet{RICH87}, {\it 8)} \citet{HAY90}, {\it 9)} \citet{DON95}, {\it 10)} \citet{MATH96}, {\it 11)} \citet{SI97}, {\it 12)} \citet{SI98}, {\it 13)} \citet{RU99},  {\it 14)} \citet{SI00}, {\it 15)} \citet{KAR04}, {\it 16)} \citet{CHUNG04}, {\it 17)} \citet{MEY04}, {\it 18)} \citet{BED06}, and {\it 19)} ATLAS3D \citep{CA11, KRA11}.
\hrule
\vspace{0.03cm}
\hrule
\vspace{0.1cm}
  \item Mean $z_{\rm c1}$: lower limit of the region were most of the light is emitted by the thick disc according to our fits (Sect.~\ref{zcs}).
  \item Mean $z_{\rm c2}$: higher limit of the region were most of the light is emitted by the thick disc according to our fits (Sect.~\ref{zcs}).
  \item Mean $z_{\rm u}$: height where the surface brightness is $\mu=26\,{\rm mag\,arcsec^{-2}}$ (Sect.~\ref{axial}).
\hrule
\end{itemize}

\subsection*{Second and third rows: Vertical surface-brightness profile fits}

Each of the plots displays the vertical surface-brightness profiles of one of the four axial bins indicated with the green lines in the {\it top-left} image (large filled circles). The extent of the axial bin is identified by the label on the top-left corner of the panel.

The dotted curve indicates the fit to the profile (Sect.~\ref{finalvertical}). The short-dashed curves indicate the thin and thick disc contributions and the long-dashed curve indicates the CMC contribution. 

The vertical dot-dashed grey line indicates the range of heights considered for each fit.

The dashed horizontal lines indicate the height, $z_{\rm u}$, where the surface brightness is $\mu=26\,{\rm mag\,arcsec^{-2}}$. This height is averaged over all the axial bins with a valid fit to obtain the mean or global $z_{\rm u}$ shown in the {\it top-right} table and indicated in the {\it top-left} image with orange horizontal dashed (Sect.~\ref{axial}). This height is sometimes found outside of the frame.

The red vertical continuous lines denote the limits of the region where most of the light is emitted by the thick disc, between the $z_{\rm c1}$ and the $z_{\rm c2}$. Those heights are averaged over all the axial bins with a valid fit to obtain the mean or global values for $z_{\rm c1}$ and $z_{\rm c2}$ shown in the {\it top-right} table and indicated in the {\it top-left} image with red horizontal continuous lines (Sect.~\ref{zcs}).

The right half of the panels contains of the numbers derived from the fits: 
\begin{itemize}
\item $\left(\left.\rho_{\rm T}\right|_{z=0}\right)/\left(\left.\rho_{\rm t}\right|_{z=0}\right)$: Fitted ratio of the thick and the thin disc mid-plane mass densities.
\item $\sigma_{\rm T}/\sigma_{\rm t}$: Fitted ratio of the thick and thin disc vertical velocity dispersions.
\item $z_{\rm T}$: Thick disc scale-height.
\item $z_{\rm t}$: Thin disc scale-height.
\item $\Sigma_{\rm T}/\Sigma_{\rm t}$: Thick to thin disc mass ratio in that axial bin.
\item $\mu_{\rm l}$: Lowest surface brightness level considered for the fit.
\item $z_{\rm l}$: Largest height considered for the fit.
\item $z_{\rm u}$: Height where the fitted surface brightness is $\mu\sim26\,{\rm mag\,arcsec^{-2}}$.
\item $z_{\rm c1}$ and $z_{\rm c2}$: Lower and upper limits of the region where most of the light is emitted by the thick disc.
\item $\Delta\mu$: Dynamical range of the fit.
\item $\mu_{\rm rms}$: Root mean square deviation of the fit.
\end{itemize}

Fits that do not fulfil our quality criteria (Sect.~\ref{actual}) are labelled as ``Not used'' with large grey letters. A few axial bins where no profile could be obtained due to bright foreground stars are labelled with ``Bright star''.

\subsection*{Fourth row: Axial surface-brightness profile fits (Sect.~\ref{finalaxial})}

The plots show the axial surface-brightness profiles for the whole height of the galaxy (yellow squares), the thin disc-dominated heights (blue circles), and the thick disc-dominated heights (red triangles).

The green vertical lines denote the limits of the four axial bins that were used to prepare the surface-brightness profiles perpendicular to the galaxy mid-plane. Those axial bins are found at $0.2\,r_{25}<\left|x\right|<0.5\,r_{25}$ and $0.5\,r_{25}<\left|x\right|<0.8\,r_{25}$.

The horizontal lines indicate the $\mu=27\,{\rm mag\,arcsec^{-2}}$ level.

The fitted disc and CMC contributions are indicated with dashed curves in the {\it left}. The CMC contribution is not marked in the {\it right} plot to avoid over-crowding. The small vertical black lines indicates the break radii. The red vertical lines indicate the outermost fitted point.

The grey region indicates the range of estimates of thick disc contributions at thin disc dominated heights based on the methods described in Sect.~\ref{truncations}.

\subsection*{First table in the fifth row: CMC properties (Sects.~\ref{axialfunctions} and \ref{axialactual})}

\begin{itemize}
\hrule
\vspace{0.1cm}
 \item $r_{\rm CMC}$: Scale size of the CMC.
 \item $n$: S\'ersic index.
\vspace{0.1cm}
\hrule
\end{itemize}

\subsection*{Second, third, and fourth tables in the fifth row: disc properties (Sects.~\ref{axialfunctions}, \ref{axialactual}, \ref{finalaxial}, and \ref{truncations})}

\begin{itemize}
\hrule
\vspace{0.1cm}
 \item Type: Galaxy axial luminosity profile type according to the \citet{FREE70}, \citet{POH06}, and \citet{ER08} classification scheme.
 \hrule
\vspace{0.1cm}
 \item $h_i$: scale-length of the $i^{\rm th}$ exponential section. The sections are counted starting at the centre of the galaxy.
 \hrule
\vspace{0.1cm}
\item $r_{i,i+1}$: radius of the break between the exponential section $i$ and the exponential section $i+1$.
\vspace{0.1cm}
\hrule
\end{itemize}

\subsection*{Fifth table in the fifth row: component masses (Sect.~\ref{components})}

\begin{itemize}
\hrule
\vspace{0.1cm}
 \item $\mathcal{M}_{\rm g}$: Gas disc mass.
 \item $\mathcal{M}_{\rm t}$: Thin disc mass.
 \item $\mathcal{M}_{\rm T}$: Thick disc mass.
 \item $\mathcal{M}_{\rm CMC}$: CMC mass.
\vspace{0.1cm}
\hrule
\end{itemize}

\subsection*{Sixth table in the fifth row: disc scale-heights (Sect.~\ref{scales})}

\begin{itemize}
\hrule
\vspace{0.1cm}
 \item $z_{\rm t}$: Thin disc scale-height.
 \item $z_{\rm T}$: Thick disc scale-height.
\vspace{0.1cm}
\hrule
\end{itemize}

\clearpage

\section{Galaxies that have a thin and a thick disc}
\label{twodiscap}

\clearpage

\newpage
\begin{center}
\begin{tabular}{c c}
\raisebox{-0.5\height}{\includegraphics[width=0.5\textwidth]{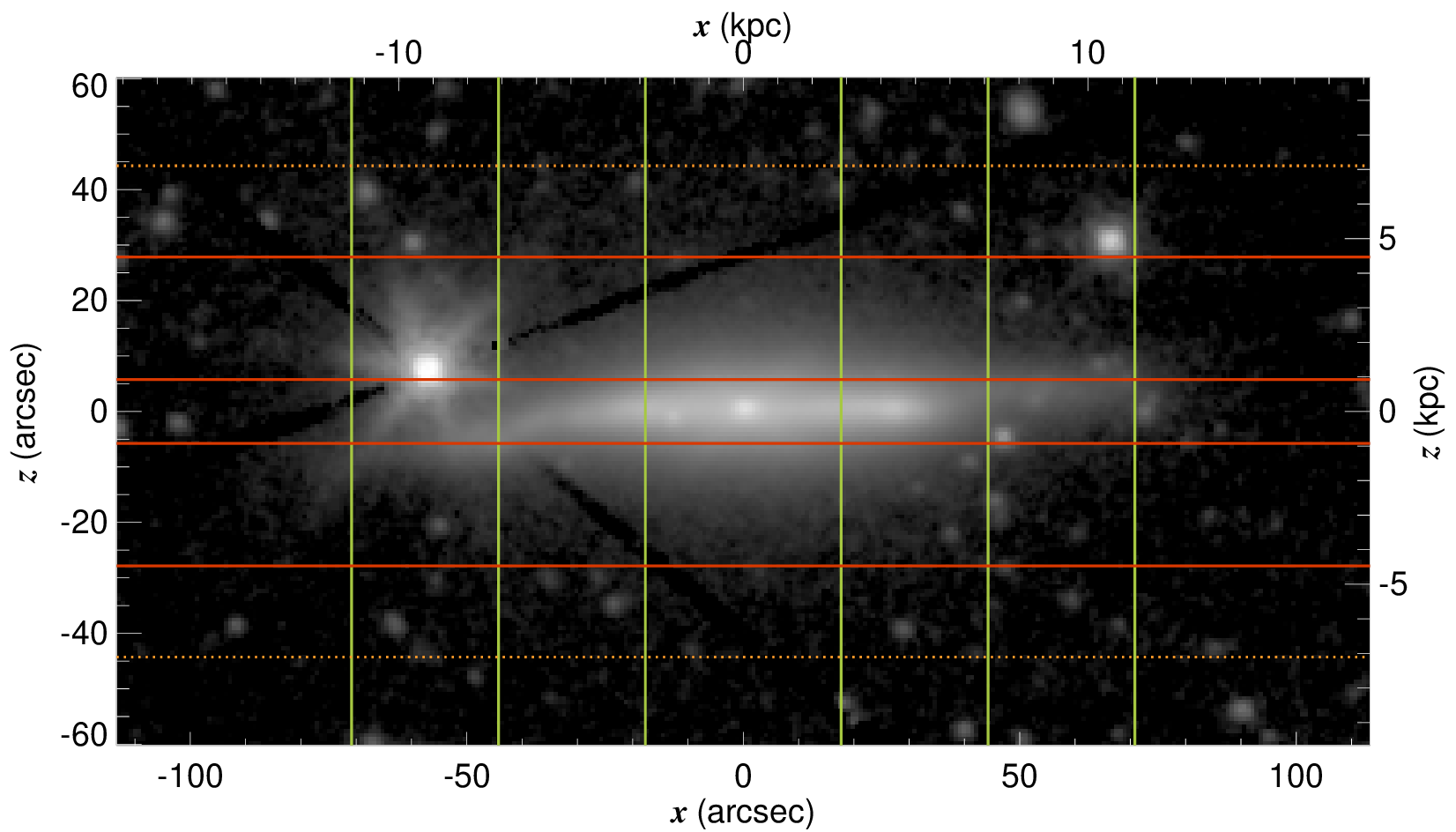}}
\begin{tabular}{l c}
\multicolumn{2}{c}{\huge{ESO~79-3}}\\
\hline\hline
RA (J2000.0) &$00^{\rm h}32^{\rm m}02\fs17$\\
Dec (J2000.0)& $-64\degr15\arcmin12\farcs2$\\
$r_{25}$&88\farcs5\\
PA&130\degr2\\
$M_{3.6\mu{\rm m}}({\rm AB})$&-20.995\\
$d$&33.10\,Mpc\,\,\,(1)\\
$v_{\rm c}$&194$\,{\rm km\,s^{-1}}$\,\,\,(10)\\
\hline\hline
Mean $z_{\rm c1}$&5\farcs8\\
Mean $z_{\rm c2}$&27\farcs8\\
Mean $z_{\rm u}$&44\farcs3\\
\hline
\end{tabular}
\end{tabular}
\vspace{-0.1cm}
\end{center}
\begin{center}
\begin{tabular}{c c}
\includegraphics[width=0.355\textwidth]{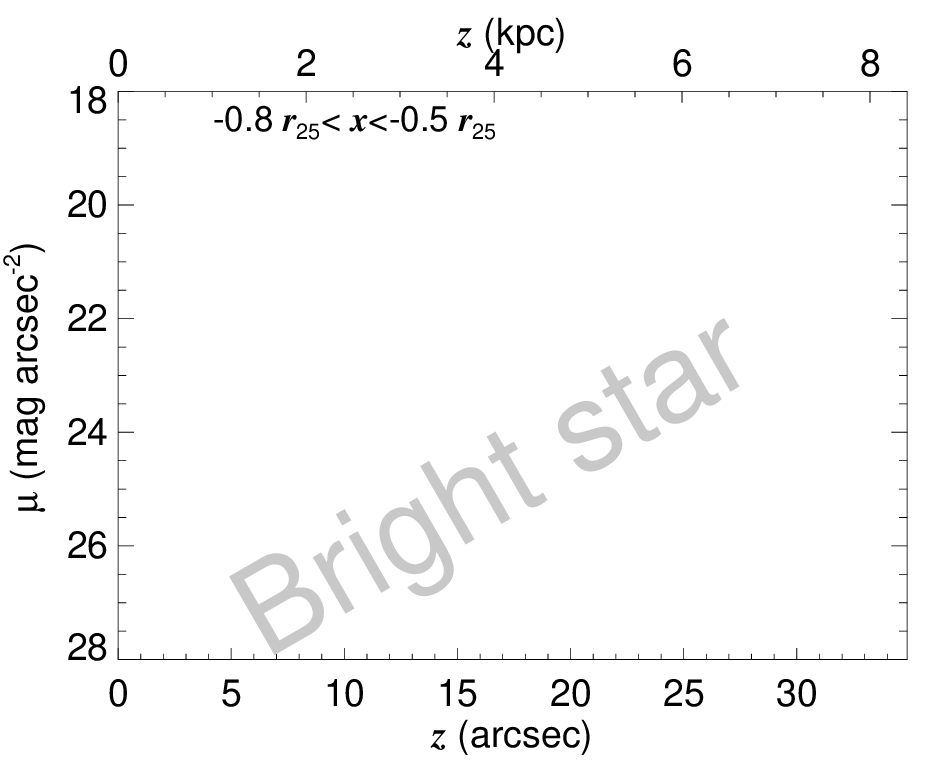}
\includegraphics[width=0.355\textwidth]{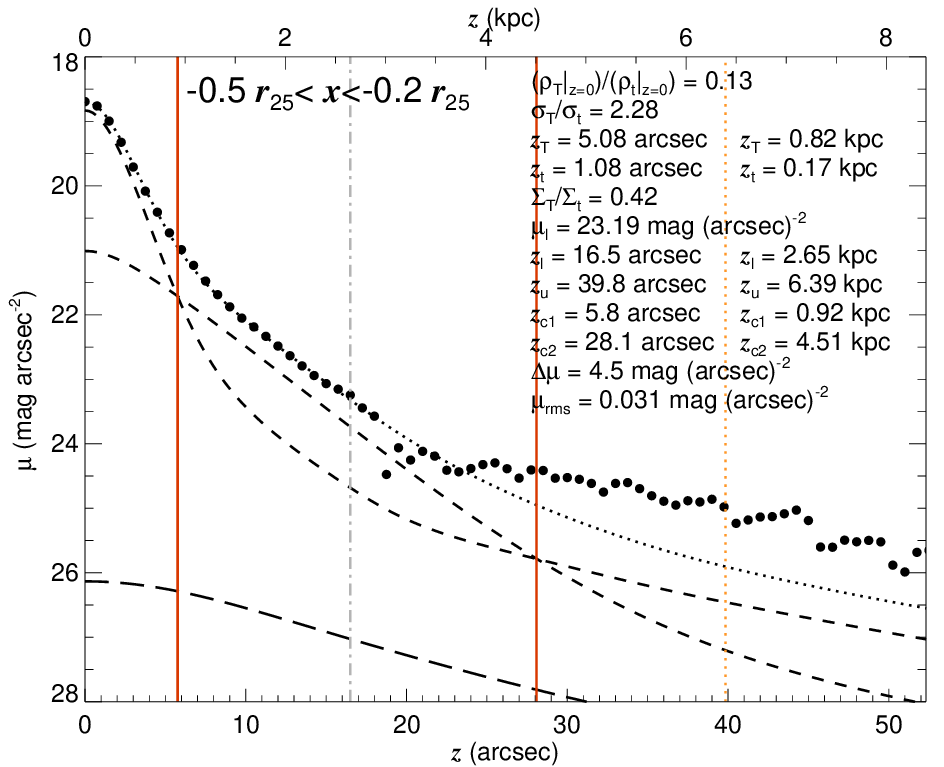}
\vspace{-0.1cm}
\\
\includegraphics[width=0.355\textwidth]{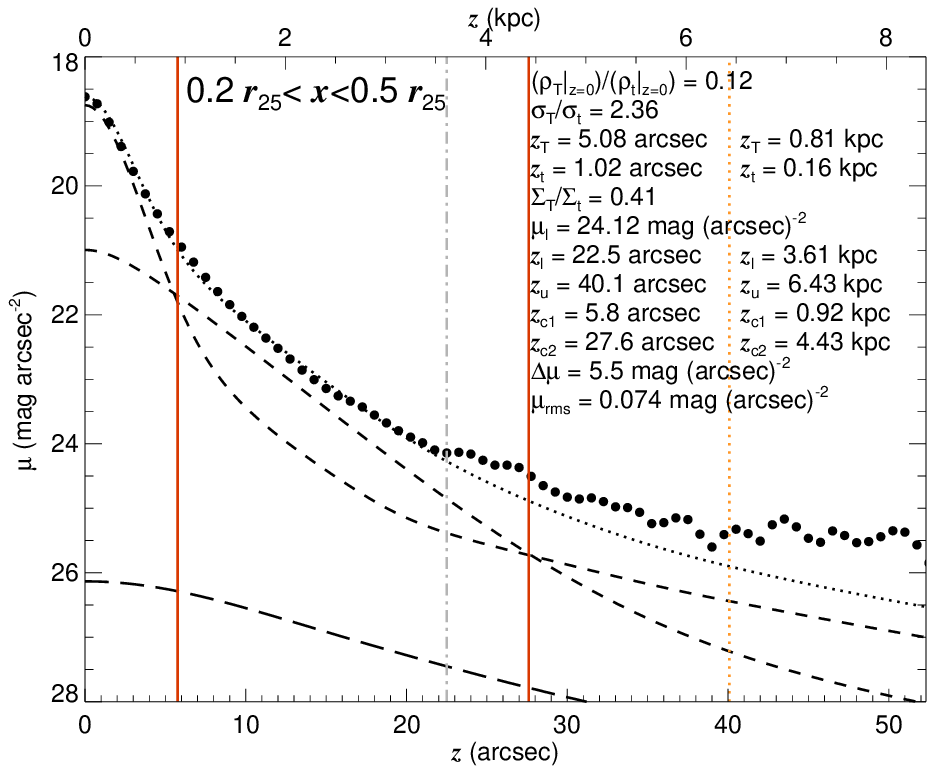}
\includegraphics[width=0.355\textwidth]{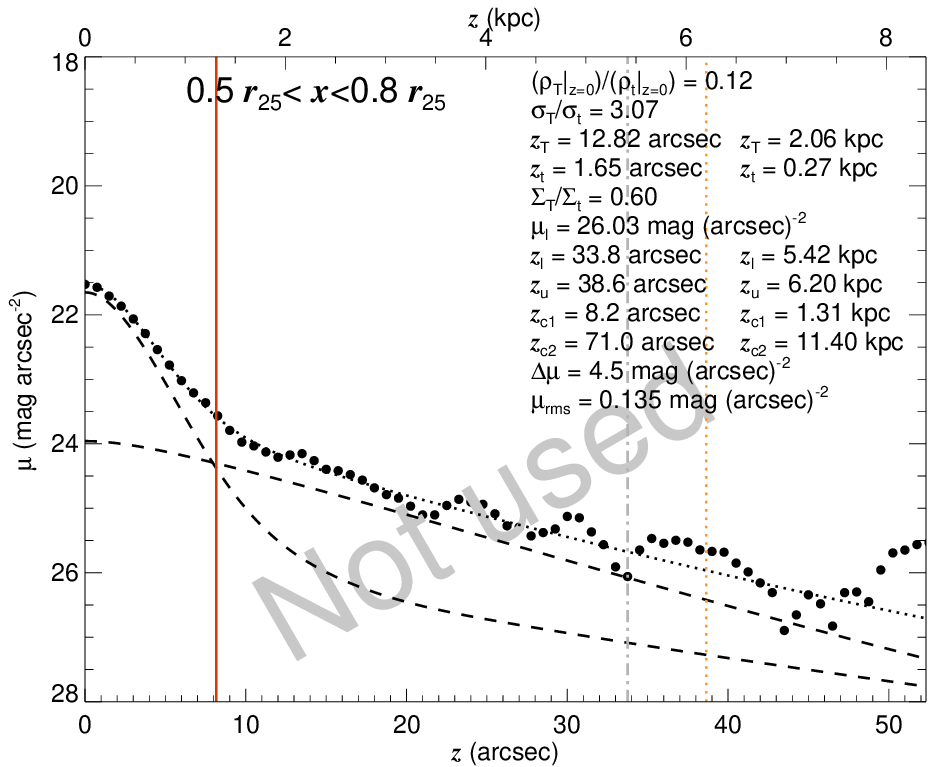}
\end{tabular}
\\
\vspace{-0.2cm}
\begin{tabular}{c c}
\includegraphics[width=0.45\textwidth]{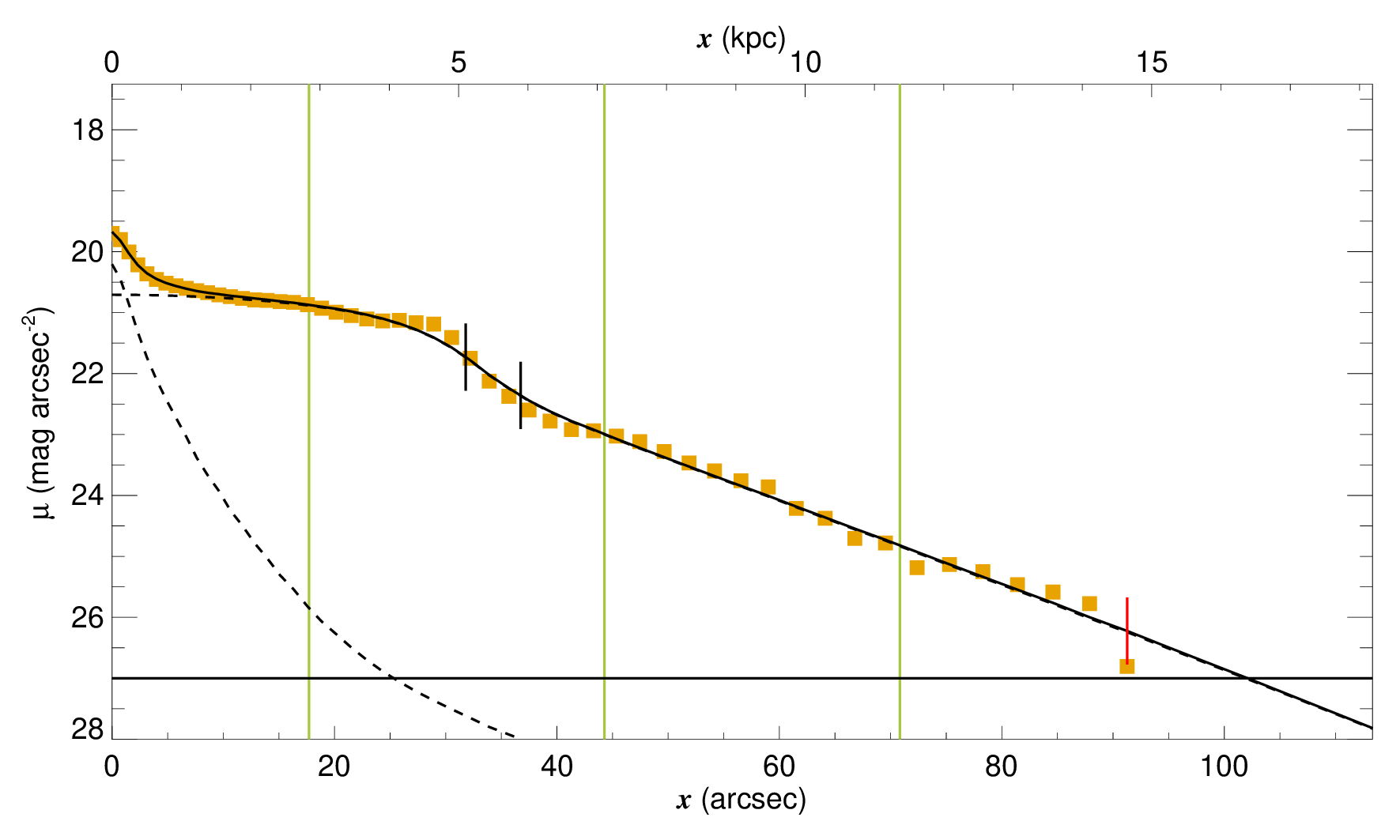}&
\includegraphics[width=0.45\textwidth]{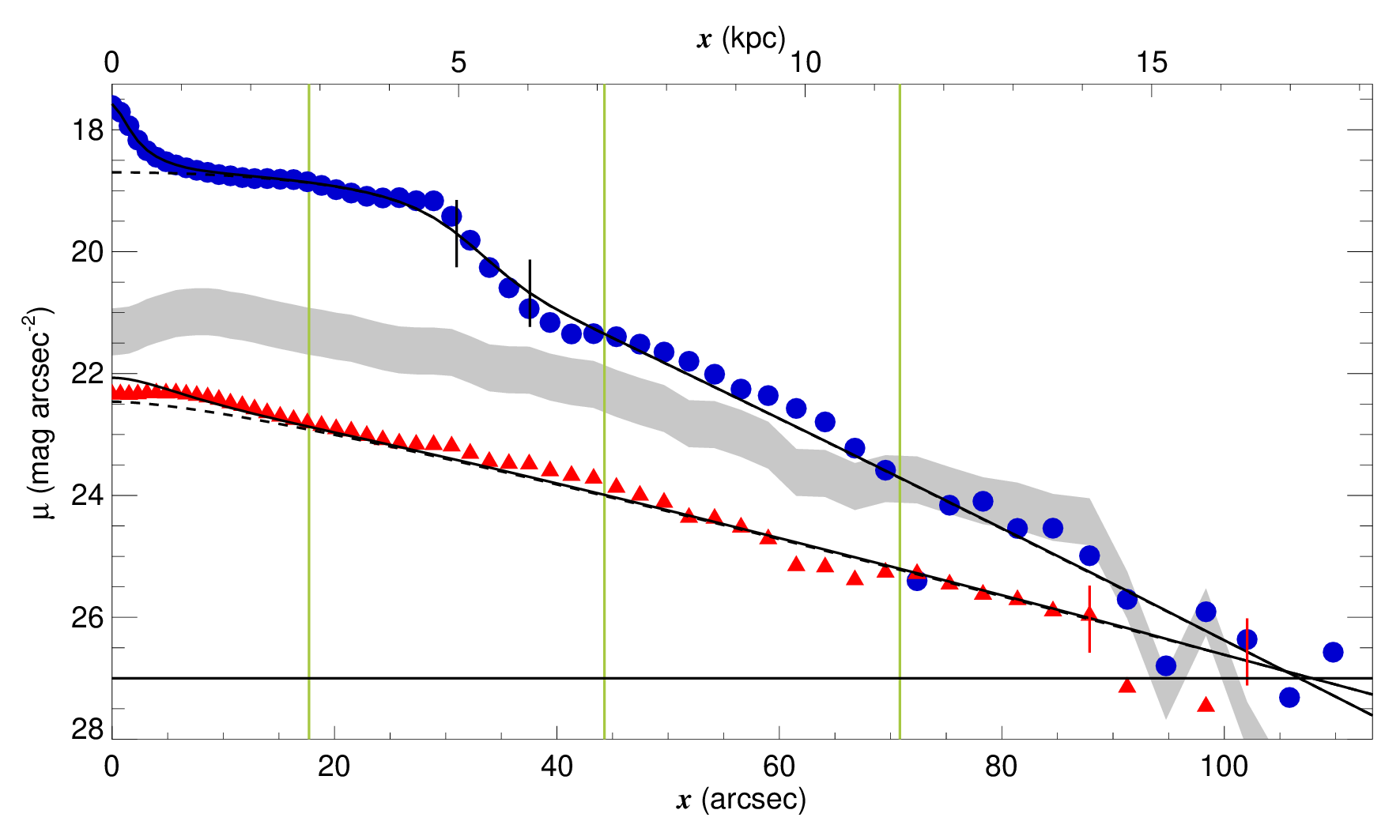}
\end{tabular}
\end{center}
\begin{center}
\vspace{-0.4cm}
\begin{tabular}{c c c c c c}
\begin{tabular}{c}
\hline\hline
CMC\\
properties\\
\hline
$r_{\rm CMC}=          0\farcs23$\\
$n=1.94$\\
\hline
\end{tabular}
\begin{tabular}{c}
\hline\hline
Total\\
profile\\
\hline
Type~II+III\\
\hline
$h_{      1}=        450\farcs5$\\
$h_{      2}=          3\farcs2$\\
$h_{      3}=         13\farcs6$\\
\hline
$r_{      1      2}=         31\farcs8$\\
$r_{      2      3}=         36\farcs7$\\
\hline
\end{tabular}
\begin{tabular}{c}
\hline\hline
Thin disc\\
profile\\
\hline
Type~II+III\\
\hline
$h_{      1}=       2052\farcs0$\\
$h_{      2}=          3\farcs9$\\
$h_{      3}=         11\farcs2$\\
\hline
$r_{      1      2}=         31\farcs0$\\
$r_{      2      3}=         37\farcs6$\\
\hline
\end{tabular}
\begin{tabular}{c}
\hline\hline
Thick disc\\
profile\\
\hline
Type~I\\
\hline
$h_{      1}=         20\farcs4$\\
\hline
\end{tabular}
\begin{tabular}{c}
\hline\hline
Component masses\\
\hline
$\mathcal{M}_{\rm g}=9.09\times10^9\,\mathcal{M}_{\bigodot}$\\
$\mathcal{M}_{\rm t}=23.60\times10^9\,\mathcal{M}_{\bigodot}$\\
$\mathcal{M}_{\rm T}=9.75\times10^9\,\mathcal{M}_{\bigodot}$\\
$\mathcal{M}_{\rm CMC}=5.74\times10^9\,\mathcal{M}_{\bigodot}$\\
\hline
\end{tabular}
\begin{tabular}{c}
\hline\hline
Disc\\
scale-heights\\
\hline
$z_{\rm t}=1\farcs0$\\
$z_{\rm T}=5\farcs1$\\
\hline
\end{tabular}
\end{tabular}
\end{center}

\clearpage

\section{Galaxies that might have a single disc}
\label{onediscap}

\clearpage

\newpage
\begin{center}
\begin{tabular}{c c}
\raisebox{-0.5\height}{\includegraphics[width=0.5\textwidth]{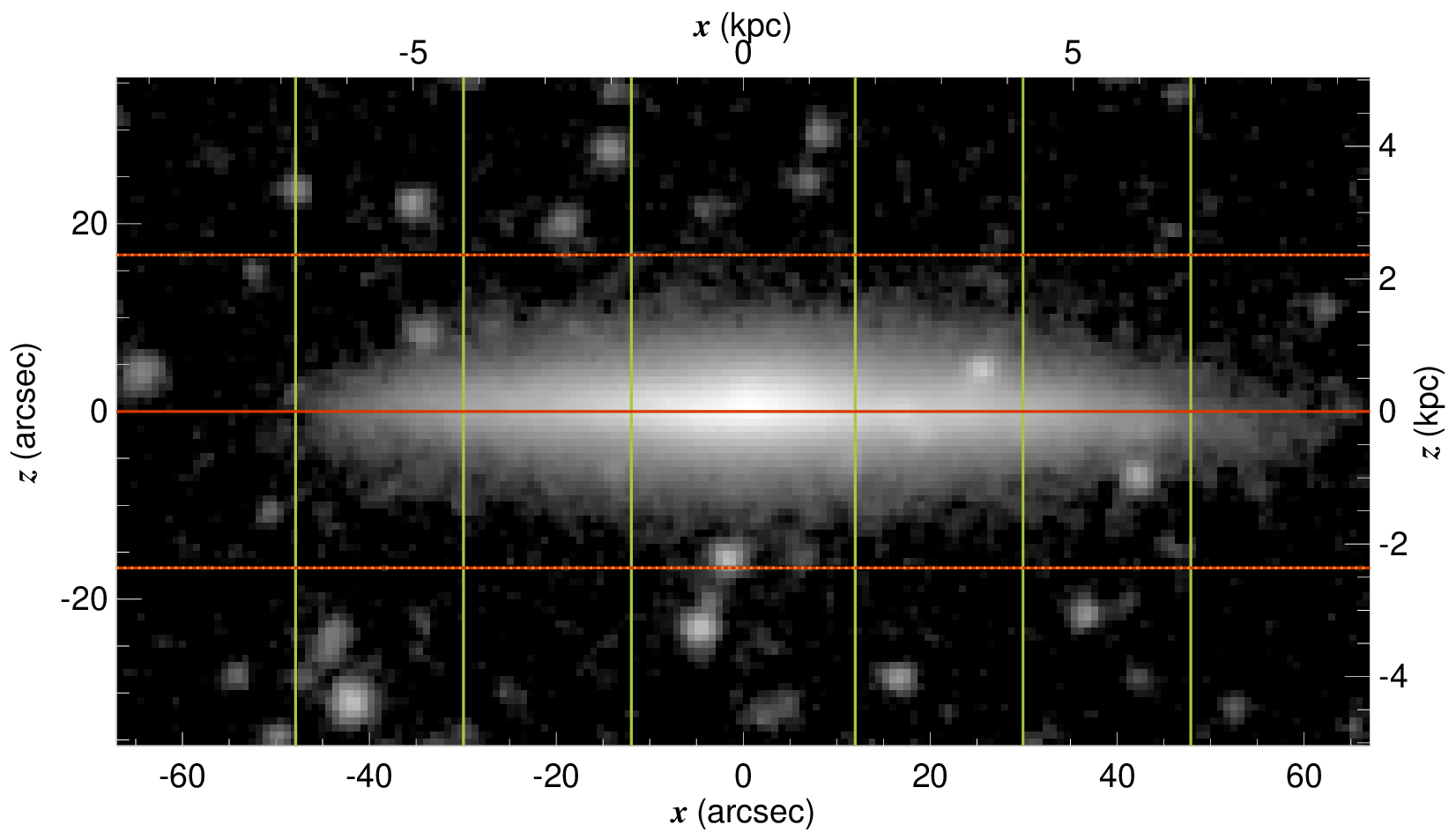}}
\begin{tabular}{l c}
\multicolumn{2}{c}{\huge{ESO~287-9}}\\
\hline\hline
RA (J2000.0) &$21^{\rm h}21^{\rm m}16\fs30$\\
Dec (J2000.0)& $-46\degr09\arcmin10\farcs0$\\
$r_{25}$&59\farcs9\\
PA&105\degr2\\
$M_{3.6\mu{\rm m}}({\rm AB})$&-18.550\\
$d$&29.20\,Mpc\,\,\,(1)\\
$v_{\rm c}$&123$\,{\rm km\,s^{-1}}$\,\,\,(4)\\
\hline\hline
Mean $z_{\rm c1}$&0\farcs0\\
Mean $z_{\rm c2}$&16\farcs7\\
Mean $z_{\rm u}$&16\farcs7\\
\hline
\end{tabular}
\end{tabular}
\vspace{-0.1cm}
\end{center}
\begin{center}
\begin{tabular}{c c}
\includegraphics[width=0.355\textwidth]{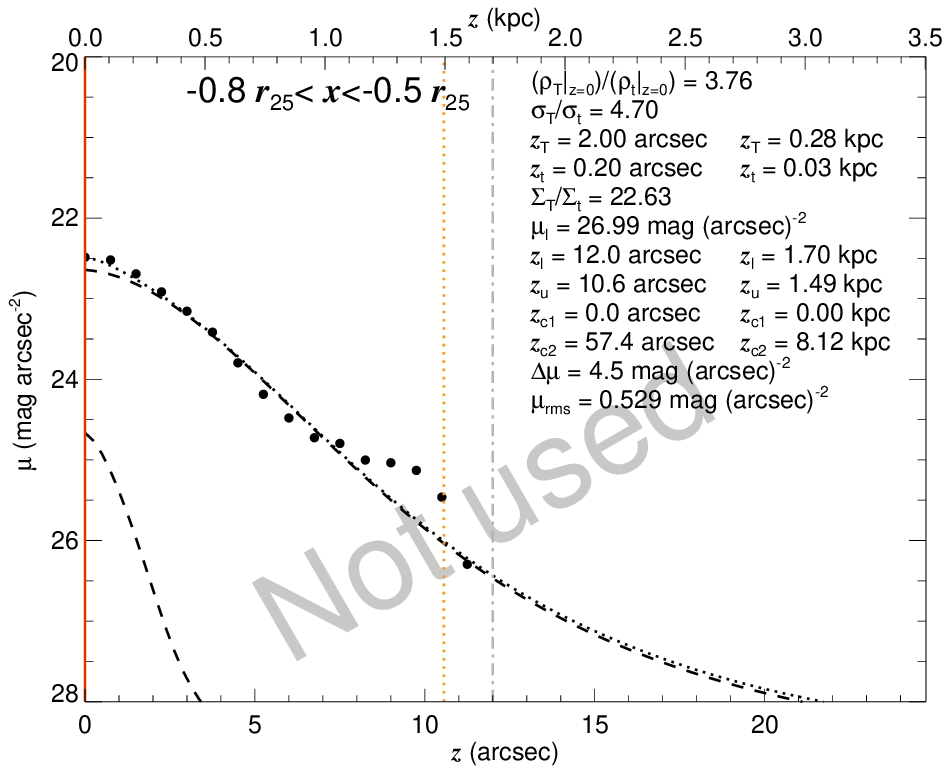}
\includegraphics[width=0.355\textwidth]{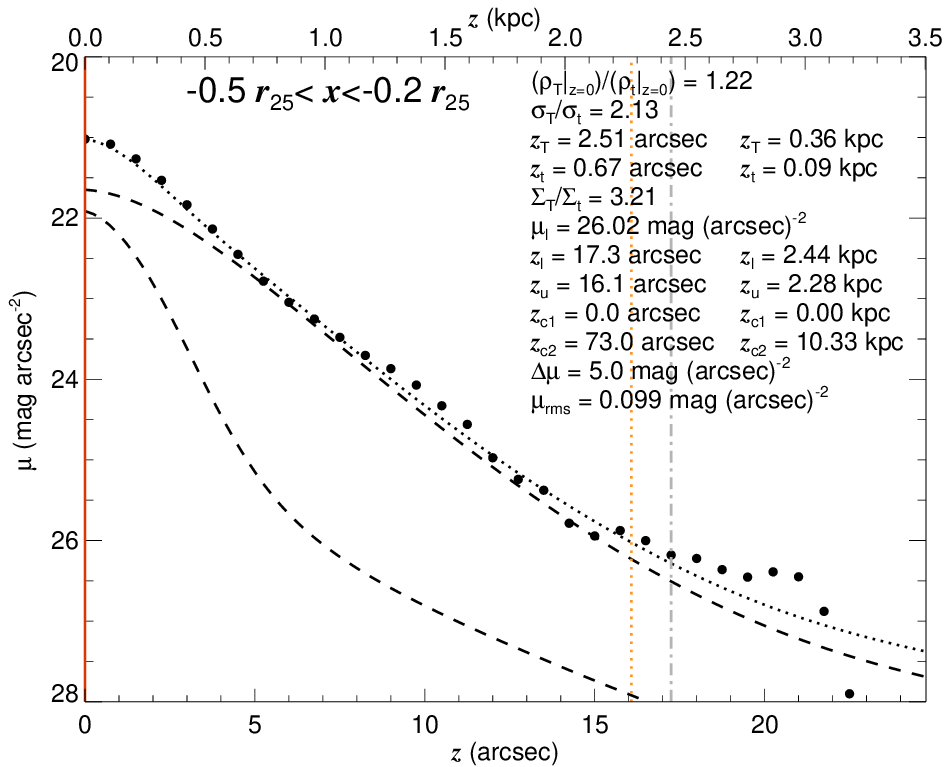}
\vspace{-0.1cm}
\\
\includegraphics[width=0.355\textwidth]{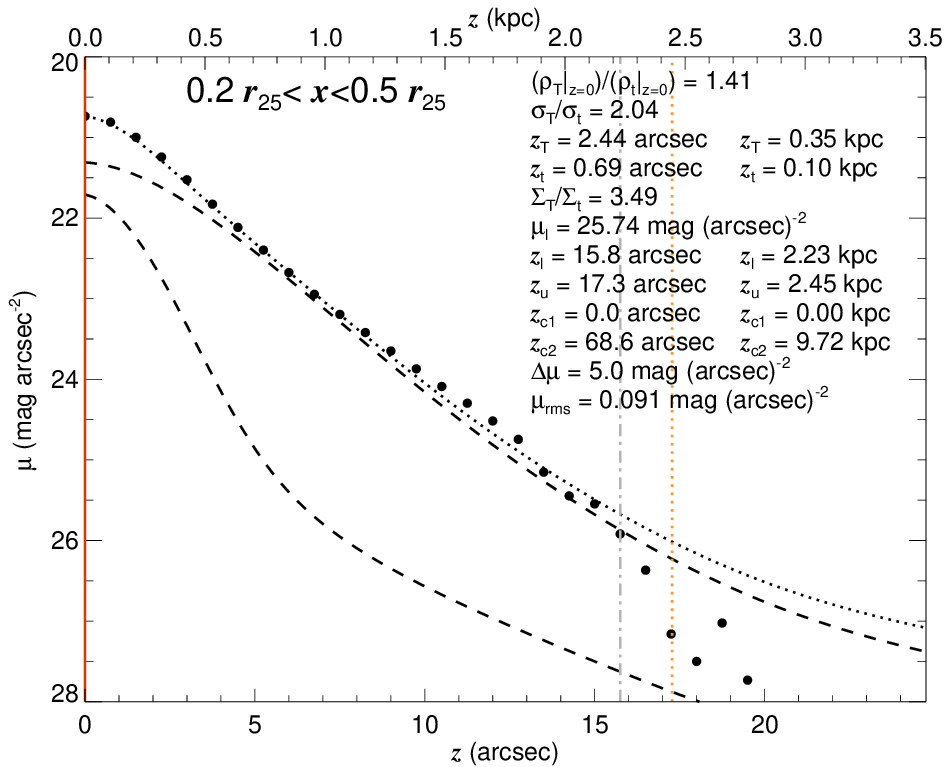}
\includegraphics[width=0.355\textwidth]{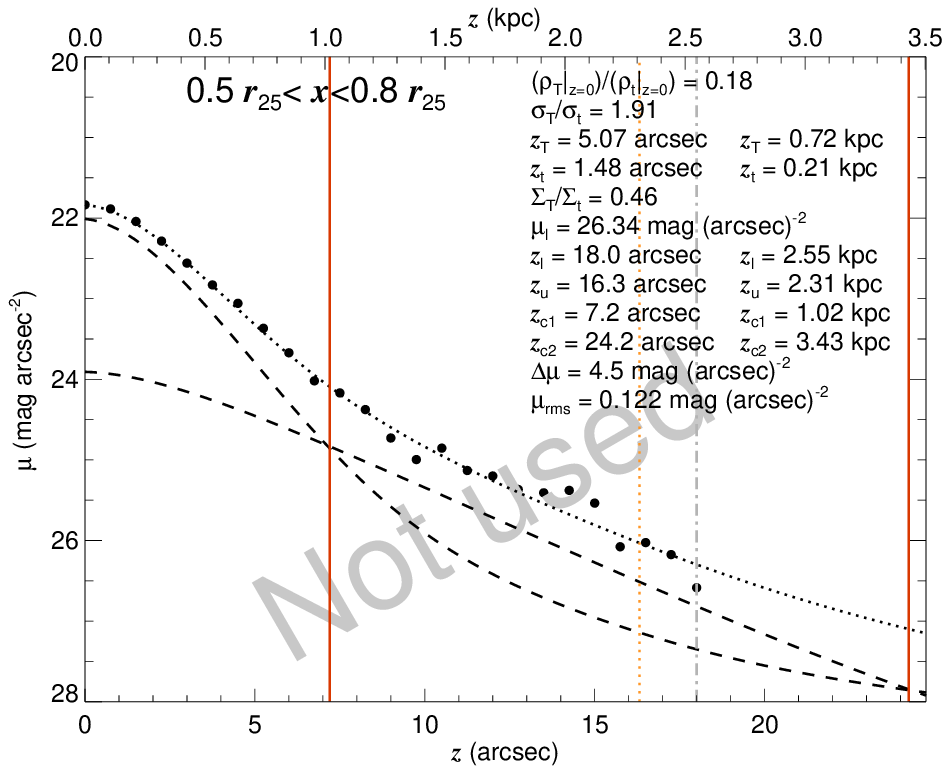}
\end{tabular}
\\
\vspace{-0.2cm}
\begin{tabular}{c c}
\includegraphics[width=0.45\textwidth]{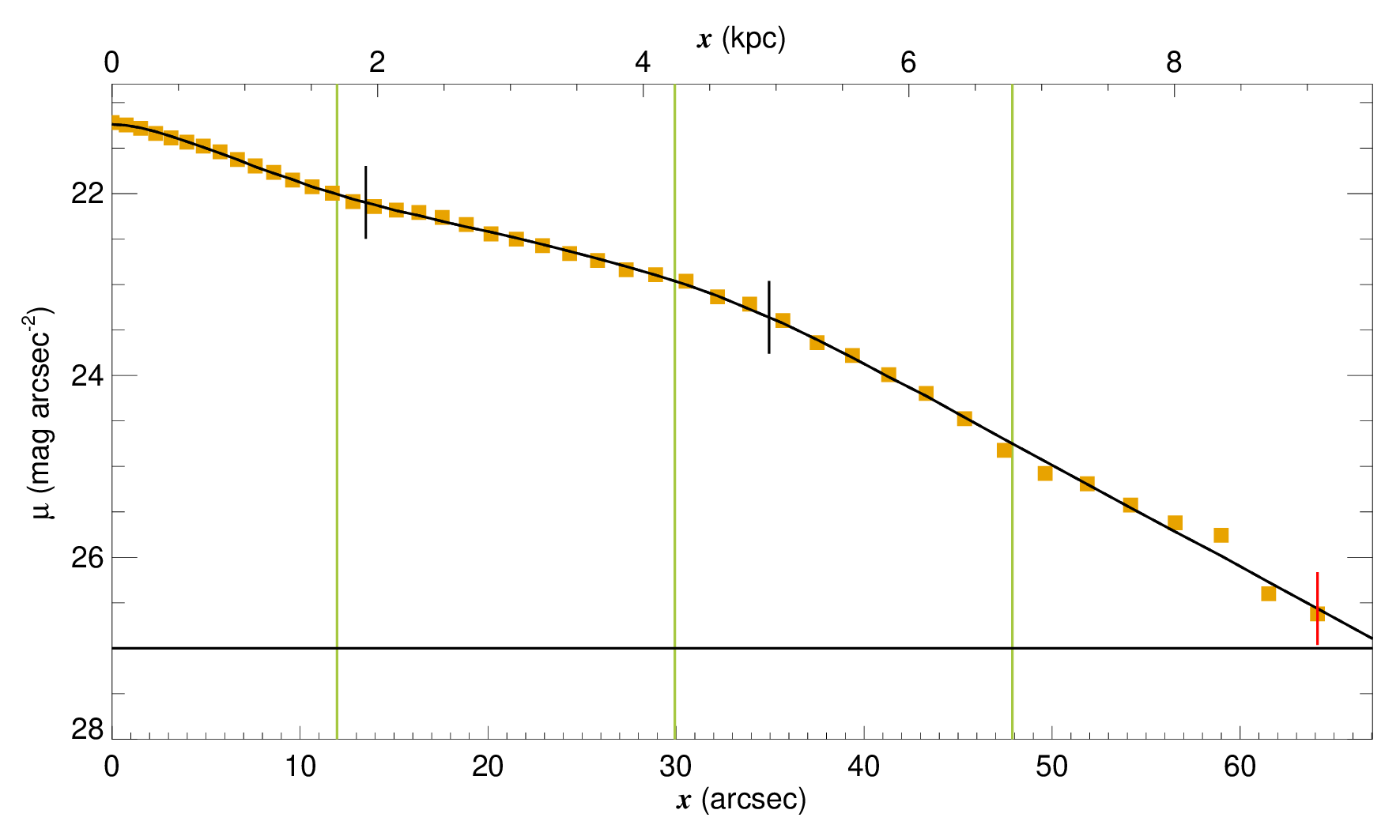}&
\includegraphics[width=0.45\textwidth]{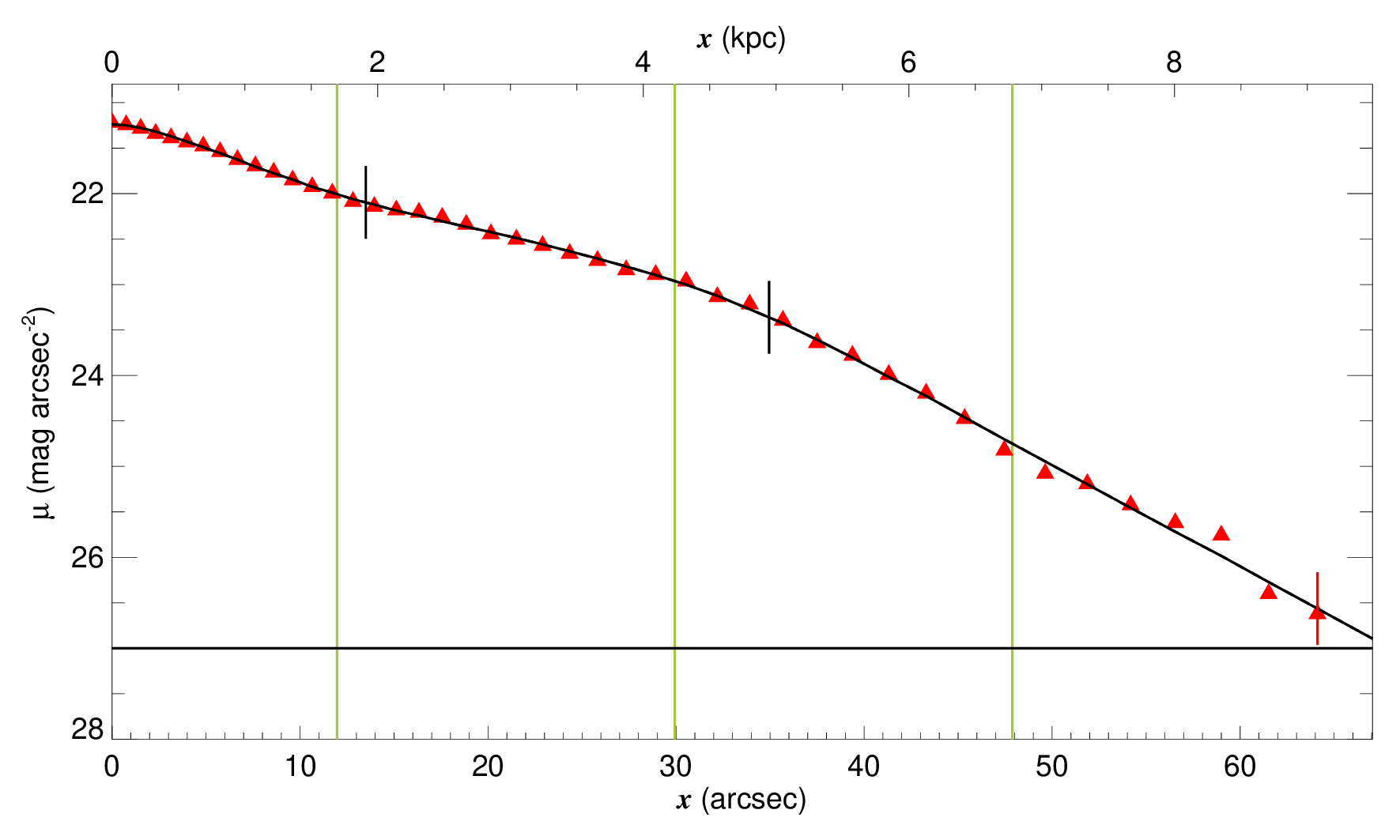}
\end{tabular}
\end{center}
\begin{center}
\vspace{-0.4cm}
\begin{tabular}{c c c c c c}
\begin{tabular}{c}
\hline\hline
CMC\\
properties\\
\hline
No CMC\\
\hline
\end{tabular}
\begin{tabular}{c}
\hline\hline
Total\\
profile\\
\hline
Type~III+II\\
\hline
$h_{      1}=          6\farcs5$\\
$h_{      2}=         21\farcs9$\\
$h_{      3}=          8\farcs6$\\
\hline
$r_{      1      2}=         13\farcs5$\\
$r_{      2      3}=         34\farcs9$\\
\hline
\end{tabular}
\begin{tabular}{c}
\hline\hline
Thin disc\\
profile\\
\hline
No profile\\
\hline
\end{tabular}
\begin{tabular}{c}
\hline\hline
Thick disc\\
profile\\
\hline
Type~III+II\\
\hline
$h_{      1}=          6\farcs5$\\
$h_{      2}=         21\farcs9$\\
$h_{      3}=          8\farcs6$\\
\hline
$r_{      1      2}=         13\farcs5$\\
$r_{      2      3}=         34\farcs9$\\
\hline
\end{tabular}
\begin{tabular}{c}
\hline\hline
Component masses\\
\hline
$\mathcal{M}_{\rm g}=4.30\times10^9\,\mathcal{M}_{\bigodot}$\\
$\mathcal{M}_{\rm t}=1.01\times10^9\,\mathcal{M}_{\bigodot}$\\
$\mathcal{M}_{\rm T}=3.40\times10^9\,\mathcal{M}_{\bigodot}$\\
$\mathcal{M}_{\rm CMC}\sim0$\\
\hline
\end{tabular}
\begin{tabular}{c}
\hline\hline
Disc\\
scale-heights\\
\hline
$z_{\rm t}=0\farcs7$\\
$z_{\rm T}=2\farcs5$\\
\hline
\end{tabular}
\end{tabular}
\end{center}

\end{document}